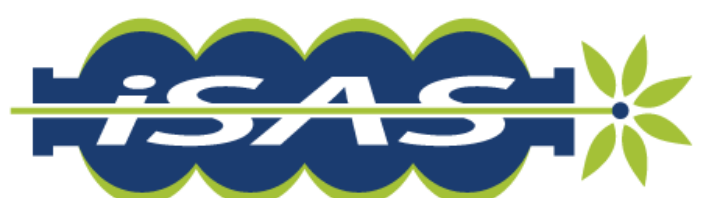

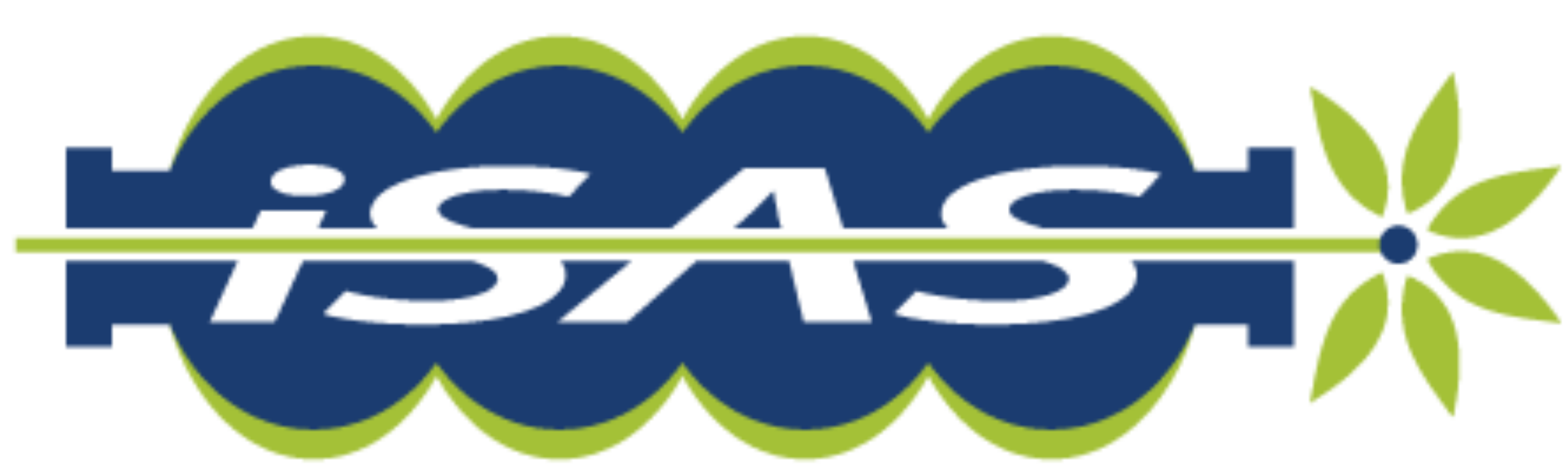

innovate for Sustainable Accelerating Systems

# D 16: Compilation of lessons learned from ESS & Benchmarking with projects in implementation phase

**Authors: K. Canderan, N. Elias, V. Parma, P. Pierini, H. Przibliski, E. Simpson, M. Skiba, M. Baylac, J. D'Hondt, J. Knobloch, C. Pira, A. Stocchi**

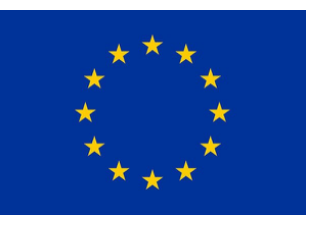

Funded by the European Union. Views and opinions expressed are however those of the authors only and do not necessarily reflect those of the European Union or the European Commission. Neither the European Union nor the granting authority can be held responsible for them.

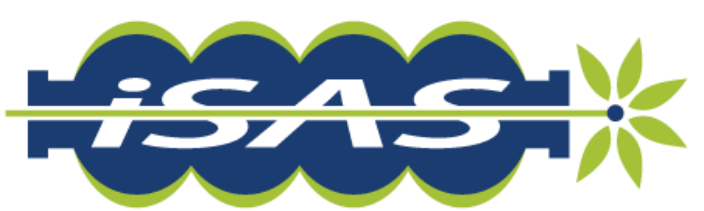

# DOCUMENT TRACKS DETAILS

| Deliverable Number | D 16 | Lead Beneficiary | ESS, CERN |
|---|---|---|---|
| Deliverable Name | Compile lessons learned from ESS and Benchmarking with Projects in implementation phase | | |
| Type | R — report | Dissemination Level | PU – Public, |
| Due Date | Project month 24 | Work Package No | WP5 |

## Deliverable Abstract

This report consolidates two complementary studies conducted within iSAS Work Package 5 (WP5) to support the development of a sustainable next-generation cryomodule.

The first study analyses the design, industrialization, commissioning, and early operation of the elliptical cryomodules developed for the European Spallation Source (ESS), capturing technical and procedural lessons learned across the full engineering lifecycle. It highlights the impact of standardization, distributed collaboration under the In-Kind Contribution model, cryogenic performance at 2 K, and industrial scalability as key enablers for long-term sustainability.

The second study benchmarks high-beta cryomodules across six leading international facilities—Spallation Neutron Source, Large Hadron Collider, PIP-II, European XFEL, Linac Coherent Light Source II, and ESS—through a structured comparative methodology covering beam and RF parameters, cryogenics, mechanical design, commissioning, and operational experience.

Together, these analyses establish a validated technical baseline and an international reference framework, forming the foundation for a WP5 roadmap toward a high-efficiency, sustainable "Cryomodule of the Future."

ANNEX I refers to the detailed report on compilation of lessons learned from ESS
ANNEX II refers to the detailed report on benchmarking of cryomodule designs for iSAS

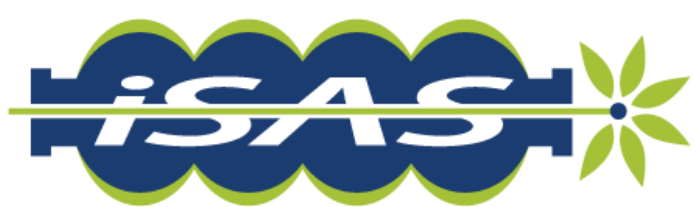

# 1) Introduction

Within the framework of the iSAS initiative, Work Package 5 (WP5) is tasked with defining and documenting a comprehensive set of requirements and technical solutions for integrating emerging technologies into a sustainable cryomodule. Establishing a solid and evidence-based foundation for this effort requires both an in-depth understanding of existing state-of-the-art systems and a structured comparison of leading international designs.

This summary consolidates the outcomes of two complementary studies conducted under WP5.

The first study, "**Evaluation of the ESS Cryomodules Design and Operation – Lessons Learned from the Current Cryomodule Technology**", examines the lifecycle of the elliptical cryomodules developed for the European Spallation Source (ESS). As ESS advances toward its goal of becoming the world's most powerful neutron source, its linear accelerator relies on sophisticated superconducting radio-frequency technology.

The study captures institutional knowledge gained throughout conceptual design, prototyping, industrialization, assembly, testing, commissioning, and initial operation, focusing primarily on the High-Beta cryomodules while referencing their shared design with the Medium-Beta systems. Attention is given to the challenges of scaling from laboratory assembly to series production, stakeholder integration within the In-Kind Contribution framework across multiple European partners, and operational performance at 2 K, including cryogenic stability, RF conditioning, and control-system reliability. The ESS experience provides a validated reference model and a practical framework for sustainability, standardization, and industrial scalability in future accelerator infrastructures.

The second study, "**Benchmarking Cryomodule Designs for iSAS WP5**", complements this internal analysis by positioning the ESS design within a broader international landscape. A systematic benchmarking exercise was conducted across six leading facilities: Spallation Neutron Source (**SNS**), Large Hadron Collider (**LHC**), Proton Improvement Plan II (**PIP-II)**, European XFEL (**XFEL**), Linac Coherent Light Source II (**LCLS-II**), and ESS.

Through a structured 17-step methodology, the benchmarking study analyses high-beta cryomodules across five key domains: beam and RF parameters, cryogenic systems and thermal management, cryomodule architecture and internal components, commissioning experience, and operational performance and availability. This comparative approach highlights how differing requirements, constraints, and strategic priorities shape design choices and lead to distinct technical configurations.

Together, these studies provide a dual perspective: a detailed case study of a large-scale European implementation and a global benchmarking assessment of alternative design philosophies. The lessons learned demonstrate the importance of standardization, modularization, industrial readiness, and collaborative governance models in achieving technical excellence and long-term sustainability. Meanwhile, the benchmarking analysis underscores the trade-offs inherent in cryomodule design and identifies the parameters that most strongly influence efficiency, reliability, and lifecycle performance.

By integrating practical experience with international best practices, this summary report establishes a consolidated knowledge base to support the next phase of WP5: the development of a roadmap toward a new generation of sustainable cryomodules.

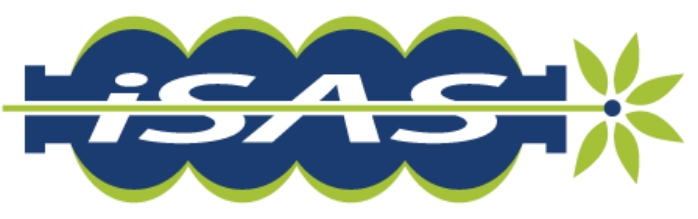

# 2) Summary of ESS cryomodules Lessons Learned for iSAS WP5

## 1. Introduction and scope

The iSAS Work Package 5 (WP5) aims to define and document a set of comprehensive requirements and technical solutions for integrating emerging technologies into a sustainable cryomodule.

To establish a robust baseline for these innovations, the ESS cryomodule will serve as reference, where the compilation of lessons learned during the design phase, cryomodule testing activities, technical commissioning, and initial operation, can lay a solid foundation towards a roadmap for a sustainable cryomodule

The report analyses the technical and procedural lessons learned during the lifecycle of the Elliptical Cryomodules for the European Spallation Source (ESS). As the ESS aims to become the world's most powerful neutron source, its linear accelerator (linac) relies on sophisticated superconducting radio frequency (SRF) cavities.

The primary goal of the document is to capture institutional knowledge gained during the design, fabrication, assembly, and testing phases, specifically focusing on High-beta cryomodules. These insights are intended to provide actionable recommendations for Work Package 5 (WP5) of the iSAS program, creating a roadmap ensuring that future accelerator infrastructures can benefit from the pioneering engineering conducted at ESS.

The scope of the analysis focuses on High-Beta elliptical systems, (though Medium-Beta modules are frequently referenced due to their shared design lineage).

The report covers the entire engineering lifecycle, from conceptual design and prototyping (M-ECCTD) to series production and final Site Acceptance Testing (SAT).

A significant portion of the study is dedicated to the challenges of industrialization, specifically the transition from laboratory-scale assembly to the high-rate production required for over 30 cryomodules.

Furthermore, the document explores the complexities of Stakeholder Integration within the In-Kind Contribution (IKC) framework, involving collaborative efforts across Sweden, France, Italy, and the UK. It addresses the logistical challenges as well as the physical installation issues and solutions encountered in the tunnel.

Finally, the report highlights operational lessons regarding cryogenic stability at 2 K, high-power RF conditioning, and the long-term reliability of control systems in a large-scale facility environment.

## 2. System overview

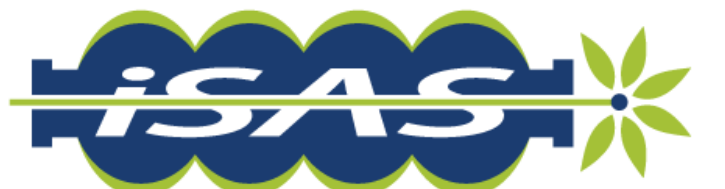

The report compiles a system overview of the European Spallation Source (ESS) superconducting linear accelerator (linac), detailing the technical infrastructure required to accelerate protons for high-flux neutron production. This section is critical because it establishes the physical and operational framework of the machine, designed to accelerate protons up to 2 GeV to trigger the neutron spallation process.

By defining the specific stages of acceleration, the report provides a roadmap of how the beam transitions from lower to higher energy states through specialized superconducting sections.

The ESS distinguishes itself from other spallation sources through its long-pulse architecture, which allows for unprecedented peak brilliance in neutron science aiming at 5 MW average beam power,

The accelerator operates at 14 Hz Repetition Rate, a frequency chosen to balance high neutron flux with the timing requirements of scientific instruments. It delivers a 2.86 ms "long pulse", which drives neutron brightness but introduces significant technical challenges regarding dynamic heat loads. The machine operates at a 4% duty cycle and a peak beam current was increased to 62.5 mA.

### 2.1. Cryogenic Scheme and thermal requirements

The accelerator cryogenic system is outlined, where a decision was taken to divide it into 3 major components: the Accelerator Cryoplant (ACCP), the Cryogenic Distribution System (CDS), and the Cryomodules (CM). This modular approach was chosen to facilitate parallel development across international partners and to ensure a streamlined timeline for testing and commissioning.

Technically, the system is driven by the rigorous demands of superconducting cavities, which require an operating temperature of 2.0 K using saturated Helium II. Maintaining stability is important, where the pressure requirement to be held within ± 0.1 mbar is necessary to prevent frequency detuning. Because the ACCP had a long procurement lead time, its sizing was established early in the conceptual phase based on estimated heat loads, including static loads (and dynamic loads from RF and beam operations).

In the design of the cryoplant additional safety margins were included, to accommodate for the uncertainty between the design and the real performance of the various components of the cryogenics system.

### 2.2. Space requirements for tunnel integration

Space limitations and requirements in the accelerator tunnel are mentioned, with allocation of specific envelopes for main systems, transportation, and conventional facilities.  A 3.85m x 3.5m region is designated for cryomodules, valve-boxes, RF distribution, and utility infrastructure. A dedicated 2.0m x 3.5m corridor facilitates the movement of all systems and dictates the maximum size of modular components. The layout philosophy places the RF distribution and cable trays furthest from the corridor, followed by cryogenic distribution, and finally the cryomodules. The transport area is strictly required to always remain clear to serve as an emergency evacuation route and access path.

### 2.3. ESS project and superconducting accelerator timeline

The project is unique as it started as a "green field" facility with no prior host institute, presenting both significant challenges and steep learning curves. The Technical Design Report (TDR), which serves as the facility's blueprint, was completed in 2013.

The first beam was accelerated through the Radio Frequency Quadrupole in 2021, followed by the first beam through the superconducting linac in 2025, after installation of the cryomodules. A second beam commissioning run is currently taking place with nominal RF pulses to achieve neutron spallation for the first time.

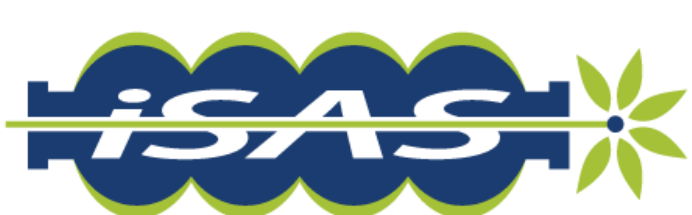

The superconducting linac (SCL) project has progressed through a structured, multi-year development driven by close collaboration between the European Spallation Source (ESS) and its in-kind partner institutes. From the outset, the project's success has depended on coordinated efforts spanning conceptual design, component qualification, production, installation, and operation.

The foundation was laid in 2013 with the publication of the Technical Design Report, which defined the SCL architecture around three cavity types: double-spoke, medium-beta elliptical, and high-beta elliptical cavities. Collaboration on cryomodule development formally began in 2014, followed in 2015 by in-kind production agreements with key partners including CEA, IJCLab (formerly IPNO), INFN, and STFC. Early technical milestones included successful vertical testing of six medium-beta prototype cavities in 2016 and assembly and testing of the first medium-beta prototype cryomodule in 2017.

From 2019 onward, the focus shifted to delivery, testing, and site integration at ESS in Lund. Prototype and first-series cryomodules were transported and installed, culminating in the commissioning of Test Stand 2 in 2021 for site acceptance testing of elliptical cryomodules. Series deliveries followed, including the first high-beta cryomodule in 2022. By 2023, installation in the linac tunnel had begun, with the first cryomodules integrated into cryoplant and cryogenic distribution system commissioning.

A major milestone was reached in 2024 with completion of all 27 cryomodules required for first beam operation and the successful integrated cooldown of the superconducting section to 2 Kelvin. The project then moved into beam commissioning: in 2025, the first beam phase achieved 860 MeV, and in 2026, a second commissioning phase is planned with nominal RF pulses. Overall, the timeline reflects a steady transition from design and prototyping to full system integration and operational readiness.

### 2.4. The importance of the in-kind model and its impact on design

The European Spallation Source strategy for construction under an In-Kind Contribution (IKC) model, means that much of the facility is built through hardware, expertise, and services provided directly by member states rather than through centralized cash procurement.

This approach allowed ESS to leverage Europe's distributed technical excellence and enabled parallel development across multiple laboratories, accelerating progress. However, it also introduced significant coordination challenges, requiring ESS to act as a strict systems integrator with strong quality assurance and interface control to avoid delays and bottlenecks.

The SRF cavity and cryomodule production chain illustrates this model where INFN (INFN-LASA) and STFC produced the medium- and high-beta cavities, which were then integrated into cryomodules at CEA in Saclay. The final qualification took place at ESS through Test Stand 2, where each cryomodule underwent comprehensive Site Acceptance Testing before installation.

## 3. Cryomodule Design Phase Lessons

### 3.1. Cryogenic architecture

The cryomodule design phase highlighted how early lattice and architectural decisions fundamentally shaped the entire accelerator and cryogenic integration strategy.

In the conceptual stage, different configurations were assessed for placing beam optics and diagnostic components either inside or outside the cryomodule vacuum envelope, as well as for defining their optimal periodicity. The selected solution established a periodic lattice structure consisting of four superconducting cavities followed by a drift space dedicated to beam correction and diagnostics.

A key outcome was the decision to locate beam optics and instrumentation in stand-alone Linac Warm Units (LWUs) outside the cryomodule envelope. This drove a segmented cryomodule concept in which each elliptical cryomodule contains four superconducting cavities, simplifying integration and clarifying functional boundaries between cold and warm systems.

The accelerator cryogenic architecture further reinforced this segmented philosophy, following the defined structure of the superconducting linac, composed of three cryomodule families (spoke, medium-beta, and high-beta cryomodules). Each cryomodule is supplied with cryogens at various thermodynamic conditions from one unique cryogenic plant (ACCP). The helium distribution is managed through a dedicated Cryogenic Distribution System (CDS), consisting of transfer lines, an endbox, and individual valve boxes, each connecting to a single cryomodule.

This creates a modular, independently controlled cryogenic structure. The staged installation sequence, first the ACCP, then the CDS, and finally the cryomodules, was deliberately aligned with subsystem qualification milestones to reduce technical and schedule risks.

Separating the cryogenic distribution system from the cryomodules, rather than integrating both within a single vacuum envelope, proved to be a strategic decision driven by technical, organizational, and risk considerations. It allowed deliverables to be distributed among in-kind partners according to expertise, reduced design and assembly complexity, and lowered performance risks. Operationally, it increased flexibility for maintenance, repair, and staged power upgrades by enabling the addition of pre-tested cryomodules.

From a project management perspective, completing and validating the CDS before installing cryomodules mitigated the risk of costly rework in case of non-conformities. The need for a temporary "dummy cryomodule return box" during early CDS commissioning further demonstrated how installation sequencing and spatial constraints must be carefully planned. Overall, the main takeaway is that modularity, functional separation, and staged validation were central principles that enhanced technical robustness, flexibility, and schedule control.

### 3.2. Cryomodule Internal design and components

The internal design of the cryomodule reflects a highly integrated and hierarchical architecture in which multiple critical subsystems are nested within a compact 6.6-meter vacuum vessel. At its core lies the cavity string: four superconducting elliptical niobium cavities (6-cell for Medium-Beta and 5-cell for High-Beta), each enclosed in individual helium tanks and mechanically linked by bellows. Titanium tie-rods provide structural stability while minimizing thermal conduction, and a magnetic shield surrounds the cavities to protect performance.

The layout is modular and repetitive, with a consistent sequence of warm-to-cold transition, cavity, power coupler region, and inter-cavity drift space. This standardized configuration simplifies integration, alignment, and beam dynamics consistency along the linac.

Each cavity is housed in a titanium helium tank connected to a slightly tilted diphasic line that functions both as a phase separator and as a communication channel between cavities. Liquid helium level is measured in a parallel circuit to ensure adequate coverage of the cavity tops and stable two-phase operation. The tanks also integrate mechanical interfaces for the tuning system and connections for cooldown and outlet circuits. This configuration ensures thermal stability, controlled cooldown, and uniform cryogenic behavior across the four-cavity string.

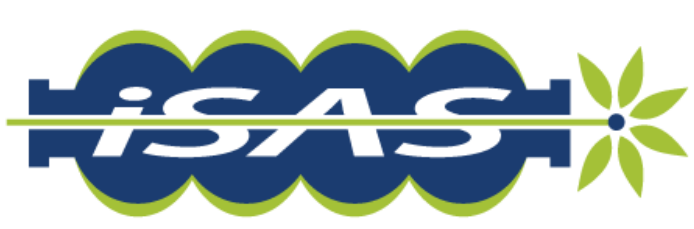

Power delivery is handled by one fundamental power coupler per cavity, capable of transferring up to 1.1 MW in pulsed operation (3.2 ms at 14 Hz). The coupler design is based on a simple 50-ohm coaxial line with a single planar alumina window, incorporating impedance matching and TiN coating to mitigate multipacting. The primary distinction between Medium- and High-Beta versions lies in antenna penetration depth. Water cooling of the antenna ensures thermal stability under high RF power loads, reflecting the balance between simplicity, robustness, and high-power performance.

Frequency control is achieved through a combined slow and fast cavity tuning system. The cold tuning system includes a motor-driven mechanical tuner for coarse adjustments after cooldown and piezoelectric actuators for rapid, dynamic compensation. The slow tuner provides a mechanical range of ±3 mm (corresponding to several hundred kHz), although only fine post-cooldown adjustments are typically required due to pre-tuning. The fast piezo tuners, integrated with the LLRF system, compensate in real time for Lorentz force detuning and beam loading effects during RF pulses, with fine adjustments on the order of 1 kHz. Together, these systems ensure precise resonance control under both static and dynamic operating conditions.

Thermal management and insulation are provided by a multi-segmented aluminium thermal shield wrapped in multilayer insulation to minimize radiative heat transfer to the 2 K cold mass.

All components are enclosed within a stainless-steel vacuum vessel that serves structural, vacuum insulation, and integration functions. The vessel includes dedicated ports for power couplers, a cryogenic jumper connecting to the valve box and distribution system, inspection hatches for maintainability, instrumentation feedthroughs, pressure relief systems, and beamline gate valves for isolation and integration with Linac Warm Units.

Overall, the design demonstrates a strong emphasis on modularity, maintainability, thermal efficiency, and precise RF control within a tightly integrated cryogenic environment.

The mechanical design was driven by the early decision to adopt a single “universal” cryomodule concept for both medium-beta and high-beta cavities. This standardization significantly simplified integration, manufacturing, procurement, logistics, and tooling, bringing major efficiency and consistency benefits across the project.

Accommodating the geometric differences between 6-cell and 5-cell cavities required targeted adaptations, such as modified intra-cavity bellows, while maintaining a common overall structure. Clear and early definition of shared and cavity-specific interfaces—especially with in-kind contributors—was essential to successfully implement this unified design approach.

The ESS elliptical cryomodule uses a “space-frame” mechanical concept in which an internal aluminium truss supports the cold mass and decouples precise cavity alignment from the vacuum vessel tolerances. This allows accurate assembly and alignment to be performed outside the vessel, before cryostating, when access is still optimal.

The space-frame maintains structural stability during cooldown to 2 K and is common to both Medium-Beta and High-Beta cryomodules, enabling a universal assembly approach. Overall, it improves alignment precision, robustness, and standardization.

The ESS elliptical cryomodules use a nested three-temperature thermal design (40–50 K, 4.5 K, and 2 K) to efficiently protect the superconducting cavities operating at 2 K.

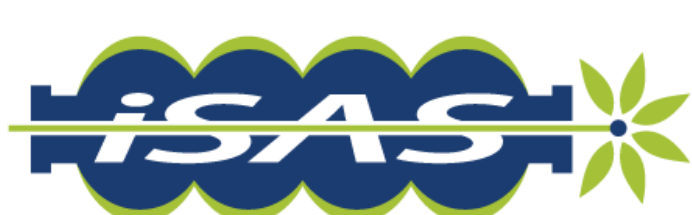

The 40–50 K thermal shield intercepts most radiative and conductive heat from ambient conditions, while the 4.5 K circuit distributes helium for cooldown, cavity filling, liquid production, and power coupler cooling.

A key efficiency feature is pre-cooling before the Joule–Thomson expansion, which enhances liquid helium production and reduces cryogenic load. The 2 K superfluid helium bath is maintained at low pressure (31mbar) via cold compressors placed in the accelerator cryoplant. The liquid helium level and pressure is stabilized at each cryomodule through redundant measurements.

Cryomodules are connected to the cryogenic distribution system via a vacuum-insulated jumper connection that transfers multiple cryogenic circuits while accommodating thermal contraction and preserving vacuum integrity.

The system allows detachable connections for testing and permanent welded connections in the tunnel. Auxiliary lines for helium recovery, purging, and pressure relief enhance operational flexibility and safety.

Overall, the design prioritizes thermal efficiency, reliability, integration, and safe operation at ultra-low temperatures.

The cryomodule uses a layered overpressure protection system to comply with the European Pressure Equipment Directive (PED), which sets a Maximum Allowable Working Pressure of 1.04 barg. Its ultimate protection consists of two burst disks that discharge helium to atmosphere; this is a last-resort measure, and the released helium is not recovered. To avoid activating these disks, two lower-pressure stages act first: a fast control valve and a spring-loaded relief valve, both directing helium to the cryoplant for full recovery.

Additionally, the isolation vacuum (Isovac) has its own low-pressure protection valve. Overall, the system balances helium recovery during normal overpressure events with atmospheric discharge only in extreme conditions.

## 4. Manufacturing and Assembly Lessons

The industrialization of the elliptical cryomodule series was led by CEA-Saclay, covering 30 series units in addition to two prototypes (Medium-Beta and High-Beta).

The prototype phase served as the final validation step before large-scale procurement, but schedule constraints and long component lead times limited the implementation of design improvements between prototyping and series production.

To shift from laboratory-based assembly to an industrial rhythm of roughly one cryomodule per month, CEA established a partnership with B&S International. Assembly activities were initially carried out by experienced CEA personnel, with a gradual transfer of responsibility to the industrial partner over several full production cycles.

The assembly process itself follows structured phases: clean room integration of cavities and couplers, installation of internal cryogenic components, cold mass assembly within a support frame and thermal shield, insertion into the vacuum vessel (cryostating), and final external integration. Each phase includes detailed sub-steps and verification checks, and success depends heavily on clear procedures, skilled personnel, and accumulated technical expertise.

Quality assurance is central to maintaining performance and minimizing non-conformities. A Digital "Traveller" system documents every assembly step and critical parameter throughout the cryomodule lifecycle, ensuring traceability and uniform quality control.

Deviations from specifications are formally recorded as non-conformities and classified by severity (minor, major, or critical). Each case follows a structured disposition process depending on its impact, characterised onto rework, repair with concession, use-as-is with approval, or rejection.

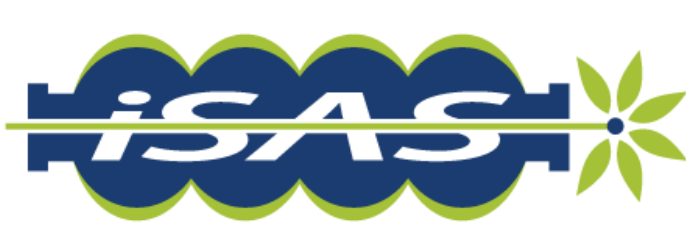

Final approval of any deviation requires formal validation before production continues. This combination of industrial partnership, structured processes, and rigorous quality control ensures that the cryomodules meet the stringent performance requirements of the ESS linear accelerator.

## 5. Testing and Commissioning Lessons

A structured and well-organized Site Acceptance Testing (SAT) phase is essential for reliable commissioning and operation phases.

The commissioning of Test Stand 2 (TS2) demonstrated that early testing campaigns must validate not only the cryomodules but also the test infrastructure itself. Using prototype and early-series units to refine cryogenic procedures, RF distribution, and data handling created a strong foundation for later efficiency.

A clear learning curve reduced the average test duration from more than two months to approximately six weeks, proving that procedural consolidation and experience significantly improve throughput.

Robust cryogenic performance validation requires redundant and complementary measurement methods. The use of multiple independent techniques to determine static and dynamic heat loads increased confidence in results and allowed early identification of discrepancies, such as higher-than-expected thermal shield loads due to design assumptions.

Continuous benchmarking against vertical cavity tests and systematic data tracking improved calibration accuracy and revealed measurable improvements in manufacturing quality over time, particularly through decreasing static heat loads.

Clean assembly processes and transport controls are critical to preserving RF performance. The strong correlation between vertical test results and fully assembled cryomodule performance confirmed the robustness of cleanroom assembly and handling procedures.

Most cavities retained their performance within a narrow margin after integration, and installed modules exceeded beam physics requirements. Strategic allocation of slightly limited cavities within the linac lattice ensured that local constraints did not affect overall accelerator performance.

Installation efficiency depends heavily on preparation, parallelization, and workflow optimization. Experience in the tunnel showed that internal components are vulnerable to transport-induced shifts and require careful inspection before integration.

Time savings were achieved by preparing piping, supports, and auxiliary systems in advance, deferring cable installation to avoid damage, and aligning two cryomodules simultaneously. Installation rates doubled once logistics, tooling, and task sequencing were optimized. Future improvements were identified in mechanical tolerances, automated welding and cutting tools, cable management design, and spare-parts tracking.

High availability requires redundancy, compensation strategies, and rapid recovery capability.

The operational target of 95% availability is supported by redundant helium level and pressure sensors, cavity heater backup strategies, and the ability to redistribute gradients among neighbouring cavities.

Cryoplant trips, however, revealed that recovery to stable 2 K conditions remains time-consuming and is a key area for improvement.

Most cavity limitations were administrative RF power constraints rather than intrinsic hardware failures, highlighting the importance of operational envelope management.

Designing for in-tunnel maintainability significantly reduces downtime risk. The ability to replace tuners, ceramic RF windows, relief devices, and other critical components without removing the cryomodule validates early design foresight.

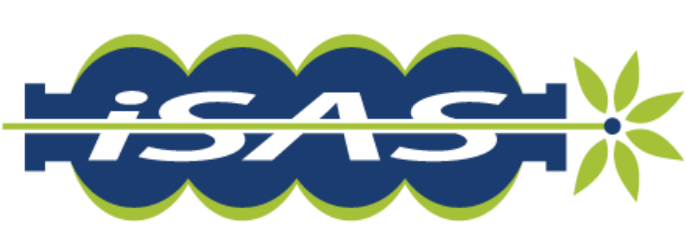

Only catastrophic beam vacuum contamination generally requires full module replacement. The demonstrated capability to warm up or cool down a single cryomodule independently of the rest of the linac further increases operational flexibility.

Comprehensive interlock logic and protective systems are indispensable for safe operation. Centralized status validation (e.g., "Cryomodule_OK") ensures that RF operation is only permitted under stable cryogenic and thermal conditions. Additional safeguards address overpressure, overfill, tuner immobilization, and cooling-water freezing risks. These layered protections confirm that performance reliability must be matched with equally robust machine protection strategies.

Overall, the combined experience from testing, installation, and commissioning shows that industrial learning curves, rigorous quality control, operational flexibility, and maintainability-by-design are decisive factors in achieving reliable large-scale superconducting accelerator operation

## 6. Sustainability Aspects

Sustainability aspects in the ESS cryomodule design are driven by a series of technical and strategic compromises made during the early design phase.

The minimizing heat loads, and thus cryogenic power consumption, is a core objective in the cryomodule design, it requires balancing key trade-offs, where designers must weigh higher upfront costs against lower operating expenses, redundancy against added static heat load, and optimized bespoke solutions against proven simplicity, always considering long-term lifecycle sustainability rather than only immediate constraints.

Energy efficiency depends on reducing static and dynamic heat inleaks through careful optimization of instrumentation, supports, and thermal intercepts. While added monitoring and thermal shielding improve reliability and lower cryoplant power consumption, they also increase complexity and integration challenges. Sustainable design therefore aims to remove heat at the highest possible temperature levels while preserving simplicity, reliability, and maintainability.

Helium management is another critical sustainability dimension. As a finite and costly resource, helium losses, especially during failure scenarios, have environmental and economic consequences. Although pressure safety regulations prioritize personnel and equipment protection, they do not directly address helium conservation. At ESS, most relief systems are designed for helium recovery, but catastrophic events can still lead to venting through burst disks. Reducing helium inventory, improving recovery capacity, and lowering the probability of critical venting events are key strategies to enhance long-term sustainability.

Exergy analysis provides a broader systems-level tool for improving cryogenic efficiency. Because cryomodules operate at extremely low temperatures, irreversible losses in compressors, turbines, throttling valves, and long transfer lines significantly affect overall performance. Exergy methods help identify where energy degradation is most severe—particularly in the final 2 K helium production stage—and support optimization of pressure ratios, valve design, and cooling distribution. Aligning cryomodule design with cryoplant architecture is therefore essential for minimizing total system inefficiencies and managing dynamic heat loads effectively.

Finally, material and component circularity is an important but often underemphasized sustainability aspect. Designing for modularity, repairability, and partial replacement—such as servicing individual cavities instead of entire cryomodules—reduces material waste, lowers environmental impact from sourcing specialized materials like niobium, and extends operational lifetime. Incorporating circular design principles strengthens both economic and environmental sustainability in large-scale superconducting accelerator facilities.

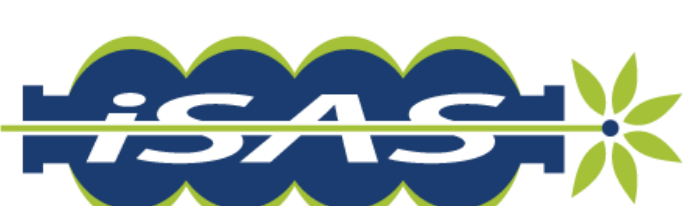

## 6. Section Conclusion

The development and commissioning of the ESS elliptical cryomodules marks a major achievement in large-scale superconducting RF technology.

The project demonstrates that complex, high-performance accelerator infrastructure can be successfully delivered through a distributed European collaboration model. The experience gained provides a practical framework for future sustainable accelerator systems, particularly within the objectives of the iSAS program.

A central lesson is the strong impact of standardization on sustainability and industrial scalability. Adopting a “universal” cryomodule design for both Medium- and High-Beta sections simplified the supply chain, reduced manufacturing waste through component commonality, and enabled harmonized assembly and testing procedures. The In-Kind Contribution model further proved effective in mobilizing distributed expertise across European laboratories, allowing parallel subsystem development despite added logistical complexity.

Key technical design choices also enhanced flexibility and long-term operability. Decoupling the cryogenic distribution system from the cryomodule vacuum vessel improved installation sequencing, maintenance access, and upgrade potential.

The internal space-frame support structure successfully separated mechanical tolerances of the vacuum vessel from the precise alignment needs of the cavity string, ensuring structural stability during cool-down to 2 K. In parallel, optimized cryogenic architecture, such as nested temperature levels and efficient pre-cooling of the 4.5 K supply, improves overall thermal efficiency and 2 K performance.

Overall, the ESS cryomodule project evolved from concept to stable operation with validated cryogenic and RF performance, while establishing strong industrialization and quality-control practices. It now serves as a reference model for sustainable, collaborative research infrastructures and a blueprint for future accelerator facilities seeking to combine technical excellence with long-term sustainability goals.

A full detailed analysis of the topics described here is presented in **Annex I:**

- **“Evaluation of the ESS Cryomodules Design and Operation - Lessons Learned from the Current Cryomodule Technology”**

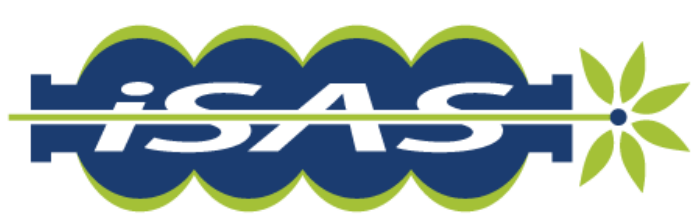

# 3) Summary of Benchmarking Cryomodule Designs for iSAS WP5

## 1. Introduction

The iSAS Work Package 5 (WP5) aims to define and document a comprehensive set of requirements and technical solutions for integrating emerging technologies into a sustainable cryomodules. To establish a robust baseline for these innovations, a benchmarking study was conducted, evaluating high-beta cryomodules across six world-leading facilities: SNS, LHC, PIP-II, XFEL, LCLS-II, and ESS.

By analysing these systems through five distinct lenses, the program aims to identify the optimal parameters for a sustainable, high-efficiency "Cryomodule of the Future".

## 2. Benchmarking Methodology

To ensure scientific rigor, WP5 follow a systematic 17-step process designed to move from theoretical data to validated expert insights:

1) Definition of key technical Areas of Interest to be included in the benchmarking study
2) Proposal of cryomodule design Concepts: Selecting specific architectures for comparison.
3) Identify Facilities: SNS, LHC, PIP-II, XFEL, LCLS-II, ESS.
4) Identify Experts: Mapping technical leads and contact persons.
5) Defining Specific Points of Interest for each area
6) Draft Questionnaire: Creating the data collection instrument.
7) Questionnaire Dry-Run: Testing the tool with 2 facilities.
8) Dry-Run Feedback: Refining the query structure.
9) Final Improvements: Finalizing the questionnaire based on feedback.
10) Literature Research: Exhaustive review of literature in published scientific papers.
11) Pre-filling Answers: Using literature to maximize the efficiency of expert time.
12) Initial Contact: Introducing the iSAS program to the facilities and the scope of the benchmarking
13) Distribution: Sending the pre-filled questionnaires for first review and completion.
14) Primary Face-to-Face Interview: First deep-dive interview with the facility experts.
15) Follow-up Face-to-Face Interview: Clarifying outstanding data and "lessons learned."
16) Compilation of Answers: Aggregating data across all six facilities.
17) Completion of Report: Drafting the report with the Benchmarking Synthesis.

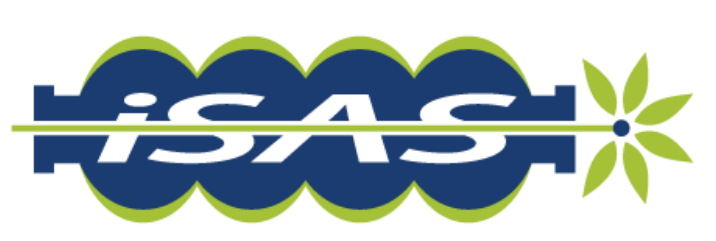

## 3. Technical Areas of Interest

The benchmarking study is structured around five core areas that capture the full lifecycle of a cryomodule from beam physics requirements to long-term operational reliability.

### 3.1. Beam & RF Parameters (The Design Drivers)

This area characterizes the fundamental machine requirements that dictate cryomodule architecture.

- Focus: Accelerating gradients, operating frequencies, and the choice between Continuous Wave (CW) and Pulsed modes.
- Goal: Understand how beam physics requirements drive the complexity and energy demands of the CM.

### 3.2. Cryogenic Systems & Thermal Management

Characterization of the "thermodynamic footprint" of the high-beta sections.

- Focus: Machine architecture, Cooling circuit layouts, temperature and pressure levels, static and dynamic heat loads, and the specific cooling capacity of the associated cryogenic plant. Management of helium inventory and ODH risk
- Goal: Characterizing design choices for cryogenic systems.

### 3.3. Cryomodule Design, Internal Components & Design Constraints

A summary of the mechanical and electromagnetic solutions within the vacuum vessel.

- Focus: Characterize Cryomodule sub-systems features, Cavity string assemblies, power couplers, tuners, and magnetic shielding.
- Goal: Map the constraints, design choices and integration aspects regarding the cryomodule design and its assembly sequence

### 3.4. Commissioning Insights

Capturing the transition from "as-designed" to "as-built."

- Focus: Knowledge gained during the assembly, testing, and initial installation phases.
- Goal: Document technical hurdles and breakthroughs that occurred during the commissioning phase, providing useful feedback for future sustainable designs.

### 3.5. Operational Experience & Machine Availability

A metric for sustainability is long-term reliability and availability.

- Focus: Factors limiting cryomodule performance, mitigation strategies, machine availability, capability for maintenance, repair and replacement.
- Goal: Use real-world operational data to ensure that energy-saving innovations do not compromise the facility's scientific output.

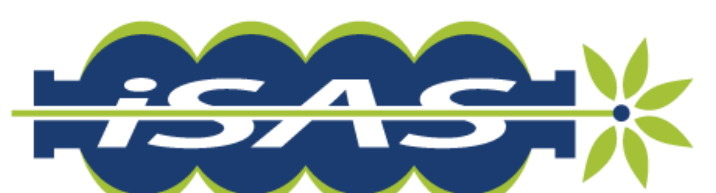

## 4. The Questionnaire

*SECTION 1: ENERGY AND RF PARAMETERS*

This section aims at characterizing the main beam and RF parameters of the machine, driving the cryomodule design choices.

Question 1.1: Provide an indication of the machine size (Number of cavities and material / number of cryomodules).

Question 1.2: What are the types of particles accelerated, the beam power and the beam energy?

Question 1.3: What is the RF power, the current amplitude and the operating frequency?

Question 1.4: Does the machine operate in continuous or pulsed mode? If pulsed, what is the duty factor?

Question 1.5: What is the cavities $Q_0$ and accelerating gradient [MV/m] in operation? What is the cavities operating temperature?

Question 1.6: What are the main features of the FPC design?

Question 1.7: Do the cavities have High Order Modes extractors (HOM couplers/BLAs)? What are the main design features? (Type, number, location)

*SECTION 2: CRYOMODULE COOLING CIRCUITS AND CRYOGENICS*

This section aims at characterizing the cryomodule cooling circuits, heat loads and the cooling capacity of the cryogenic plant.

Question 2.1: Characterize the machine architecture (segmented/continuous, cryogenic distribution line, quadrupoles).

Question 2.2: List the temperature and pressure levels of the cryomodule (cold and warm) circuits and their stability levels.

Question 2.3: What is the cooling strategy for the FPC (Antenna and outer conductor)? What are the reasons for these design choices?

Question 2.4: What is the cooling scheme for the HOM couplers? (Port, antenna and jacket if applicable) and the reasons for these design choices.

Question 2.5: Does the cryomodule have a thermal shield? At what temperature?

Question 2.6: What are the maximum dynamic and static heat loads, at the different temperature levels, for which the cryomodule has been designed? And what are the current operational values?

Question 2.7: What devices are installed to protect the equipment from overpressure (burst disks, pressure relief valves)? Where is the helium discharged? Is there a system for helium recovery?

Question 2.8: What is the installed cooling capacity of the cryoplant at the different temperature levels (if applicable)? What is the installed liquefaction capacity?

*SECTION 3: CRYOMODULE DESIGN*

This section aims at summarizing the main characteristics of the cryomodule internal components, and the main constraints leading to the described design solutions.

Question 3.1: What are the main features of the helium tank?

Question 3.2: How are the cavities supported and aligned in the cryomodule? What is the fixed longitudinal reference point during thermal contraction? How are vacuum forces compensated?

Question 3.3: What are the main features of the cavity tuning system?

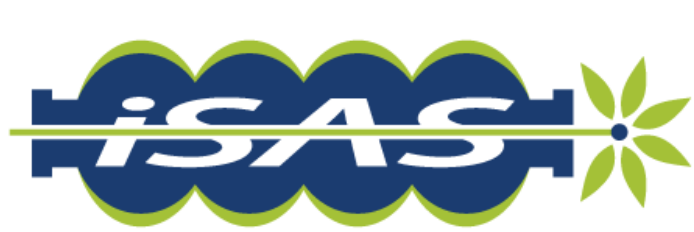

Question 3.4: Does the cavity have magnetic shielding? If yes, how many layers? Cold or warm? Where is it /are they located?

Question 3.5: Does the cryomodule have access hatches? What components can be accessed?

Question 3.6: Are there any design compromises implemented in favour of assembly/alignment/cost which impacted the performances of the cryomodule?

*SECTION 4: LESSONS LEARNED IN THE COMMISSIONING PHASE*

This section aims at collecting the insights and the knowledge gained during the commissioning phase of the facility.

Some questions are proposed as guidelines but any input on the topic is considered extremely valuable.

Keywords can be inserted in this questionnaire, a proper in-person discussion is meant to be scheduled to address these aspects in details.

Question 4.1: What are the main differences between the design values and the cryomodule performances measured in the test stand? What was the identified cause of these differences?

Question 4.2: Describe the key points of the experience gained in the installation phase?

Question 4.3: What design modifications could have facilitated the installation and commissioning phases?

Question 4.4: Describe the procedure for in situ warm/cold RF conditioning? How much time does it take? How could it be made faster (automatized?) to reduce commissioning and machine down-time?

*SECTION 5: STATUS OF THE MACHINE IN OPERATION AND CURRENT AVAILABILITY*

This section aims at collecting the experience in operation and the main factors limiting the machine availability.

Some questions are proposed as guidelines but any input on the topic is considered extremely valuable.

Keywords can be inserted in this questionnaire, a proper in-person discussion will be scheduled to discuss these aspects.

Question 5.1: What is the target availability of the machine? What are the redundant components/systems or overhead installed to compensate in case of failures? What is the real availability achieved in the last years?

Question 5.2: What are the mechanisms limiting the performances of the cavities? How were they addressed to continue the operations?

Question 5.3: What components can be accessed and maintained in the tunnel? (FPC, tuner, HOM cables/connectors etc.)

Question 5.4: If any, what were the cryomodule failure modes? Were they foreseen during the design stage? Could the cryomodule be repaired in the tunnel or was a replacement needed?

Question 5.5: In case of maintenance in the tunnel, is it possible to warm-up/cool-down a single cryomodule? Is the LHe inventory emptied in the adjacent cryomodules? If yes, how is the temperature rise controlled?

Question 5.6: What is the instrumentation installed for hardware (cavities, FPC and helium tank) protection? What values are used as interlocks?

Question 5.7: What is the current status of the facility? What is the foreseen next upgrade and what are the main challenges associated? What is the driver for the upgrade?

## 5. Literature Research and Pre-Filling The Questionnaire

The literature research phase serves as the analytical foundation of the benchmarking study. Rather than starting from a blank slate, the iSAS team reviewed existing documentation to "pre-fill" the benchmarking questionnaires. This approach serves three purposes:

- Optimized Discussion: Instead of spending time reviewing design parameters (e.g., operating frequency or cavity count), the discussion immediately dives into technical nuances and "lessons learned."
- Reduced Expert Burden: Allowing them to focus on correcting inaccuracies or providing context that isn't found in published papers.
- Higher Accuracy: This "correction-based" feedback loop often reveals discrepancies between the "as-designed" specifications and the "as-operated" reality.

### 5.1. Information Sources

To provide a comprehensive pre-filled baseline, data is collected from a wide spectrum of technical repositories:

- Technical Design Reports (TDRs): These serve as record of the initial machine architecture, cooling circuit logic, and design constraints.
- Scientific Papers & Published Articles: Peer-reviewed journals (e.g., Physical Review Accelerators and Beams) provide validated data on cavity performance and cryogenic efficiency.
- Conference Presentations (JACoW): Proceedings from IPAC and SRF workshops are valuable sources as they capture real-time technical challenges and machine availability data.
- Internal Technical Notes: Where available, internal facility documentation provided the internal details on sub-components like couplers and tuners.

It is important to note that the literature research does not provide an “homogeneous” set of information to be comparable. Therefore, the interview process became essential part of the process, specifically for the part of commissioning and operation, the set of questions is not sufficient to cover all the possible scenarios faced by the single facilities.

## 6. The Interview Process

The execution of the interview phase represented one of the most intensive periods of the benchmarking study.

Planning a series of technically rich meetings across the six world-leading facilities (SNS, LHC, PIP-II, XFEL, LCLS-II, and ESS) presented to be a complex endeavour, requiring coordinating the calendars of high-level experts across multiple time zones as the team had to adapt the availability of several key stakeholders simultaneously.

These unavoidable scheduling constraints and the resulting synchronization efforts led to minor deviations from the original project timeline.

The use of video-conferencing tools for virtual meetings as a primary communication platform provided the necessary flexibility to bridge the geographic gap, allowing for a structured yet adaptable dialogue with each facility.

A critical factor for the success of these sessions was the pre-filled questionnaire. By presenting experts with data already harvested from scientific literature and their initial feedback, the iSAS team transformed the interview from a standard data-gathering exercise into a high-level conversation.

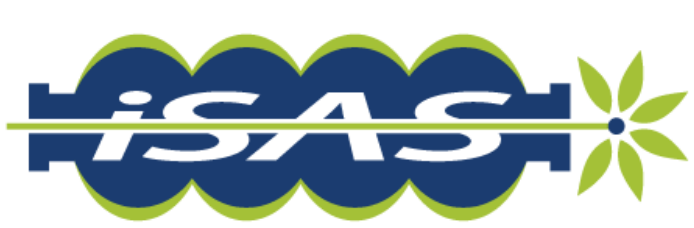

This approach was instrumental in driving the discussions forward, as it bypassed the need for experts to provide basic, publicly available information. Instead, it empowered them to focus on nuance, correction, and the provision of context that is rarely captured in formal documentation. This pre-filling strategy proved to be a highly effective asset, ensuring that every minute of the scheduled meetings was spent on high-value technical exchange.

The atmosphere of the interviews was defined by an exceptionally amicable, honest, and open spirit of collaboration. Despite the formal nature of the iSAS program—and the fact that many of the participating institutions are not official project members—the experts shared their insights with remarkable transparency.

This openness yielded a "fruitful" dialogue that frequently moved beyond the initial scope, uncovering unexpected information regarding commissioning hurdles and operational constraints that were not originally anticipated.

Given the high volume of detailed information shared, the need for additional follow-up meetings became a necessity. These subsequent sessions allowed for thorough “deep-dives” into specific subsystems, ensuring that the final benchmarking report captures the operational reality of high-beta cryomodules at the selected facilities.

## 7. Compilation of Results

The compilation phase marks the transition from individual facility data collection to cross-facility synthesis and an exhaustive analysis will create solid basis for proposing a roadmap for new, sustainable, cryomodule design.

After completion of the interviews and validation of the pre-filled questionnaires, the data from the six facilities (SNS, LHC, PIP-II, XFEL, LCLS-II, and ESS) is being consolidated into a structured comparison framework aligned with the five Technical Areas of Interest.

The final objective is to identify recurring design patterns, critical divergences, performance trade-offs and compromises that directly impact cryomodule sustainability.

Although there are some clear lines that can already be drawn, a complete and thorough analysis will continue.

## 8. Section Conclusion

As a summary from the benchmarking analysis, we can list high-level conclusions:

*Beam & RF Parameters – Design Drivers*

The benchmarking revealed that beam requirements fundamentally dictate the cryomodule architecture.

- Continuous Wave (CW) machines prioritize high intrinsic $Q_0$ and minimized dynamic losses, as cryogenic efficiency directly impacts operational cost.
- Pulsed machines, while operating at lower duty factors, tend to push higher peak gradients, placing stronger emphasis on mechanical robustness and Lorentz force detuning compensation.
- The type of machine and particles to be accelerated also dictate different requirements (alignment, losses, etc.)
- FPCs usually have horizontal orientations when they are supposed to deliver low average RF power (<40/50 kW)

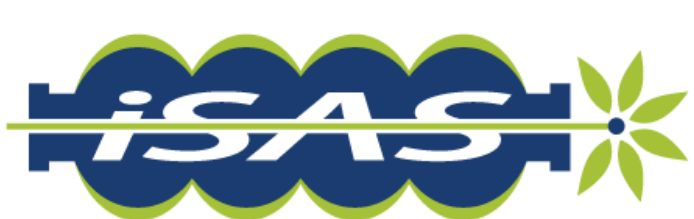

- For proton linear accelerators, the beam is less sensitive to trapped HOM perturbations, thus HOM couplers are non-essential and may be omitted at low currents, avoiding the associated design complexity. This does not apply for circular colliders, and other aspects like current amplitude, beam emittance, beam stability limit, bunch intensity, etc. must be evaluated.

*Cryogenic Architecture and Thermal Management*

Significant diversity exists in cooling circuit segmentation, temperature levels, and thermal shield configurations. However, common trends emerged:

- Multi-level temperature architecture (2 K /4–5 K/ 8-10K / 40–80 K) is standard across all facilities. Only LHC presents a single temperature level and floating thermal shield (compromise in favour of the assembly sequence) which is strongly increasing the static heat load to the helium bath.
- Static heat loads represent a substantial fraction of total cryogenic consumption, especially in long accelerators.
- FPC internal conductor cooling plays a major role in designs with fixed antenna. Active cooling with helium for the FPC outer conductor is considered more reliable than fixed thermal intercepts especially in operation with high average RF power (> 40/50 kW).
- Cryogenic distribution strategy (continuous vs segmented) strongly affects maintainability and failure recovery capability.
- Facilities with more modular cryogenic segmentation demonstrate improved operational flexibility, particularly for maintenance scenarios.
- The benchmarking confirms that cryogenic design is one of the dominant drivers of overall sustainability.

*Cryomodule Mechanical Design and Integration*

While cavity technology is broadly similar (elliptical SRF cavities), integration philosophies differ:

- Variations in cavity supporting schemes, vacuum forces compensation, longitudinal references and alignment strategies have been identified.
- Helium tank design influences mechanical stiffness and microphonics sensitivity.
- Variations in Magnetic shielding implementation (cold vs warm, single vs multi-layer) with impact on achievable $Q_0$ have also been identified
- Some facilities reported design compromises made to facilitate assembly, reduce cost, or drivers related to schedule, with related impact on operational performance. These trade-offs will continue to be analysed in WP5.
- The integration of a second thermal shield, when in presence of the 8-10K heat intercept, was evaluated but not considered convenient in comparison to its required capital cost.
- It is now a standard to include access hatches in the helium tank design, but it is fundamental to maintain the accessibility when the cryomodule is installed with all the services in the tunnel.

*Commissioning Experience*

Important to note is that the facilities used for benchmarking find themselves at very different phases. Some have not fully entered the commissioning phase; others have decades of experience.

An important theme across facilities is the aim to minimise discrepancy between "as-designed" and "as-operated" performance. Factors contributing to performance deviation include:

- Different experiences have been described regarding performance limitations
- Cryoplant limitations during ramp-up have been mentioned

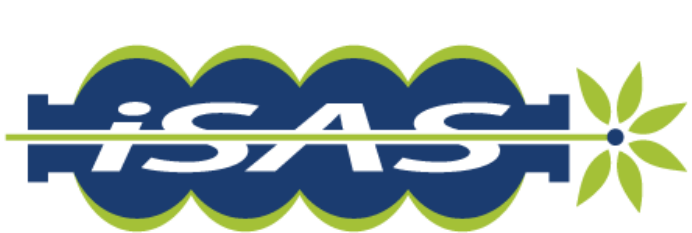

- Significant differences in heat loads from design to operation, become complex to manage when the cryoplant design is fixed.
- Integration aspects and mismatching interfaces, with emphasis in continuous quality controls, have been identified as important lessons to reduce re-work.
- Commissioning strategies, including various approaches for RF conditioning and automation, in conjunction with experts' oversight, significantly reduce machine downtime.

*Operational Reliability and Availability*

Long-term availability emerged as a critical sustainability metric. Facilities emphasized:

- Redundancy of internal equipment, critical for operation
- Modular cryogenic systems and replacement strategies
- Importance of accessibility was emphasized, more specifically couplers and tuners.
- Failure modes reported include coupler degradation, tuner actuator failure, and vacuum-related incidents.
- The ability or inability to isolate a single cryomodule for maintenance without affecting adjacent units strongly impacts overall machine uptime
-

**Final Remarks**

This study has highlighted various compromises made during the cryomodule design phase, where different bias to requirements or constraints can drive the design towards different configurations.

The amount of information shared by the participating facilities has exceeded our expectations, triggering many discussions of great importance to be also included in scope of the study.

A thorough analysis of the results will continue as WP5 enters the phase for development of a roadmap towards a new sustainable cryomodule design, where we hope to maintain a fruitful and open collaboration with the participating experts and attract other actors.

A full detailed analysis of the topics described here is presented in **Annex II:**

- **"Benchmarking Cryomodule Designs for iSAS WP5"**

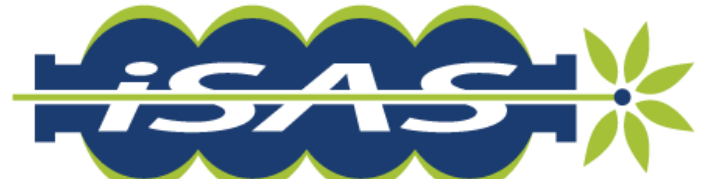

## 9. Acknowledgements

The authors wish to extend their sincere gratitude to everyone who participated in the benchmarking process, specifically the experts from world-leading facilities across the globe who contributed to this endeavour. It is noteworthy that while most of these institutions are not formal members of the iSAS project, their involvement was secured through the dedicated effort and collaborative spirit of the iSAS WP5 team. Below we list the important actors on this endeavour.

CERN (LHC, Geneva, Switzerland)

- Vittorio Parma
- Karin Canderan
- Katarzyna Turaj
- Krzysztof Brodzinski
- David Smekens

ESS (European Spallation Source, Lund, Sweden)

- Paolo Pierini
- Nuno Elias
- Emma Simpson
- Henry Przibliski
- Marek Skiba

SNS (Spallation Neutron Source, Oak Ridge, USA)

- John Mammouser
- Paolo Pizzol

XFEL (European X-ray Free Electon Laser, Hamburg, Germany)

- Serena Barbanotti
- Kay Jensch
- Julien Branlard
- Joshua Einstein-Curtis

LCLS-II (Linac Coherent Light Source II, SLAC, California, USA)

- Thomas Petersen
- Sebastian Aderhold

PIP-II (Fermi National Laboratory, Batavia, Illinois, USA)

- Donato Passarelli
- Vincent Roger

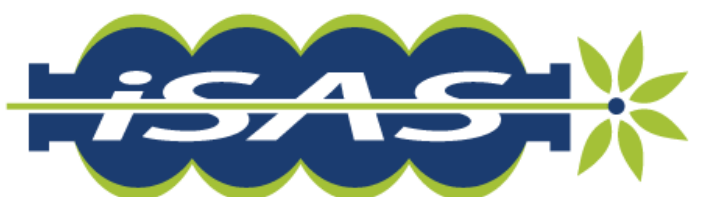

# 4) Conclusion

The work conducted within iSAS WP5 establishes a foundation for the development of a next-generation sustainable cryomodule. By combining an in-depth lifecycle analysis of the elliptical cryomodules developed for the European Spallation Source (ESS) with a structured international benchmarking across leading facilities—including Spallation Neutron Source, Large Hadron Collider, PIP-II, European XFEL, and Linac Coherent Light Source II—the study delivers both a validated reference model and a comparative framework to guide future design choices.

Together, these complementary perspectives bridge practical implementation experience with international best practice, forming the basis for a roadmap toward a sustainable Cryomodule Design.

The ESS experience demonstrates that large-scale superconducting RF infrastructure can be successfully realized through a distributed collaboration model under an In-Kind Contribution framework. A central lesson is the decisive impact of early architectural choices.

The adoption of a "universal" cryomodule concept for both medium- and high-beta sections, combined with a modular cryogenic architecture separating cryomodules from the cryogenic distribution system, significantly enhanced industrial scalability, integration clarity, and long-term maintainability.

Standardization and reduced supply-chain complexity facilitated quality assurance and enabled a smooth transition from prototype validation to series production. The project further highlights the importance of staged validation, rigorous interface control, and strong systems engineering to manage the complexity inherent in multi-partner infrastructures.

From a technical perspective, the ESS cryomodules validate the effectiveness of a segmented, multi-temperature cryogenic architecture operating at 2 K, with careful management of static and dynamic heat loads. The nested thermal design, efficient pre-cooling strategies, and redundancy in cryogenic instrumentation underpin stable operation and high availability.

Equally important are mechanical and integration choices, such as the internal space-frame concept, modular cavity strings, and maintainability-oriented access provisions, which decouple alignment precision from vessel tolerances and enable in-tunnel servicing of critical components. The progressive reduction of testing time, improvement of heat-load performance, and preservation of cavity RF characteristics through industrialization confirm the value of structured quality control, digital traceability, and accumulated production learning curves.

The benchmarking exercise through world leading facilities reinforces and contextualizes these findings. Across facilities with differing missions, continuous wave versus pulsed operation, proton versus electron beams, it becomes clear that beam and RF parameters fundamentally drive the cryomodule architecture, cooling strategies, and operational constraints.

Nevertheless, common design patterns emerge, multi-level temperature schemes are standard; cryogenic efficiency is a dominant determinant of sustainability; and modular segmentation strongly enhances operational flexibility and failure recovery.

Trade-offs are unavoidable: between simplicity and optimization, capital cost and operational efficiency, redundancy and added static load, where an efficient sustainable design depends on making these compromises explicit and lifecycle oriented.

Operational experience across facilities confirms that long-term availability is one of the primary sustainability metrics. Redundancy of critical sensors and actuators, the ability to redistribute accelerating gradients, modular cryogenic isolation, and maintainability-by-design directly influence uptime. Failure modes such as coupler degradation, tuner malfunctions, and vacuum incidents underline the need for robust protection systems and accessible layouts.

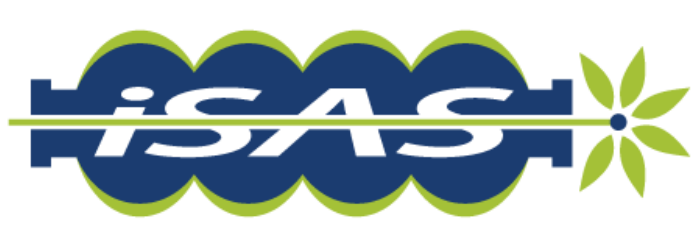

Commissioning insights further emphasize the importance of minimizing discrepancies between “as-designed” and “as-operated” performance through early test-stand validation, automation of RF conditioning, and continuous performance benchmarking.

Finally, sustainability emerges as a multidimensional objective extending beyond energy efficiency alone. Minimizing cryogenic exergy losses, optimizing helium management and recovery, reducing static heat loads, and embedding modularity and repairability into the design all contribute to lowering lifecycle environmental and economic impact.

The ESS case demonstrates that sustainability is not achieved through a single innovation, but through coherent system-level integration of thermal efficiency, standardization, industrial readiness, operational flexibility, and circular design principles.

In conclusion, the combined ESS lessons learned, and international benchmarking provide a consolidated knowledge base for WP5.

They confirm that the path toward a sustainable cryomodule lies in early architectural clarity, disciplined systems engineering, modular cryogenic segmentation, industrial standardization, and design-for-maintainability, supported by continuous dialogue with the global accelerator community.

These insights now form the strategic and technical backbone for the forthcoming roadmap toward a new generation of high-efficiency, resilient, and future-proof cryomodules.

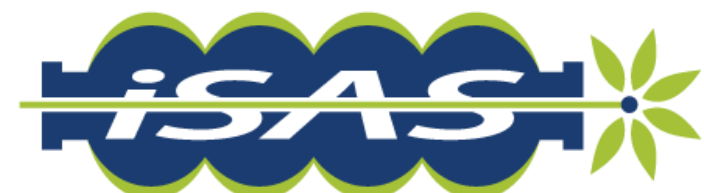

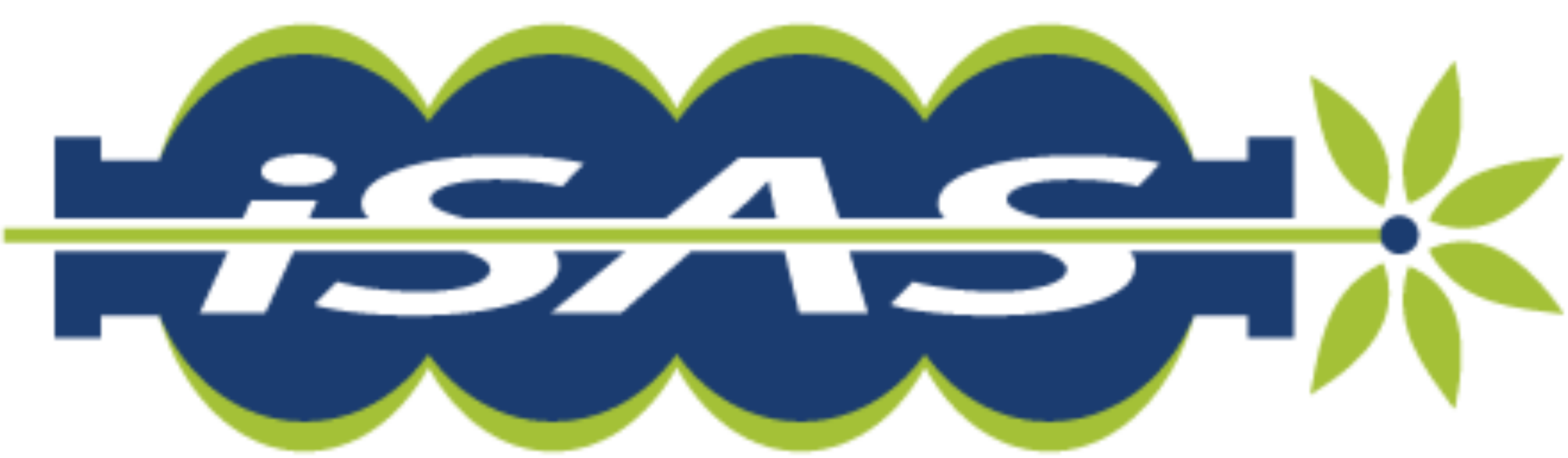

innovate for Sustainable Accelerating Systems

# ANNEX I
# Compilation of lessons learned from ESS

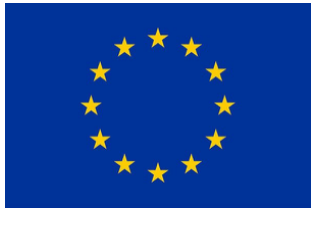

Funded by the European Union. Views and opinions expressed are however those of the authors only and do not necessarily reflect those of the European Union or the European Commission. Neither the European Union nor the granting authority can be held responsible for them.

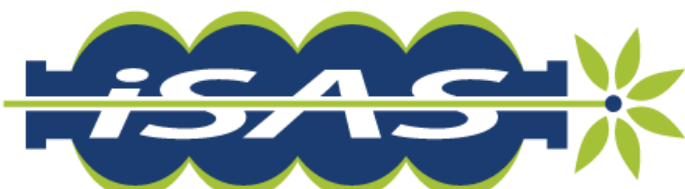

# DOCUMENT TRACKS DETAILS

| Deliverable Number | D 16 | Lead Beneficiary | ESS, CERN |
|---|---|---|---|
| Deliverable Name | Compile lessons learned from ESS and Benchmarking with Projects in implementation phase (**ANNEX I**) | | |
| Type | R — report | Dissemination Level | PU – Public, |
| Due Date | Project month 24 | Work Package No | WP5 |

## Deliverable Abstract

This report consolidates two complementary studies conducted within iSAS Work Package 5 (WP5) to support the development of a sustainable next-generation cryomodule.

The first study analyses the design, industrialization, commissioning, and early operation of the elliptical cryomodules developed for the European Spallation Source (ESS), capturing technical and procedural lessons learned across the full engineering lifecycle. It highlights the impact of standardization, distributed collaboration under the In-Kind Contribution model, cryogenic performance at 2 K, and industrial scalability as key enablers for long-term sustainability.

The second study benchmarks high-beta cryomodules across six leading international facilities—Spallation Neutron Source, Large Hadron Collider, PIP-II, European XFEL, Linac Coherent Light Source II, and ESS—through a structured comparative methodology covering beam and RF parameters, cryogenics, mechanical design, commissioning, and operational experience.

Together, these analyses establish a validated technical baseline and an international reference framework, forming the foundation for a WP5 roadmap toward a high-efficiency, sustainable “Cryomodule of the Future.”

ANNEX I refers to the detailed report on compilation of lessons learned from ESS

ANNEX II refers to the detailed report on benchmarking of cryomodule designs for iSAS

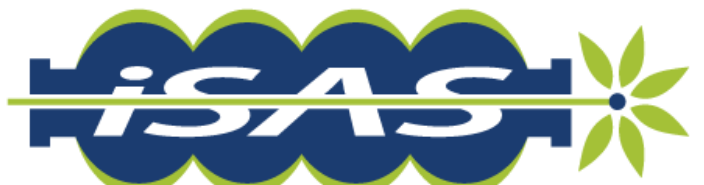

TABLE OF CONTENTS PAGE

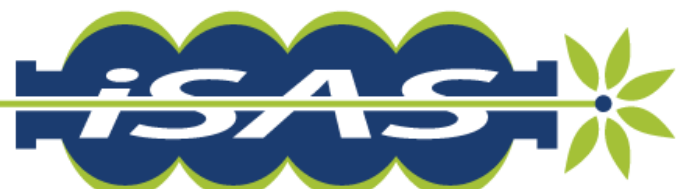

# 1. INTRODUCTION AND CONTEXT

## 1.1. Purpose

This report provides a comprehensive analysis of the technical and procedural lessons learned during the design, fabrication, assembly, and testing of the Elliptical Cryomodules for the European Spallation Source (ESS) superconducting cryomodules.

ESS relies on a state-of-the-art linear accelerator (linac) that utilizes both spoke and elliptical superconducting radio frequency (SRF) cavities, with the goal to become the world's most powerful neutron source.

The primary **objective** of this document is to capture institutional knowledge gained across the engineering lifecycle of the **Elliptical Cryomodules with focus on High-beta cryomodules**.

These experiences can then be translated into actionable recommendations to be explored within **Work Package 5 (WP5) of the iSAS (Innovate for Sustainable Accelerating Systems) program**.

By documenting both **technical successes and operational challenges**, this report will help to establish a roadmap for future accelerator systems, ensuring that next-generation infrastructures leverage the pioneering engineering conducted at ESS.

## 1.2. Scope

The scope of this report focuses specifically on the High-Beta elliptical cryomodules; however, Medium-Beta cryomodules will also be mentioned frequently as the cryomodule design was developed to encompass both types of cavities.

While the ESS linac also incorporates spoke cryomodules, the lessons detailed here are derived from the unique challenges of the elliptical systems, including:

- **Lifecycle Phases:** Conceptual design, prototyping (M-ECCTD), series production, and Site Acceptance Testing (SAT).
- **Industrialization:** Transitioning from laboratory-scale assembly to high-rate production for over 30 cryomodules.
- **Stakeholder Integration**: The collaborative workflow involving the ESS (Sweden), CEA (France), INFN LASA (Italy), and STFC (UK) within the In-Kind Contribution (IKC) framework.
- **Transport and Logistics:** The challenges associated with the trans-European transport of fully assembled cryomodules, including vacuum integrity monitoring, shock-absorption during transit, and the environmental impact of long-distance logistics.
- **Installation Challenges:** Managing the tunnel integration, precise alignment of the cryomodule string, and the physical connection to the Cryogenic Distribution Line (CDL).
- **Operational Challenges:** Lessons learned during cool-down and warm-up cycles as well as long-term cryogenic stability at 2 K, high-power RF conditioning, and the reliability of the control systems in a facility-scale environment.

Note: Some details might be left out to protect ownership, authorship and intellectual property.

### 1.3. Alignment with iSAS objectives covered in WP5

The **iSAS (Innovate for Sustainable Accelerator Systems) program** seeks to pioneer a new era of particle accelerators that prioritize energy efficiency, resource optimization, and industrial scalability.

This report serves as one of the contributions within the Work Package 5 (WP5), titled "Integrate into the design of a sustainable LINAC cryomodule" by utilizing the ESS elliptical cryomodule experience as a primary dataset to address the following topics:

- **Energy Efficiency & Heat Load Reduction:** Analysing the static and dynamic heat loads from design to operation.
- **Resource Management and Circularity:** Evaluating the lifecycle of critical materials and components and the management of Helium as a finite resource. Lessons from ESS cool-down/warm-up cycles and recovery systems are essential for designing sustainable infrastructures.
- **Industrialization & the "Parametric Design" Study:** A key deliverable of iSAS WP5 is the development of a Parametric Cryomodule Design. The ESS "common design" approach (using one vessel for both Medium and High-Beta cavities) provides the baseline for this study, offering insights into how standardization reduces manufacturing waste, lowers costs, and simplifies the industrial supply chain.
- **Improved availability and Cost reduction:** By documenting operational challenges such as coupler conditioning, expected component limitation, or maintenance access, this report helps define design choices that minimize long-term intervention costs and maximize the "uptime" of sustainable research infrastructures.

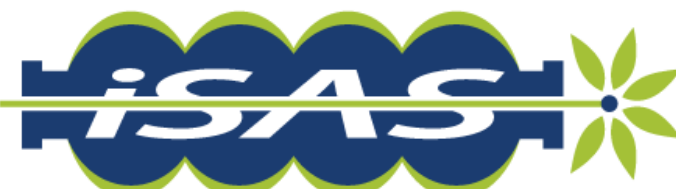

# 2. SYSTEM OVERVIEW

## 2.1. The ESS Superconducting Linac

The ESS linear accelerator is designed to accelerate protons up to 2 GeV to strike a tungsten target, thus producing a high-flux neutron beam via spallation.

The superconducting section of the linac, which represents the bulk of the machine's length and energy gain, is composed of three distinct types of cryomodules:

**The Spoke Section:** Following the normal-conducting front-end, the Spoke section is composed of 13 cryomodules.
Each cryomodule contains two double-spoke cavities operated at 352.14 MHz, with an optimum $\beta$ of 0.5, and a nominal accelerating gradient of 9 MV/m.
These handle the initial superconducting acceleration at lower energies from 90 MeV to 215 MeV.

**The Medium-Beta Section:** The first part of the elliptical linac, consists of 9 cryomodules.
Each cryomodule houses four 6-cell elliptical cavities operated at 704.42 MHz optimized for a geometric $\beta$ of 0.67 and nominal accelerating gradient of 16.7 MV/m.
This section accelerates the proton beam from 215 MeV to 571 MeV.

**The High-Beta Section:** The final and largest acceleration stage, consisting of 21 cryomodules.
Each cryomodule houses four 5-cell elliptical cavities also operated at 704.42 MHz optimized for a geometric $\beta$ of 0.86 and nominal accelerating gradient of 19.9 MV/m.
This section accelerates the proton beam from 571 MeV to 2 GeV.

The tunnel is designed to accommodate future upgrades or performance contingency. This is done by allocating a drift space corresponding to 10 elliptical cryomodules in the High-energy beam transport section of the tunnel.
Ultimately this could, in the future, allow for an extension up to ~7MW proton beams

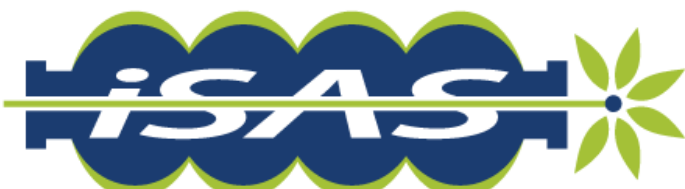

## 2.2. Technical Requirements of the Elliptical Cryomodules

### *2.2.1. High-Level beam requirements*

The ESS is a "game-changer" in neutron science specifically because of its long-pulse architecture, which allows it to deliver unprecedented peak brilliance. The main requirement of **5 MW average beam power** on target is roughly five times higher than existing spallation sources like the SNS (USA) or J-PARC (Japan). This power is achieved by striking the target with 14 pulses per second. Unlike "short-pulse" sources that use an accumulator ring to compress the beam into microseconds, the ESS delivers a **2.86 ms "long pulse** providing a neutron pulse that is perfectly timed for the next generation of "chopper" based instruments**.**

- **Repetition Rate:** 14 Hz; This frequency was selected as an optimal compromise between high average neutron flux and the "frame overlap" requirements of the neutron instruments (the time it takes for the slowest neutrons of one pulse to reach the detector before the fastest neutrons of the next pulse arrive).
- **Pulse length:** 2.86 ms**;** One of the longest in the international spallation community, a driver for neutron brightness, but with added technical challenges as the dynamic heat loads become considerably high.
- **Duty Cycle:** 4 %; Calculated as the pulse length (2.86 ms) multiplied by the repetition rate (14 Hz).
- **Peak Beam Current:** 62.5 mA**;** This is the instantaneous current flowing during the 2.86 ms window. Originally designed for 50 mA at 2.5 GeV, the requirement was increased to 62.5mA when the final energy was "descoped" to 2.0 GeV to maintain the 5 MW power target.
- **Final Beam Energy:** 2.0 GeV; This energy is sufficient to ensure high spallation efficiency (the number of neutrons produced per incident proton).

### *2.2.2. Cryogenic Scheme and thermal requirements*

Early on it became apparent that the **accelerator cryogenic system** should be split into three main components:

- Accelerator Cryoplant (ACCP),
- Cryogenic Distribution system (CDS)
- and the Cryomodules (CM).

This scheme made it possible to conduct parallel development of sub-systems and assign the development, construction and testing to various companies of In-Kind partners (see following section). Moreover, by splitting the cryogenic scheme into various deliverables, ESS managed to sequentially align the testing, installation and commissioning of each component, such that the project timeline can be efficiently managed.

The need of superconducting cavities to deliver such high accelerating gradients in a compact manner comes with requirements on operating temperature, stability and heat load management.

- **Operating Temperature:** 2.0 K (Saturated Helium II).
- **Pressure Stability:** ± 0.1 mbar to prevent frequency detuning caused by pressure fluctuations in the helium bath
- **Heat loads:** The importance of **sizing the Accelerator Cryoplant** early on arose from the long lead time for delivery and its need for commissioning the other parts of the cryogenic system.

In the conceptual design phase, where the elliptical cryomodule design was still in a preliminary state, the estimated requirements were as follows:

- Static heat loads to 2K: **13.23 W** (1 cryomodule)
- Static heat loads to 50K: **46.23 W** (1 cryomodule)
- Total Liquefaction for coupler cooling: **92 mg/s** at 5K (4 couplers)
- Dynamic heat loads from RF to 2K: **24.4 W** (4 High-beta cavities)
- Dynamic heat loads from beam to 2K: **3.25 W** (1 cryomodule)

The ACCP sizing took into consideration the possibility for a staged ramp up of operation with the use of additional contingency space as a future upgrade. To accommodate this, additional heat loads were included in the ACCP design in the following quantities:

- 3.0 kW at 2K
- 11.4 kW at 40-50K
- and additional 9g/s liquefaction (FPC),

As a rough breakdown, the ACCP cryogenic capacity accommodates heat loads with origin on the CDS (~28%), SPK CM (14%), MB CM (14%) and HB CM (48%).

The cryogenic capacity includes safety margins on static heat loads 1.3, and operational safety margin 1.15.

### *2.2.3. Space requirements*

High-level space requirements in the tunnel, have also been defined early on, stipulating the cross-section for accelerator systems, transportation and conventional facilities.

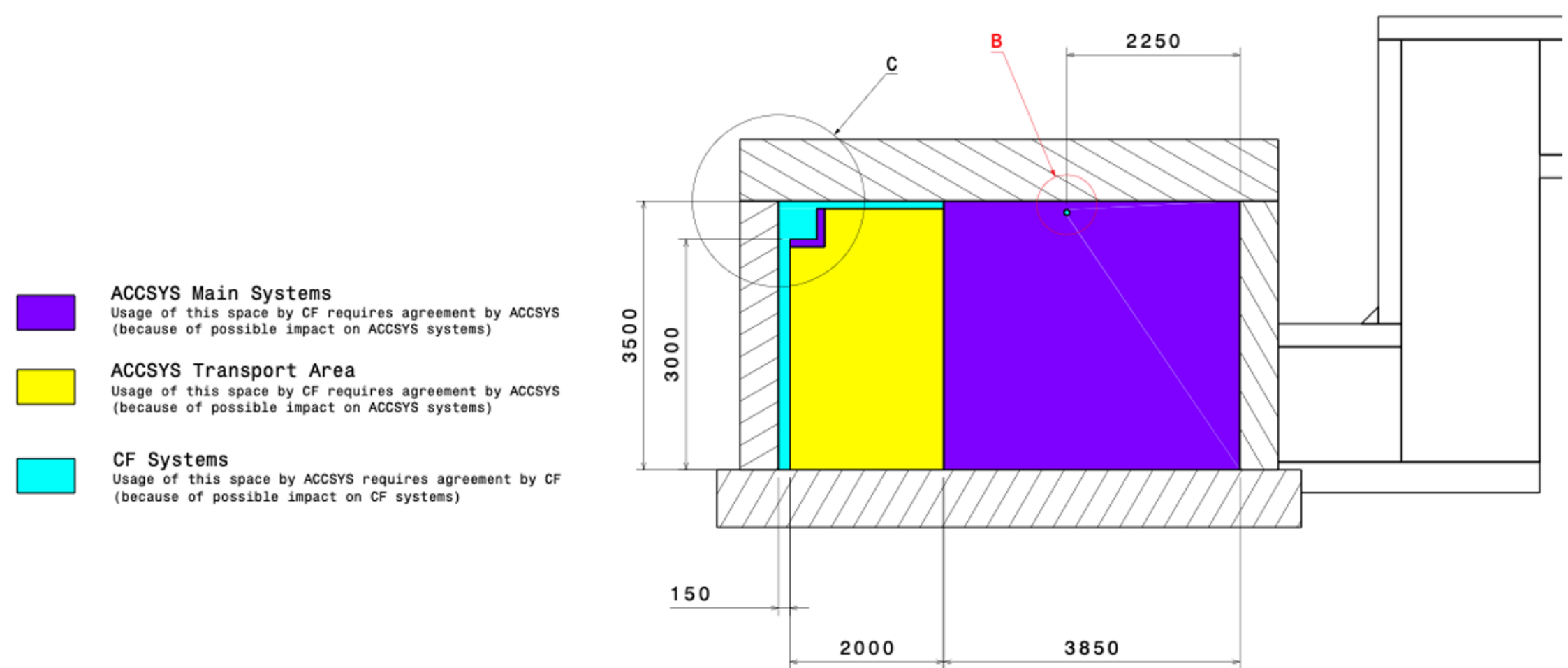

**Figure 1: Cross-section schematic of the space allocation in the accelerator tunnel.**

The cross-section of the Transport Area defined envelope is roughly 2.0 meter-by 3.5 meter, used for transporting all systems at different periods of the facility lifetime. This sets a maximum limit on the size of the components to be broken down.

The space reserved for the accelerator main systems is roughly 3.85 meter-by-3.5 meter, and designates the area to accommodate cryomodules, valve-boxes, supporting structures, RF distribution systems, cable trays, water cooling, instrument air, etc.

The efficient use of the tunnel space allows for sequential installation, maintenance, repair, and removal of components in case of need.

The preferred generic philosophy was to assign the space further away from the transportation areas for RF distribution and cable trays, followed by allocation of space for the cryogenic distribution system, and finally the space for the cryomodules. The transportation area is to be always cleared so as to allow for access and evacuation purposes.

### 2.3. ESS Project Timeline to date

The ESS project has the particularity of starting from a green field facility, with no prior institute or lab as a host. This represents both a major challenge and an opportunity, however it requires a steep learning curve to enable successful achievement of project deliverables. Below is a list of selected key points of the ESS project timeline:

- **2002:** The first technical proposal for a European Spallation Source is presented. Following the OECD recommendation that new high-intensity neutron sources be built in North America, Asia, and Europe.
- **2009:** Following an international site evaluation, **Lund, Sweden**, is officially selected as the host site, with the Data Management and Software Centre (DMSC) allocated to Copenhagen, Denmark.
- **2010:** The project transitions into its legal identity as **ESS AB**, a limited liability company owned primarily by Sweden and Denmark.
- **2013:** Completion of the **Technical Design Report (TDR)**, which serves as the blueprint for the current facility.
- **2013:** Initial funding commitments are secured from member countries, and the project begins establishing its international In-Kind partnership network.
- **2014:** The formal **Groundbreaking Ceremony** takes place in September, marking the start of civil construction.
- **2015:** ESS officially becomes a **European Research Infrastructure Consortium (ERIC)**, providing a legal framework for international governance.
- **2021:** First beam is successfully accelerated through the Radio Frequency Quadrupole (RFQ).
- **2025:** First beam is successfully accelerated through the **Superconducting linac** to the test beam dump.
- **2026:** Second beam commissioning run takes place with nominal RF pulses.

ESS aims to soon start sending proton beam pulses to the Target to achieve neutron Spallation for the first time, and with that the lead into Steady State Operations and science output, facilitating world-class scientific experiments to be performed at a number of different specialized beamlines.

### 2.4. Superconducting Linac Project Timeline to date

The superconducting linac project timeline comprising the different cavities and cryomodules is listed below. The project success relies on a strong collaboration between ESS and its in-kind partners, from the conceptual phase, qualification of components, installation and operation.

- **2013:** Publication of the **Technical Design Report (TDR)**, defining the three sections of the SCL: Double-spoke, Medium-beta elliptical, and High-beta elliptical cavities.
- **2014:** Start of Cryomodule Collaboration with partner institutes.
- **2015: In-kind agreements** with identified partners for Cryomodule and cavity production (CEA, IPNO later turned IJCLab, INFN and STFC).
- **2016:** Vertical test of 6 medium-beta prototype cavities at INFN-LASA.
- **2017: Medium-beta prototype cryomodule** (M-ECCTD) assembled at CEA and first tests.
- **2019:** Arrival of the **Medium-Beta prototype cryomodule** at the ESS site after road transport test from CEA-Saclay.
- **2020:** Arrival of the First Series Elliptical cryomodule at ESS.
- **2021:** Final Commissioning of **Test Stand 2 (TS2,** the dedicated facility for the Site Acceptance Testing (SAT) of all elliptical cryomodules), in Lund with the First Series cryomodule.
- **2022:** First **High-beta CM** arrives at ESS.
- **2023: Start of Series Installation:** The first two cryomodules were officially installed in the linac tunnel (one spoke and one medium-beta) and took part in the integrated commissioning of the cryoplant and cryogenic distribution system.
- **2024:** Completion of the **Spokes, Medium-Beta and High-Beta** cryomodule installations required for the first beam milestones (a total of 27 Cryomodules). In November, the first integrated cooldown of the superconducting section to **2 Kelvin** was successfully achieved.
- **2025:** First **beam commissioning phase** of the superconducting linac, achieving 860MeV.
- **2026:** Second **beam commissioning phase** of the superconducting linac with nominal RF pulses.

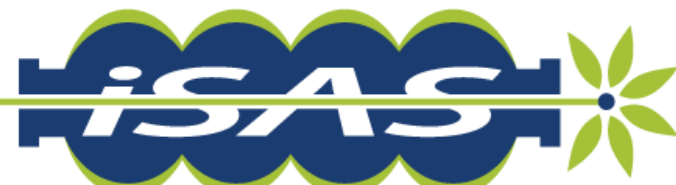

## 2.5. The In-kind Contribution Model

The European Spallation Source (ESS) distinguishes itself from traditional international scientific ventures through its reliance on the **In-Kind Contribution (IKC) model**. Rather than operating as a centralized procurement agency funded by liquid capital, the ESS functions as a collaborative hub that coordinates the technical outputs of its member states.

In this framework, a large fraction of the construction budget is satisfied by the direct delivery of hardware, personnel, and specialized services. The adoption of the In-Kind model was driven by both economic pragmatism and the pursuit of technical excellence allowing the project to tap into the "distributed intelligence" of Europe’s most renowned research institutions.

While the IKC model offers **unparalleled access to expertise**, it introduces significant logistical and managerial complexity. The primary **advantage lies in the ability to conduct parallel development**; multiple systems can be designed and prototyped simultaneously in different labs, accelerating the early phases of the project.

However, these benefits are balanced against the **challenges of maintaining uniformity across diverse organizational cultures**.

The ESS team act as a rigorous systems integrator, enforcing strict Quality Assurance (QA) and **standardized interface controls** to ensure that the various cryomodule components can be assembled, installed and operated as a single system.

Any **delay at a partner institute can create a "bottleneck" effect**, threatening the critical path of the entire project.

The production of the SRF cavities and their subsequent integration into cryomodules required a highly synchronized pipeline involving several of Europe’s leading physics labs.

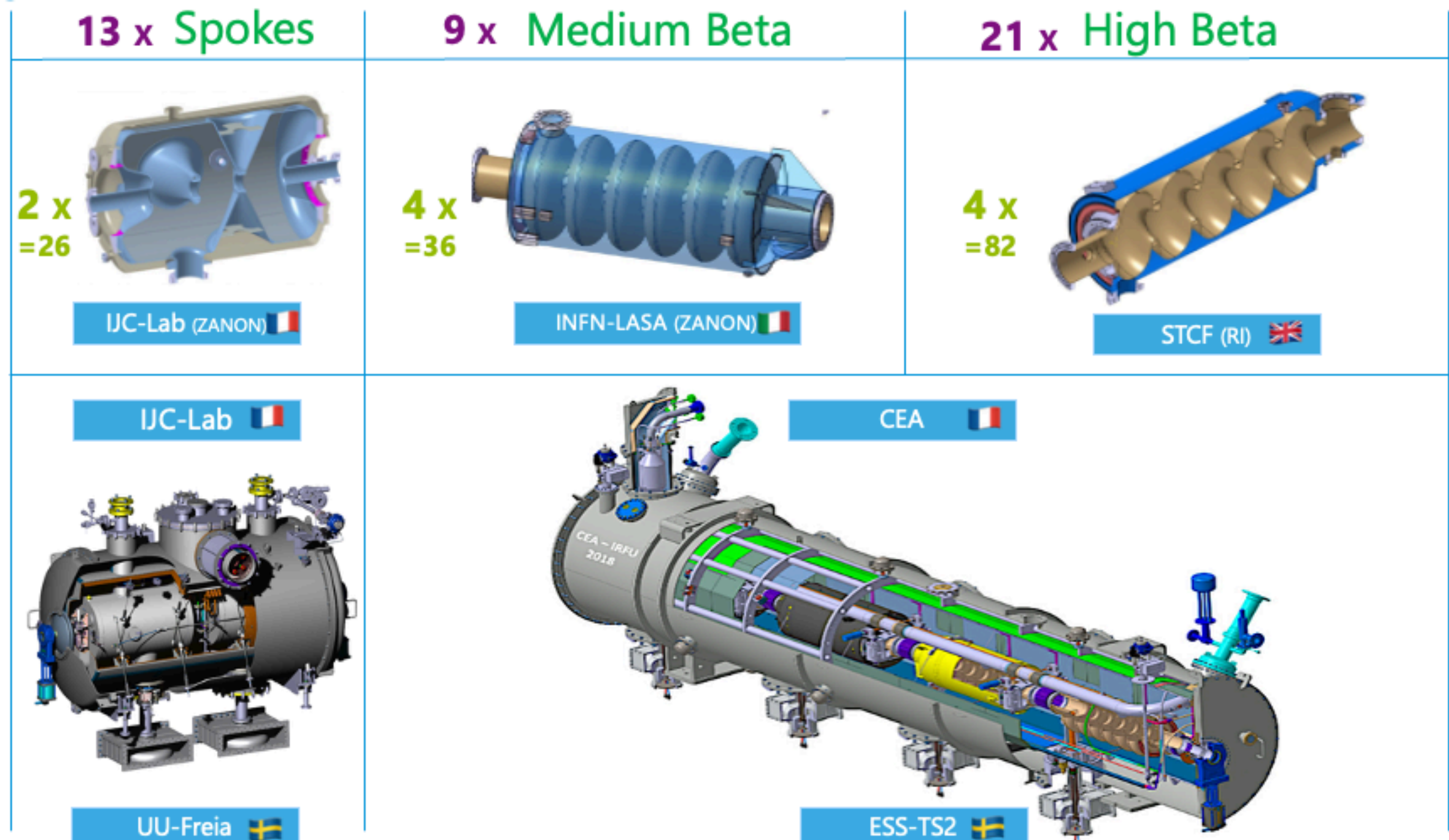

**Figure 2: Table with 3D representations of the different cryomodule designs used at the ESS, and the In-Kind partners responsible for delivering and testing those components.**

In this workflow, the Italian National Institute for Nuclear Physics (**INFN-LASA**) and the United Kingdom's Science and Technology Facilities Council (**STFC**) took responsibility for the medium-beta and high-beta elliptical cavities, respectively.

Once these niobium structures were fabricated and validated, they were transported to the Commissariat à l'énergie atomique (**CEA**) in Saclay, France. CEA acted as the central integration node, housing the massive cleanroom and other assembly operations required to integrate the cavities and other critical components into their complex cryomodule vessels.

The final process of qualification takes place at **ESS Test Stand 2** (TS2) where Site Acceptance Tests (SAT) are performed. In this process each module is subject to strict inspection, verifications and qualification tests covering multiple disciplines, making sure it is ready to be installed in the ESS accelerator tunnel and able to meet the requirements.

A visual summary for the different phases of the lifecycle of the cryomodules and where the responsibility for each phase lies can be found in Table 1

**Table 1: A summary for the responsible institutions for the different steps in the cryomodule development lifecycle**

| | SPOKES | MEDIUM-BETA | HIGH-BETA |
|---|---|---|---|
| Nb Procurement* | IPNO | INFN | STFC |
| Nb Eddy scanning* (location) | - | ESS (DESY) | ESS (DESY) |
| Cavity manufacturer* | IPNO | INFN (ZANON) | STFC (RI) |
| Cavity Testing* (Location) | IPNO (IPNO) | INFN (LASA and DESY) | STFC (STFC and DESY) |
| Cryomodule Assembly | IPNO | CEA | CEA |
| CM Testing (prototype) | IPNO | CEA | CEA |
| CM Testing (series) | UU (FREIA) | ESS (TS2) | ESS (TS2) |

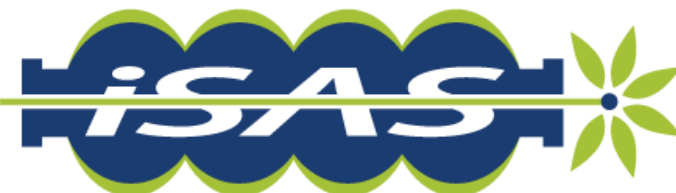

# 3. CRYOMODULE DESIGN PHASE LESSONS

## 3.1. Cryogenic architecture of cryomodules within the Accelerator Cryogenic System

### *3.1.1. Superconducting accelerator lattice*

The superconducting accelerator layout and lattice design requires beam optics components and beam instrumentation to allow for the focusing/defocussing, correction, and measurement of the beam position and phase. In the early conceptual design phase, various concepts were analysed: including having these components positioned inside or outside the cryomodule vacuum envelope, but also the required periodicity of these components for optimized beam transport.

The result of this analysis proposed a lattice with a periodic structure comprising four superconducting accelerating cavities followed by a drift space for beam dynamics correction and diagnostics.

A further analysis proposed as best configuration the positioning of the beam optics and diagnostics in stand-alone units (called Linac Warm Units, or LWUs) outside of the cryomodule envelope.

This choice largely dictated the architecture of the cryomodule, thus having a segmented approach where each elliptical cryomodule encompasses 4 superconducting cavities.

### *3.1.2. Accelerator Cryogenic System Architecture*

The superconducting linac is composed of three families of cryomodules, comprising 13 spokes, 9 medium-beta and 21 high-beta cryomodules.

- The cooling power required for operating the cryomodules is provided by the accelerator cryogenic plant (ACCP) enabling the operation of the superconducting cavities at 2K.
- Helium is supplied and returned to the ACCP via a cryogenic distribution system (CDS), composed of a multi-channel transfer line, an endbox, and multiple valve boxes (VBox)
- Each Valve Box interconnects to a single cryomodule forming a unified system.

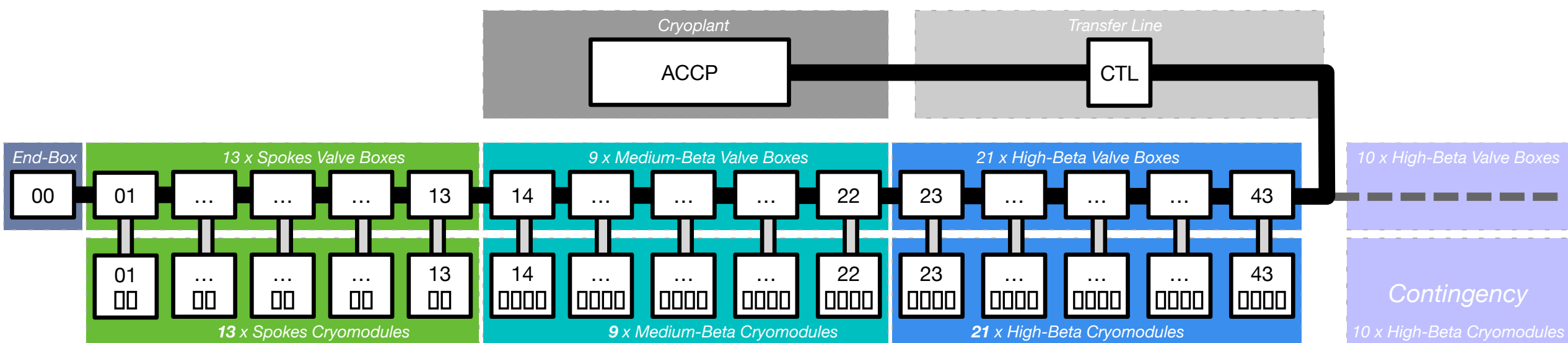

**Figure 3: Schematic of the cryogenic infrastructure supporting 2K operation of the series cryomodules, highlighting the segmented approach with independent cryomodule cryogenic control**

The grouping of the cryogenic architecture somewhat reflects the planned milestones for installation, with a breakdown in subsystems able to simplify the installation in a sequential way, and qualification of each system prior to interconnecting to the next one.

The ACCP was the first system to be installed, followed by the Cryogenics Distribution System (CDS) and finally the installation of cryomodules in a staged approach.

The choice of having a separate cryogenic distribution system from the cryomodules (external to the cryomodule vacuum vessel), arises from a number of constraints:

- First, the necessity to break down deliverables into more manageable IK contributions, aligning with the expertise of each collaborator.
- Second, the added complexity of technically combining both functions within one single envelope would require longer design time, more complexity on assembly and added risks in performance.
- Third, the advantage of a separated CDS and fully segmented system gives the opportunity for more flexibility in maintenance, repair and replacement as well as having the capability of ramping of the power of the machine in a staged manned by simply adding (pre-tested) cryomodules.

Following this strategy, the cryomodule installation could only start after the full installation and successful commissioning and validation of the Cryogenic Distribution System.

This was considered a schedule mitigation strategy, to avoid the complex removal of any systems placed in front of the cryogenic distribution system and to allow full access for repair, or removal, of any CDS components, if necessary, in case of unforeseen non-conformities found during the CDS qualification.

If the installation of Cryomodules or any other infrastructure (mechanical support stands, water circuits, instrument air piping, cabling, etc.) would have started in parallel with the CDS installation, this would have resulted in a large re-work and further delays to resolve any CDS non-conformities.

In addition, during the initial commissioning of the CDS Valve Boxes, and in the absence of a cryomodule, a “dummy CM return box” was required to allow for the cryogens to be circulated from the inlet circuits to the return circuits. The space necessary for the “dummy CM return box” limited the possibility to install any of the cryomodules into their final position.

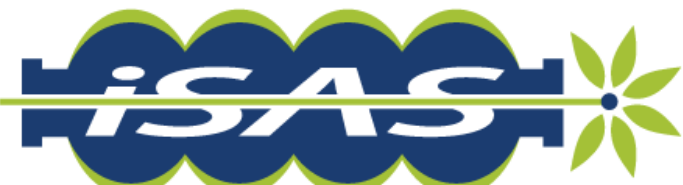

## 3.2. Cryomodule main Internal components

The cryomodule internal layout and integration involves a complex nesting of several critical sub-systems:

### *3.2.1. Cavity String*

A chain of four superconducting elliptical cavities made of Niobium (6-cell for Medium-Beta, 5-cell for High-Beta) in individual helium tanks are connected by bellows and supported by Titanium tie-rods. These tie-rods provide high mechanical strength with extremely low thermal conductivity. The cavities are covered by a magnetic shield. The beam line is centred at ~1.5m from the tunnel floor.

The cavity string is organized in such a way that the same cavity configuration and orientation is repeated, this means that the beam first enters a drift space in the so-called warm-cold transition of the vessel, then it enters the first cavity, followed by the power coupler antenna region and finally an inter cavity bellow drift region before the next cavity.

### *3.2.2. Cavity helium tank and diphasic line*

The superconducting cavities are each surrounded by a helium tank made of titanium. The helium tank has a cylindrical shape with a connection on the top (slightly tilted) for connection (welded) with the diphasic line. At the bottom each helium tank features a connection to the outlet of the "cooldown valve".

The helium tank also includes features for attachment of the Cavity tuning system on the upstream side (opposite side to the fundamental power coupler). The Helium level is measured in a parallel circuit to the tanks, ensuring that the top of the cavities are covered by liquid and that the liquid level is reaching the diphasic line, acting as a phase separator and communicating channel between each of the cavities in the individual cryomodule.

### *3.2.3. Fundamental Power Coupler*

One fundamental power coupler is positioned per cavity, able to transfer up to 1.1 MW power over a 3.2 ms pulse width at a repetition rate of 14Hz. The design relies on a simple coaxial line at 50 ohm and one single planar lumina window including impedance-matching and TiN anti-multipacting coating. The main difference between the high-beta and medium beta couplers is the penetration depth of the antenna. The antenna is water cooled, and the power coupler is helium cooled using a double walled tube

### *3.2.4. Cavity Tuning system*

Each cavity is equipped with a Cold Tuning System (CTS) consisting of a stepper motor for coarse tuning and piezo actuators for fast compensation. The mechanical tuners are set to compensate for frequency mismatch after the cavity is cooled, and the piezoelectric tuners provide real-time dynamic control from the supplied RF to fine-tune the cavity to compensate for the Lorentz force detuning (LFD).

The slow-tuner combines lever arms and excentric shafts actuated by a motor gearbox and main screw assembly to provide the slow mechanical tuning action with a range of +/-3mm used for cavity elongation (pull force) corresponding to a tuning range of 650kHz and 590kHz for MB and HB cavities. The cavities are pre-tuned, however, requiring only about 200kHz final adjustment.

The fast tuner is comprised of two piezo elements that act with the LLRF to provide the final tuning during the pulse and beam compensation, with a maximum adjustment of 1kHz.

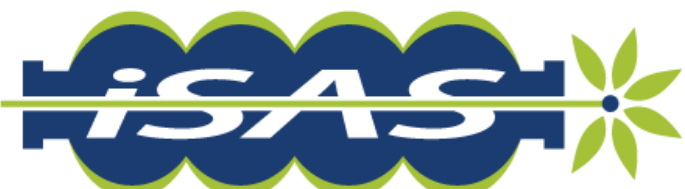

### *3.2.5. The Thermal Shield:*

A multi-segmented aluminium shield that surrounds the cold mass. The thermal shield is covered in Multi-Layer Insulation to minimize radiative heat transfer from the vacuum vessel.

### *3.2.6. The Vacuum Vessel:*

A 6.6-meter long, 1.3-meter diameter stainless steel cylinder.

- It features four ports for the Fundamental Power Couplers at the bottom.
- It has one cryogenic jumper at one of the extremities facing away from the transport side that provides vacuum insulation and interconnection of all the cold lines to the Valve Box on the Cryogenics Distribution.
- It features eight access hatches (4 on each side of the vessel), allowing for inspection, access and the possibility for repair of some of the internal components.
- It features several flanges with mounted instrumentation feedthroughs (high vacuum-grade), which are distributed along the top of the vessel.
- It features a pressure relief parallel plate and vacuum valve for interconnection with the insulation vacuum system pumping manifold.
- It features 2 beam line all-metal gate valves for beam line isolation and interconnection with LWUs, where particle-free pumping groups are present.

## 3.3. Mechanical Design and Component Integration

As described previously, each elliptical cryomodule comprises 4 superconducting cavities.

An important decision taken early on was to have a "universal" Cryomodule design concept, for both the medium-beta and high-beta sections of the cryomodule. This decision turns out to have major advantages as most of the internal components can be integrated in a standardized way. Moreover, it takes advantage of standardization of the components design as well as manufacturing, procurement, and logistics, as well as the minimization of changes in tooling.

To enable this decision, an important effort had to be made to accommodate the main differences in geometry between the medium-beta and the high-beta cavities (6-cell vs. 5-cell). To note, there was a need to have different intra-cavity bellows to compensate the change in length of the different cavity designs.

Moreover, this decision had to be communicated early on to the various IK contributors (as main suppliers of the cavity package), and a proper definition of common and different interfaces needed to be clearly established.

## 3.4. Internal integration of cryomodule components

The mechanical design of the ESS elliptical cryomodule uses a **"Space-frame"** concept.
Unlike other cryomodule designs where the cold mass is suspended directly from the vacuum vessel or uses a single supporting system from the bottom, the ESS design utilizes an internal skeletal structure that decouples the vacuum vessel's mechanical tolerances from the high-precision alignment of the cavity string. It allows for most of the components to be integrated before cryostating into the vacuum vessel, where from then on, access to components becomes limited.

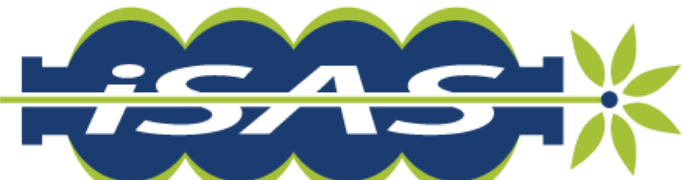

#### *3.4.1. The Space-frame Support System*

The "Space-frame" is a longitudinal aluminium truss structure that serves as the backbone of the cryomodule. The truss structure is made of longitudinal and circular sub-components and acts as a supporting element to the cavities, decoupled from the cold mass (meaning it does not experience relevant thermal contraction).

The space-frame provides several key advantages for the iSAS sustainability and industrialization goals:

- **Decoupled Alignment:** The cavity string is aligned to the space-frame outside of the vacuum vessel, allowing for easier access and higher precision during assembly.
- **Structural Stability:** It supports the entire cold mass (cavities, magnets, and thermal shield) and maintains their relative positions during the thermal contraction associated with 2 K cooling.
- **Standardization:** The same space-frame geometry is used for both Medium-Beta and High-Beta cryomodules, facilitating a "universal" assembly procedure.

### 3.5. Cryomodule Thermal Design and cryogenic circuits

The thermal management of the ESS elliptical cryomodules is based on a "nested" temperature approach designed to minimize heat ingress to the 2 K superconducting environment.

The design relies on three temperature levels, each serving a specific thermal and mechanical purpose.

To achieve high thermal efficiency, the design aims at integrating components while minimizing heat inleak in an effective way.

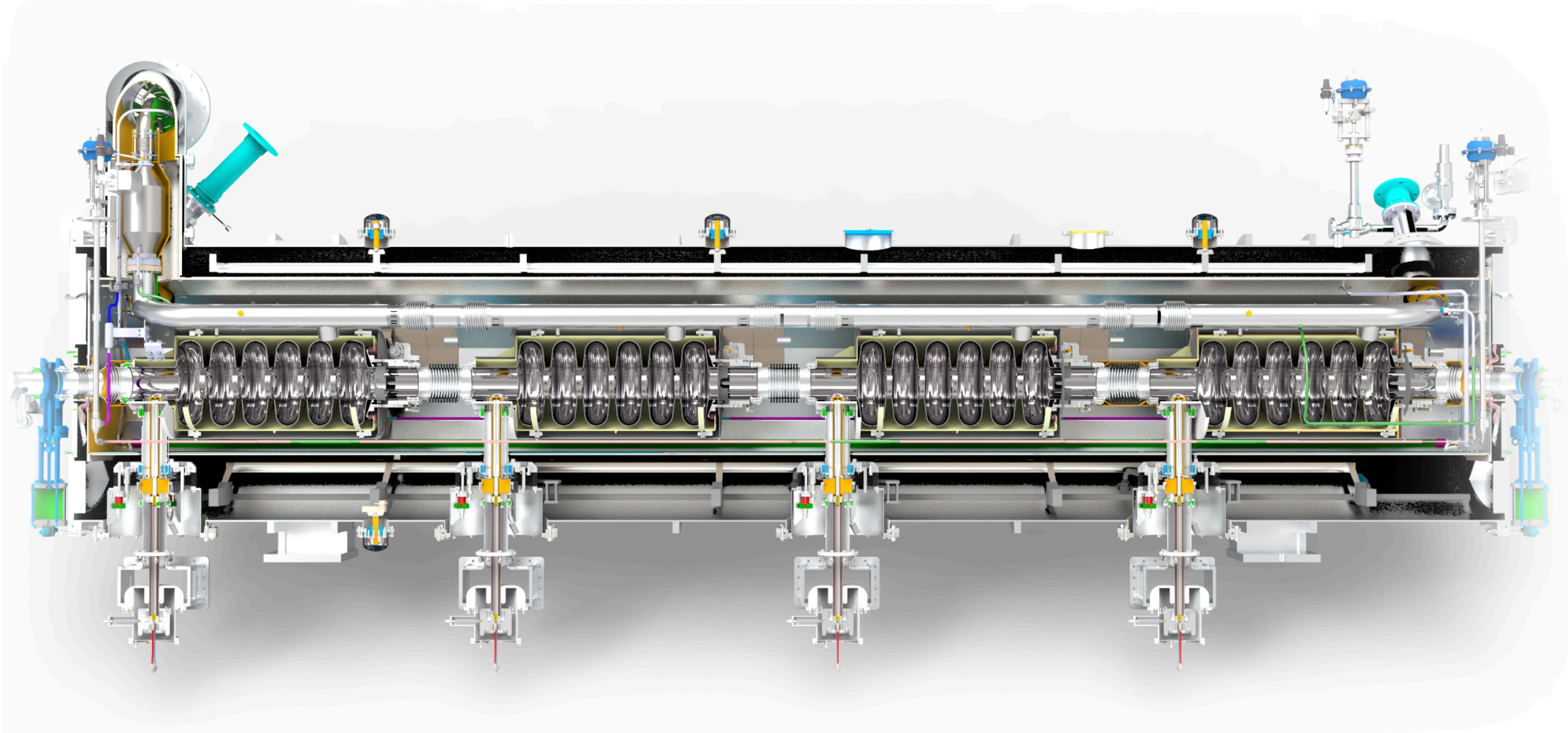

**Figure 4: Artistic display of the elliptical cryomodule rendering showing the major components and circuits.**

### *3.5.1. Cold process lines*

**The 40–50 K Thermal Shield Circuit:** This circuit uses high-pressure (19.5 bara) helium gas to intercept most of the radiant heat from the room-temperature vacuum vessel. These circuits cool the thermal shield envelope (made of aluminium). The thermal shield acts as a thermal intercept of any heat transported by thermal conduction from ambient temperature (power couplers, support structure, warm-to-cold transitions, wiring of instrumentation, etc.).

**The 4.5 K cooling Line:** This circuit mainly provides the helium supply used to fill the cavity vessels. However, before reaching the cavities, the helium circuit is split in 3:

- One branch supplies the expansion valve (CV-02) used for cooldown, this valve discharges at the bottom of the helium tanks.
- The second branch is pre-cooled via a heat exchanger before expansion through the main Joule-Thomson valve (CV-01) feeding the upper part of the helium tank for liquid production at 2K. The pre-cooling maximizes the liquefaction efficiency.
- A third branch supplies cooling to each of the power couplers to remove the heat transported by conduction between the ambient outer shell and the cryogenically cooled internal components of the power coupler. The helium flows from the base of each coupler near the cavity, following the double wall internal structure of the power coupler, before the exhaust helium exits the cryomodule vessel and is recovered at the cryoplant at room temperature.

**The 2 K Saturated Superfluid Bath:** This is the primary cooling circuit for the SRF cavities. Liquid helium at 4.5 K is expanded through a Joule-Thomson (JT) valve into a **Phase Separator** (commonly referred to as the diphasic line), where it reaches a saturated state at approximately 31 mbar to maintain the cavities at **2.0 K**. Under nominal conditions, the superconducting cavities operate fully immersed in the superfluid liquid helium bath, with the helium level stabilized inside the diphasic line which interconnects the four helium tanks at the top. The evaporated flow is then pumped out through the VLP line (vapour low pressure circuit) towards the ACCP cold compressors.

In addition, it is worth referring to the importance of pre-cooling of the 4.5K cooling line via the **heat exchanger when operating at 2K**, it allows recovery of the available frigories from the LP line and significantly enhances the efficiency of the “Joule-Thomson” (JT) expansion valve feeding the 2 K bath.

Moreover, for each cryomodule, the helium level in each tank is measured via redundant superconducting wire level gauges placed in a parallel circuit to the helium tanks, allowing communication with both the bottom and the top of the tanks. A simplified schematic of the main circuits inside a cryomodule can be found in Figure 5, and a full process and flow diagram is presented in Figure 6.

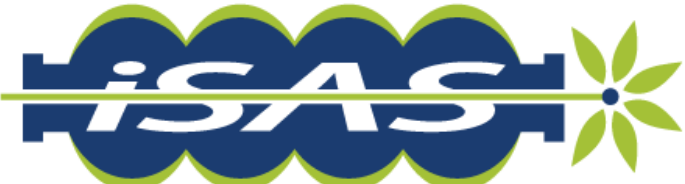

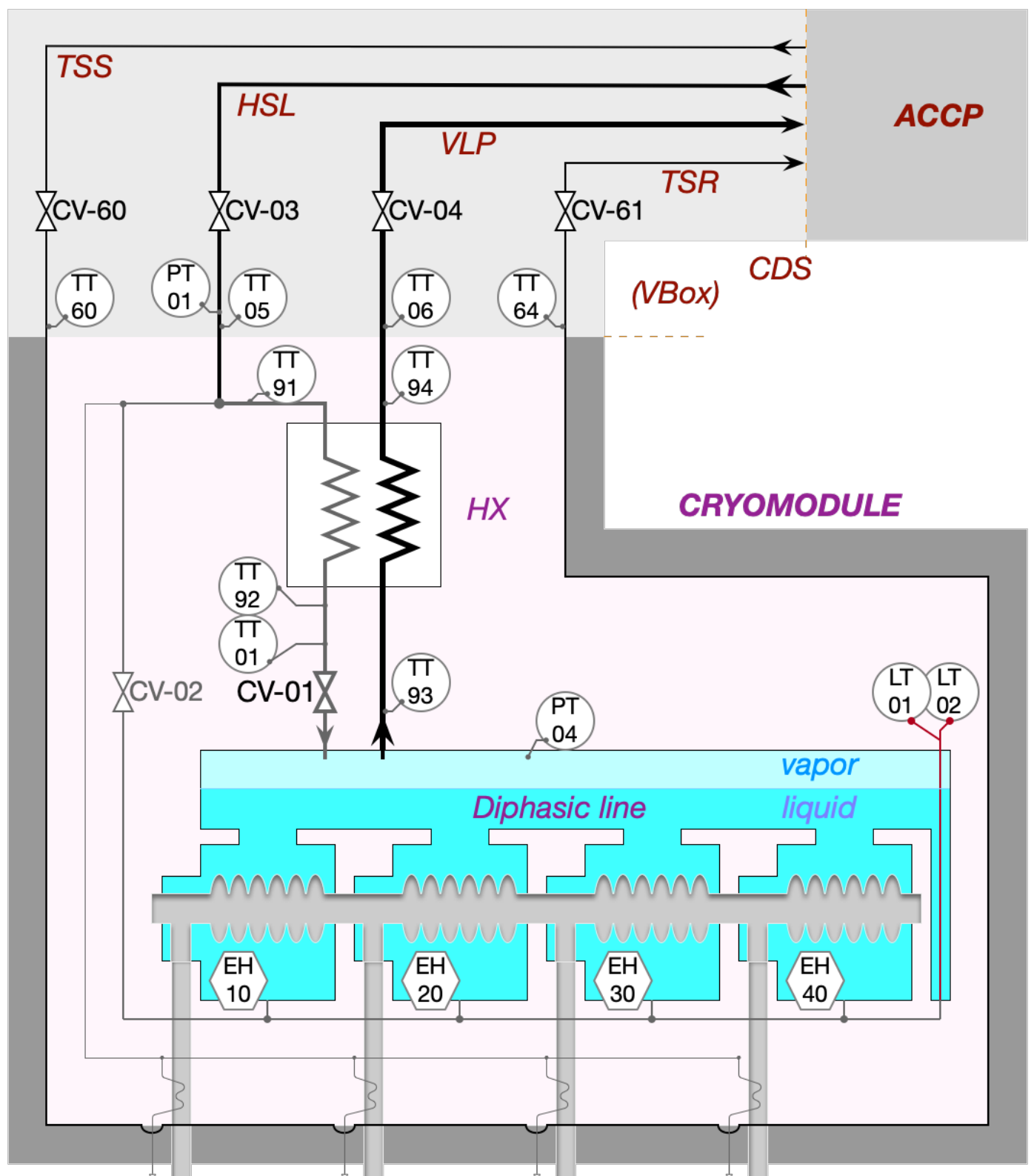

TSS : Thermal Shield Supply
TSR : Thermal Shield Return
HSL : Helium Supply Line
VLP : Vapour Low Pressure Line
PCDWC : Power Coupler Double Wall Cooling
ACCP : Accelerator Cryoplant
CDS : Cryogenic Distribution System
VBox : Valve Box
HX : Heat Exchanger
LT : Level Transmitter (level gauge)
TT : Temperature Transmitter (temperature sensor)
CV : Control Valve
EH : Electrical Heater

Figure 5: A highly simplified schematic of the main circuits found inside a cryomodule, and nomenclature used.

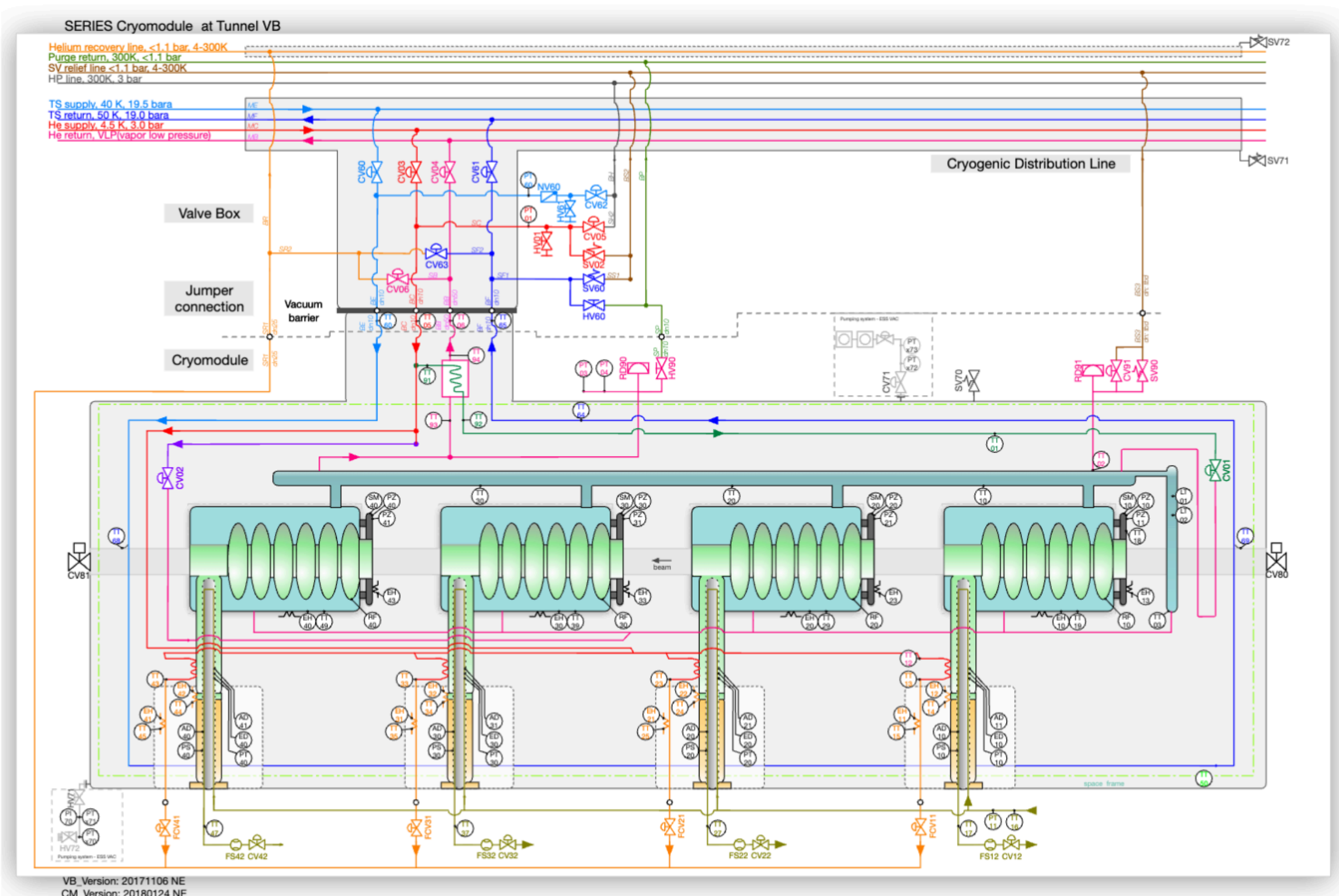

Figure 6: A process and flow diagram representation for an elliptical cryomodule is shown.

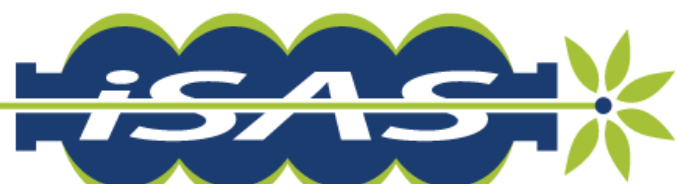

#### *3.5.2. The Cryogenic "Jumper connection"*

The superconducting elliptical cryomodules are integrated with the Cryogenic Distribution Line using a sophisticated cryogenic interface known as the Jumper Connection. This assembly serves as the functional bridge between the Cryogenic Distribution Line and the cryomodule (CM), ensuring the delivery of refrigerants while maintaining the stringent thermal and vacuum isolation.

The jumper connection is a vacuum-insulated, multi-channel transfer branch with 4 cryogenic circuits crossing the interface (TSS, TSR, HSL, VLP). The Valve Boxes part of the Cryogenic distribution System contain additional valves and circuits to allow for off-normal operations (see next section). The valve Box also contains a vacuum barrier that separates the insulating vacuum of the CDS from the insulating Vacuum of the Cryomodules.

**Mechanical coupling and thermal compensation:** Because the internal process lines shrink significantly when cooled the jumper connection incorporates **hydroformed bellows**. These bellows provide the necessary mechanical flexibility to absorb axial and lateral displacements without compromising the integrity of the vacuum or the internal piping.

**Multi-purpose interface configuration:** Since all Cryomodules undergo SAT (Site acceptance testing) TS2 prior to installation in the tunnel, it was opted for the capability for mounting and dismounting the cryogenic circuit interconnections for the SAT, while having welded interconnections in the tunnel. This resulted in 2 different interfaces depending on the location. For the configuration during SAT the cryogenic interconnection is done with Swagelock VCR (for the TSS, TSR, HSL circuits) and Conflat (for the VLP circuit), while for the tunnel this interface is cut away and final orbital welds are performed.

**The physical connection of the jumper** is established through a sequence of high-precision operations:

- **Internal Welding/ or tightening of compressible joint:** The process lines are joined using orbital TIG welding. Each weld undergoes rigorous Non-Destructive Testing (NDT), including X-ray inspection and helium leak mass spectrometry.
- **Shield Continuity:** The jumper's thermal shield is mechanically and thermally bridged to the cryomodule's shield to ensure no "gaps" and covered with MLI blankets.
- **The Vacuum Sleeve:** A "sliding sleeve" on the outer vacuum jacket is positioned over the junction and welded to the cryomodule's vacuum vessel.

#### *3.5.3. Auxiliary process lines*

Important to notice is the existence of 4 auxiliary process lines that constitute part of the CDS with external interconnections to the cryomodule, namely:

- **Helium Recovery Line**: This circuit is used for collecting the helium from each PCDWC (Power Coupler Double Wall Cooling) circuit and it also has the function of returning the flow to the cryoplant in case of need to warm-up/ cool-down a single cryomodule.
- **Purge Return Line**: This circuit is used prior to cool-down and to enable pumping, purging and purification of the circuits, allowing removal of most of the air contaminants before cold operations.
- **SV relief line**: This circuit is used as a collector for the discharge of pressure relief devices.
- **HP relief line**: This circuit is used to enable the warm-up of a single cryomodule or to enable additional gas handling capabilities at warm.

Additionally, 2 helium collector lines are positioned, each connected to a cryomodule rupture disk and used in case of emergency discharge:

- **Safety helium collector**: (missing from the diagram): Is used in the accelerator tunnel as a helium collector in case of major accidental scenarios, when personnel need to access the tunnel and helium is vented through rupture disks. This helium is not recovered to the cryoplant, and is instead lost.

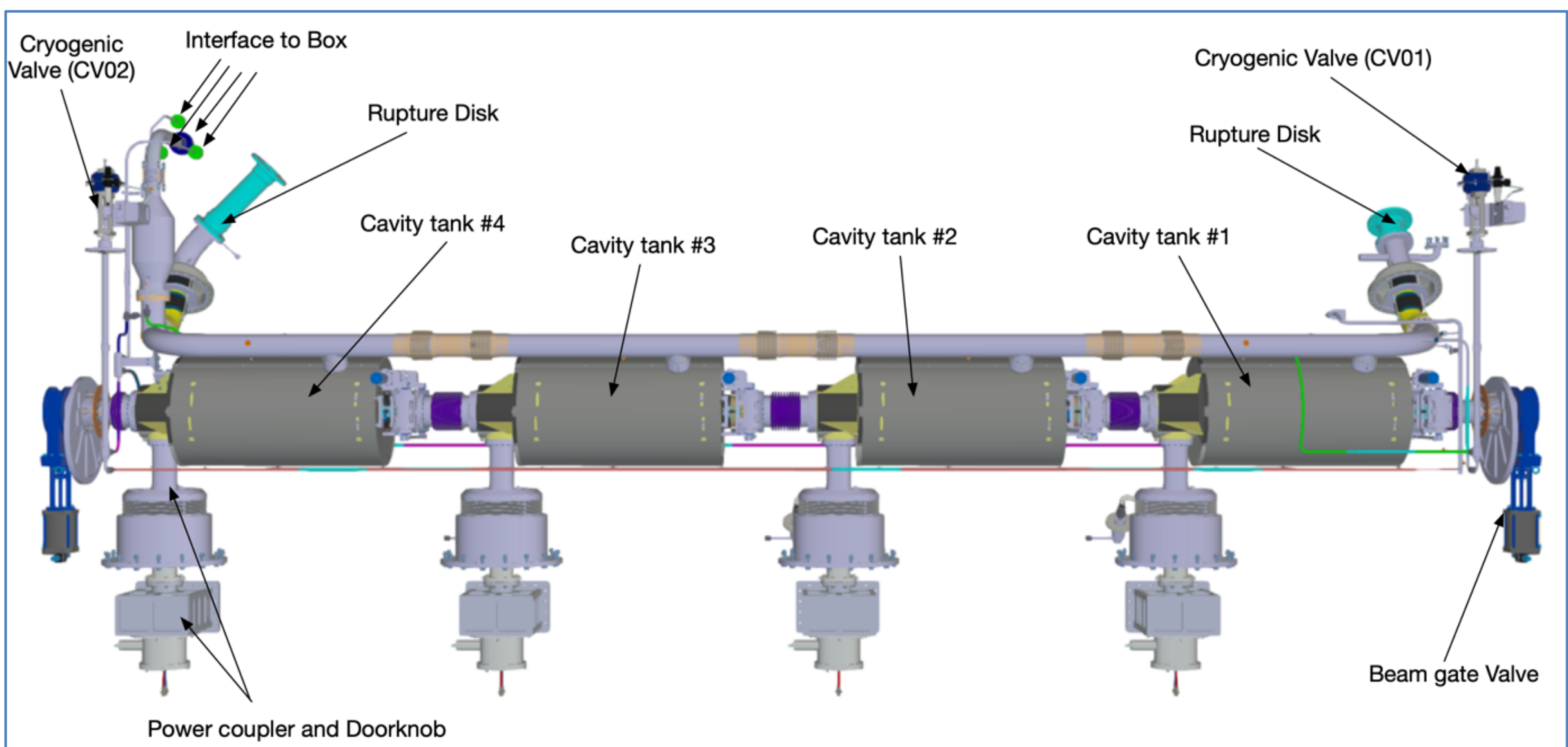

**Figure 7: Front side view of the dressed cavity string that sits inside the vacuum vessel.**

### 3.6. Pressure Equipment Protection

To protect the cryomodule equipment from overpressure, a multi-layered safety strategy is employed, primarily focused on the helium tank circuits.

The cryomodule pressure equipment and circuits are classified under the European Pressure Equipment Directive 2014/68/EU PED, and according to Article 4.3, which result in a Maximum Allowable Working Pressure (MAWP) of 1.04 barg.

To adhere to this, the primary ultimate safety devices are two burst disks (RD90 and RD91), both calibrated to the same burst pressure, set just below the MAWP. These disks discharge directly into a dedicated "Safety Helium collector" maintained at atmospheric pressure. Any helium released through this specific path is not recovered.

To prevent the activation of these non-reclosable burst disks and minimize helium inventory loss, the system utilizes two lower-pressure relief tiers that prioritize helium recovery. The first line of defense is a cryogenic fast-acting control valve (CV91) that opens at a set pressure defined in the control system and typically equal to 1.5 bara. If pressure continues to rise, a spring-loaded pressure relief valve (SV90) acts as a second tier with a crack pressure of 0.67 barg. Both of these devices discharge into a "Safety Relief line" connected to the cryoplant, where the helium is fully recovered.

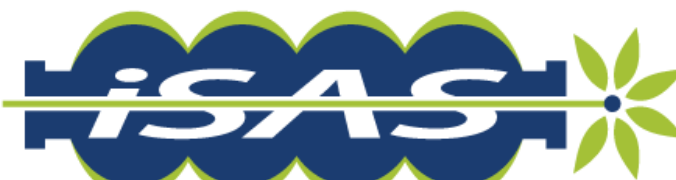

Beyond the main helium circuits, the Isovac (isolation vacuum) system is equipped with its own dedicated overprotection valve. This valve is set to a much lower pressure of 0.1 barg to protect the vacuum vessel itself. While the valve box contains the safety devices for all other auxiliary circuits, the cryomodule’s primary protection remains a balance between controlled recovery via relief valves and ultimate atmospheric discharge via burst disks.

All flanges belonging to the cryomodules sub-atmospheric circuits are helium guarded, this is achieved by having a double o-ring seal, where the interspace between the 2 o-rings in kept with an atmosphere of helium. This prevents direct air ingress in case of the o-ring seal degradation.

# 4. MANUFACTURING AND ASSEMBLY LESSONS

## 4.1. Industrialization of Series Production

The assembly of all elliptical cryomodules is under the IK contribution of **CEA-Saclay**.

A total of 30 series cryomodules were specified in the contribution, with additional agreements for the prototyping and testing of one Medium-Beta prototype cryomodule and one High-Beta prototype cryomodule.

The prototyping phase served as the final verification step prior to large scale purchase of the different components.

Due to planning and schedule challenges the number of improvements from prototyping to series production was limited, mainly due to the large lead time of many components that needed to be purchased in advance and prior to the final integration and validation in a prototype configuration.

## 4.2. Assembly phases

To transition from a laboratory base prototype assembly into a "factory" output of roughly **one cryomodule per month**, CEA implemented an industrial partnership with B&S International company**.**

In the early stages of the partnership, the assembly steps were performed by expert CEA personnel and used guidance for improving and defining the detailed assembly procedures. The gradual transition to the company performing all the assembly steps happened under guidance of CEA and lasted for at least 3 cryomodule full assembly cycles.

The cryomodule assembly procedure consists of the following phases:

- **Clean Room Operations:** The process begins with the assembly of power couplers onto the clean cavities, followed by the integration of the four-cavity string within an ISO 4 environment to prevent particulate contamination.
- **Internal Component Integration:** This phase involves the mounting of cryogenic equipment and other internal hardware directly onto the cavity string.
- **Cold Mass Assembly:** The prepared string is installed into a specialized spaceframe, secured by titanium support rods, and integrated with the thermal shield.
- **Cryostating:** This step involves the precision insertion and alignment of the complete cold mass assembly into the vacuum vessel.
- **Final External Integration:** The assembly is completed with the installation of coupler bells and external subsystems.

There are many details at each assembly step, requiring very clear instructions but also relying on the transferrable knowledge that is acquired when dealing with cryomodule technologies. The outcome, which can be measured in the number of non-conformities or ultimately the reduced cavity or cryomodule performance, strongly relies on having the correct competencies and expertise, and very clear and tested assembly procedures.

Each phase is comprised of assembly sub-steps and includes important verification checks dedicated to proper maintenance of uniform quality assurance depending on specific quality controls.

### 4.3. Quality Assurance/ Quality Control

The success of this series production is measurable through the minimization of non-conformities and the preservation of cavity performance. This relies on highly detailed assembly instructions and the implementation of a robust quality framework:

#### *4.3.1. The "Traveller" System*

Each phase is broken down into specific sub-steps, each requiring mandatory verification checks. These are recorded in a **Digital Traveller**—a comprehensive lifecycle record that ensures uniform quality control across the production series. The traveller tracks critical parameters (such as bolt torque values for vacuum flanges; results of intermediate helium leak tests; results from survey and alignment, clean room particulate counts during string assembly, etc.).

#### *4.3.2. Non-Conformity Management*

**Deviation:** When a deviation from the established procedure or a technical specification occurs, it is documented as a **Non-Conformity (NC)**. The outcome of the assembly strongly depends on clear and tested procedures for managing these deviations.

Typically, non-conformities can be classified at different levels (minor, major, critical), classified by the different levels of impact in ascending order of severity. Some examples of minor, major and critical non-conformities are as follows:

- Does not affect performance, interfaces, or safety. (e.g. a cosmetic scratch on a non-functional surface). **Minor.**
- Affects a physical interface (e.g. a flange misalignment) or a performance parameter. **Major.**
- Fundamental violation of requirements or irreparable damage to a primary component. **Critical.**

**Mitigation:** When an NC is raised, the workflow follows a rigorous disposition path. The technical team experts evaluate the impact on cryomodule performance and propose actions, typically:

- Rework: Bringing the part back into full compliance with the original drawing (e.g., re-welding a joint).
- Repair: Fixing the part so it is functional, even if it no longer meets the original design exactly (requires a Concession Request).
- Use-As-Is: Accept the deviation because the impact on the overall linac performance is deemed negligible (requires high-level approval from ESS).
- Scrap/Reject: The component is removed from the series and replaced.

**Approval:** Depending on the severity, ESS must approve the "as-built" configuration through a formal concession process before the assembly moves to the next workstation.

This rigorous combination of industrial expertise and stringent quality controls ensures that the final cryomodules meet the demanding operational requirements of the ESS linear accelerator.

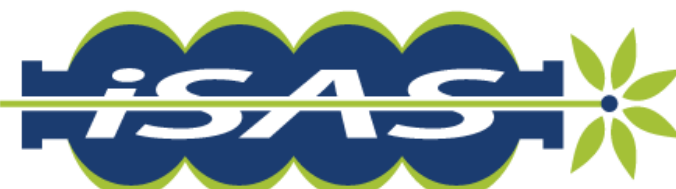

# 5. TESTING AND COMMISSIONING

The cryomodule qualification Test Stand (TS2) at ESS has been commissioned fully since 2021 and is performing Site Acceptance Testing of all Elliptical Cryomodules (CMs) before installation in the linac tunnel.

## 5.1. Site Acceptance Tests

The final site acceptance (SAT) tests at ESS allow for the determination of the readiness for installation of all cryomodules. A thorough inspection protocol is followed to assess that all requirements are met. This allows for the correct validation of previously measured parameters as well as to validate the correct functioning of other systems (e.g. dimensional control and final integration of components, absence of damage during transport, cold tuning system operation associated with tuning sensitivity measurements, power coupler conditioning, cryogenic heat loads, absence of leaks, etc.).

Once the final site acceptance is granted, the cryomodule status becomes ready for installation (RFI) and will follow the ESS installation, testing and commissioning plan.

The SAT of each Elliptical Cryomodule is the responsibility of ESS, with important participation by the IFJ-PAN team as Poland In-Kind collaborators to the project.

The workflow for cryomodule site acceptance testing at ESS TS2 is divided into 9 distinct phases where the cryomodule is moved through various locations of the TS2 area, see Figure 8. Dedicated transportation and transfer equipment as well as a positioning stand allow for the cryomodule to be safely transferred and positioned at each of the various phases.

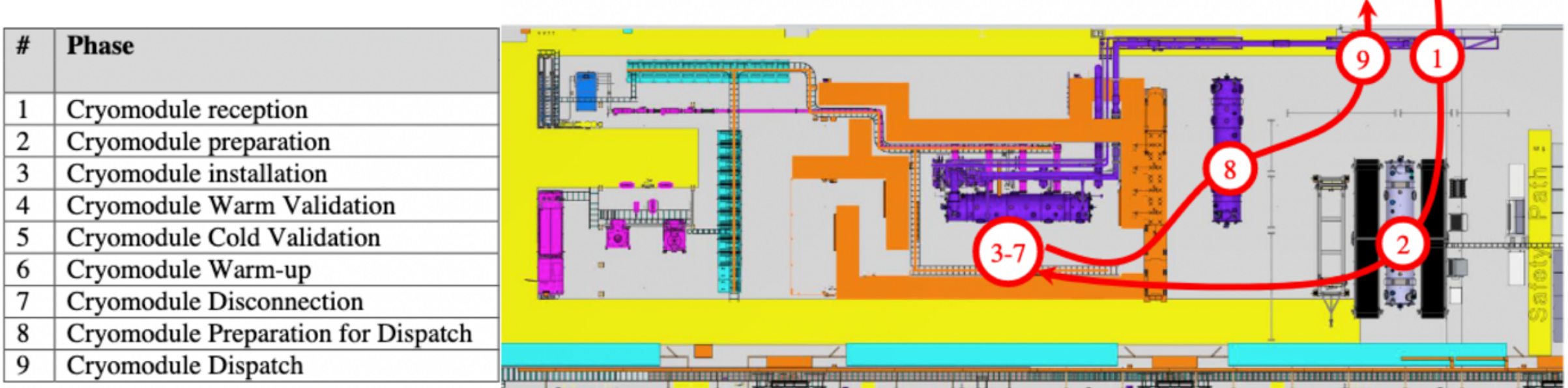

| # | Phase |
|---|---|
| 1 | Cryomodule reception |
| 2 | Cryomodule preparation |
| 3 | Cryomodule installation |
| 4 | Cryomodule Warm Validation |
| 5 | Cryomodule Cold Validation |
| 6 | Cryomodule Warm-up |
| 7 | Cryomodule Disconnection |
| 8 | Cryomodule Preparation for Dispatch |
| 9 | Cryomodule Dispatch |

**Figure 8:** Phases and workflow of the Cryomodule at Test Stand 2 for Site Acceptance Tests (SAT).

Figure 9 **Erreur ! Source du renvoi introuvable.**summarizes the main activities occurring at each of the phases during SAT. The bulk of the testing occurs when the cryomodule is inside the bunker (phases 3 through 7).

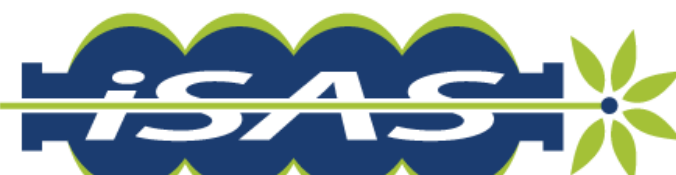

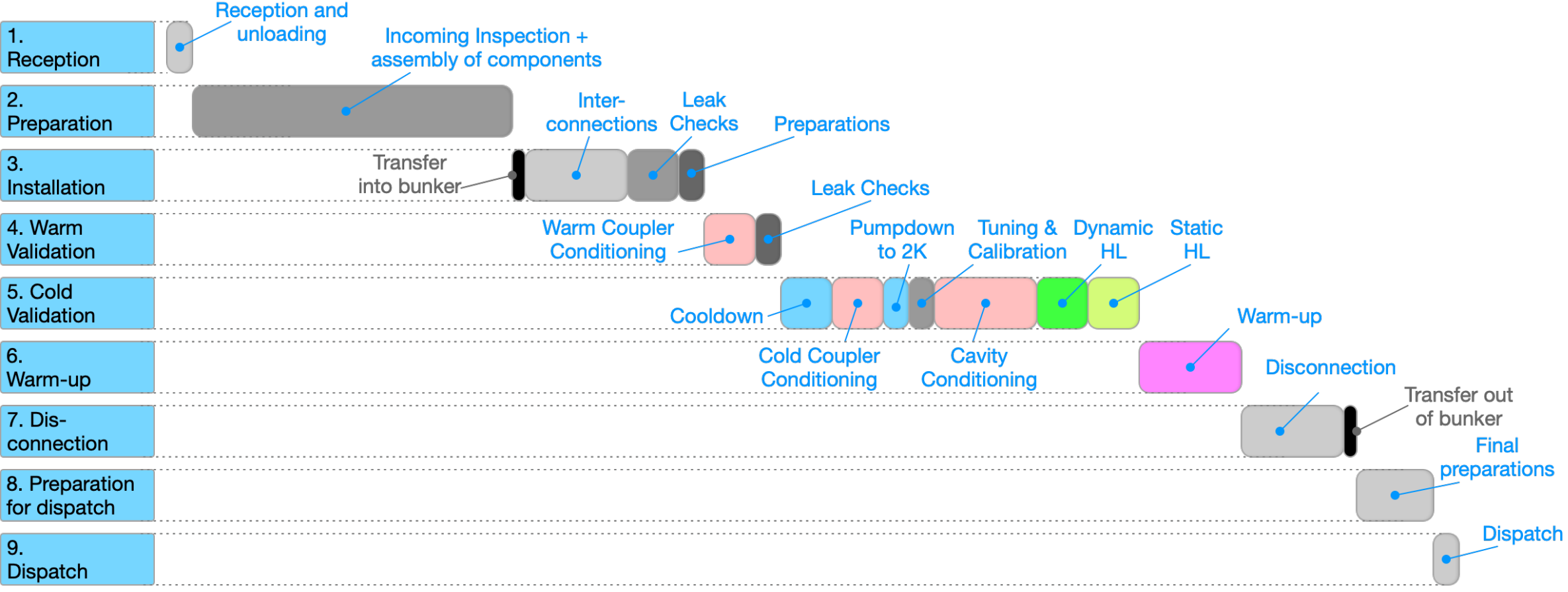

Figure 9: Net plan at each phase of the cryomodule SAT and main activities.

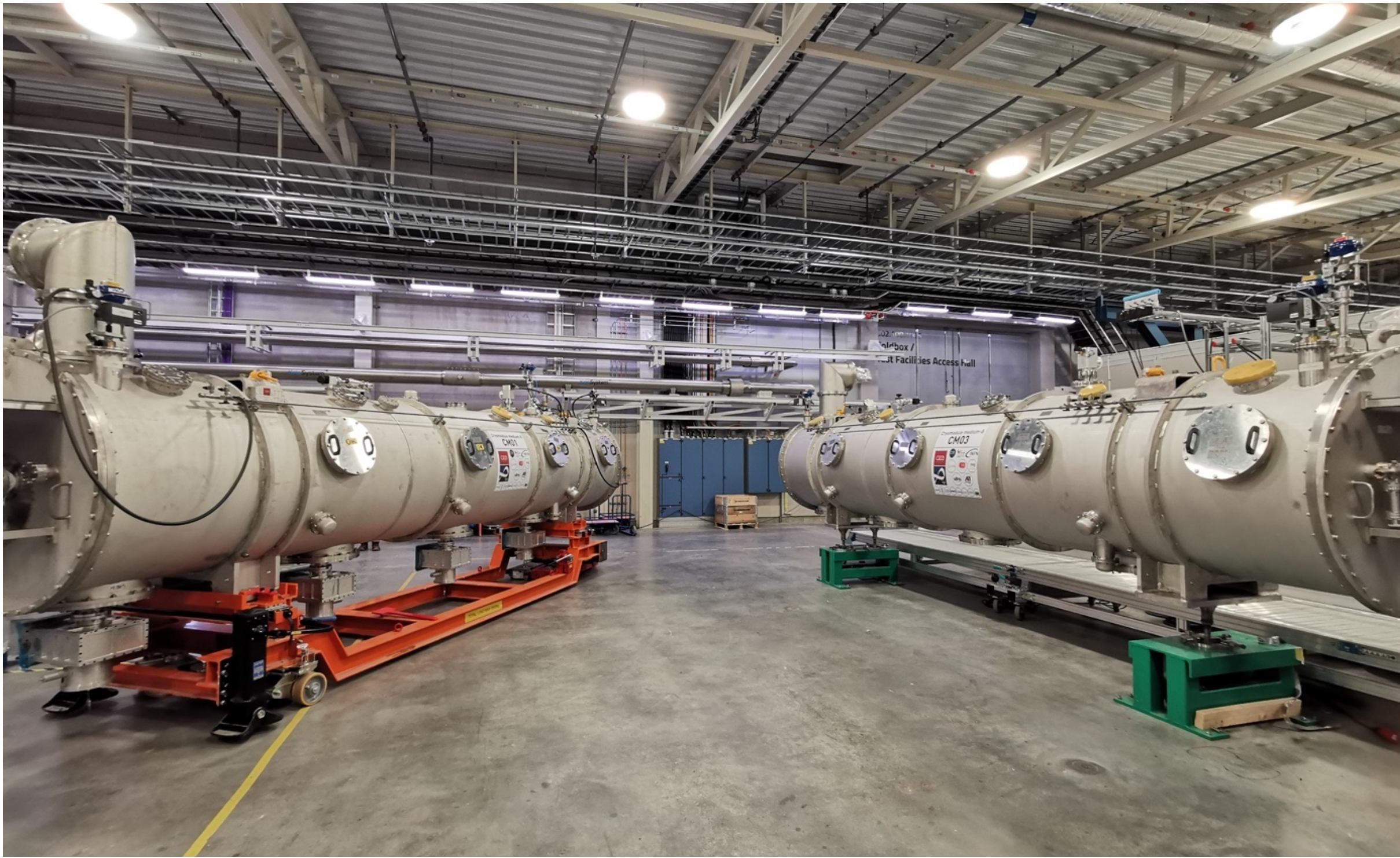

Figure 10: Picture of 2 cryomodules in the preparation area. One mounted on top of the transportation cart and one mounted on the preparation stands.

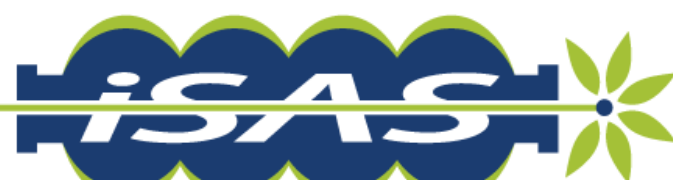

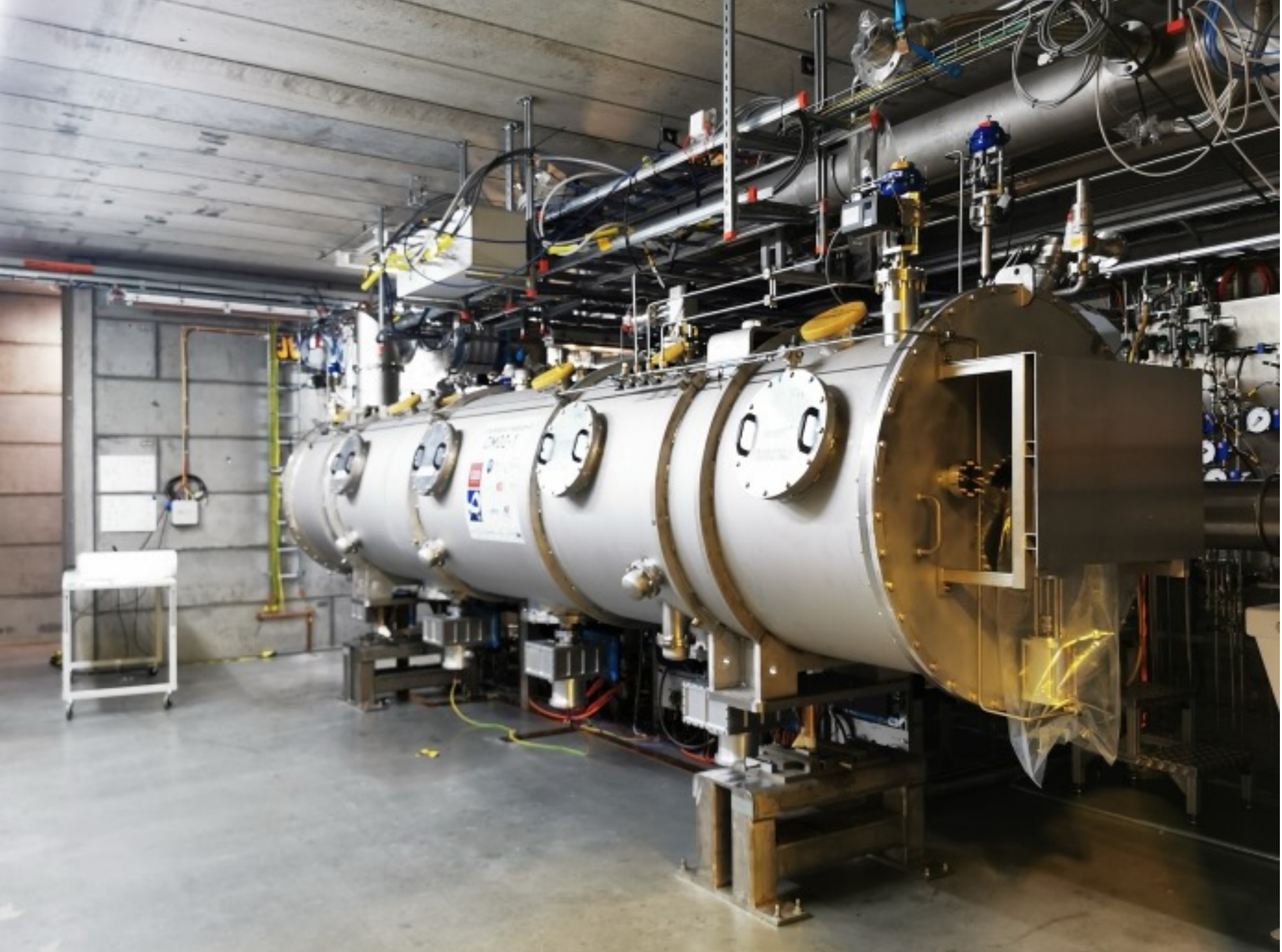

Figure 11: Picture of a cryomodule installed inside the radio-protection test bunker on top of the stands.

### 5.2. Cryogenic Performance

The cold test takes place inside the radio-protection bunker where the CM is connected to various auxiliary systems, such as radiofrequency distribution, beam vacuum, isolation vacuum, water cooling, and cryogenic distribution.

The TS2 cryomodule cryogenics system is designed and constructed such that the nominal operating conditions for the accelerating cavities can be achieved. It also allows for the cryomodule to safely go through a series of operating modes necessary for, for example, the cool down, warm up or response to accidental states.

At the TS2 a dedicated cryogenic plant, TICP, and distribution system, CDS, supplies refrigeration to the cryomodule at conditions like those present in the accelerator tunnel where each single cryomodule operates together with a dedicated valve box (VBox).

The learning curve during the Site Acceptance Tests (SAT) at Test Stand 2 (TS2) was a defining factor in transitioning from the commissioning of the infrastructure to a steady-state production testing environment.

Initially, the process was characterized by the intensive validation of the test stand itself, using the prototype cryomodule (CM00) to refine cryogenic sequences and RF power distribution. The early series modules (such as CM01 and CM03) encountered durations of over 130 days in the bunker as the joint ESS and IFJ PAN teams developed the necessary tools and consolidated and refined the testing procedures.

As the campaign progressed, a distinct "training curve" emerged over approximately the first six testing cycles. During this period, the average testing duration dropped significantly from an initial 63 calendar days to a nominal target of approximately 42 days (6 weeks).

By the time high-beta (HB) cryomodule testing began, the team had reached a consolidated state where the "Ready For Installation" (RFI) status could be achieved reliably within the 6-week timeframe unless specific non-conformities, such as cold leaks or cryoplant trips, occurred.

Beyond just speed, the learning curve involved a qualitative shift in data management and system integration. The implementation of a dedicated ESS Cavity Database allowed the team to benchmark SAT results against previous vertical tests (VT) in real-time, helping to identify and correct calibration uncertainties.

During the Site Acceptance Tests (SAT) at Test Stand 2 (TS2), the evaluation of cryogenic performance is a cornerstone for qualifying the ESS elliptical cryomodules. The measurement methods and results provide critical insights into the quality of assembly and the operational limits of the cavities.

### *5.2.1. Heat Load Measurement Methods*

At TS2, static and dynamic heat loads are measured using independent techniques to ensure high confidence in the results, below are a few of the most important methods:

- Flowmeter method: This method provides a direct measurement of the heat loads by including the latent heat, however, its accuracy is directly affected by the flow measurement accuracy.
- Helium Level Decrease Method (2 K Bath): This is the primary method used for evaluating the 2 K heat load. Once the cryomodule reaches thermal equilibrium and the 2 K bath is stabilized, the helium supply is closed. The rate of liquid level drop (measured by superconducting level sensors) is recorded over a specific time. By knowing the geometry of the helium vessels, the volume change is converted into a mass flow rate, which multiplied by the latent heat of helium yields the heat load.
- Pressure increase method: In this method, the cryomodule’s 2 K helium circuit is completely isolated by closing both the liquid helium inlet valve and the gas return valve to the VLP line. This thermodynamic prosses follows an isochoric line. Since the volume of the helium vessels are fixed, any heat entering the system causes an increase of energy into the system resulting in a pressure increase over time.
- Dynamic Heat Loads: Before (or after) RF power is sent into the cavities the cryogenic system is stabilized and referenced. A steady helium flow is supplied to the bath by fixing the opening of the inlet valve. Heaters placed on the cavity tanks are powered and precisely adjusted to maintain the helium level constant; the power supplied to the heaters is noted. The cavities are then supplied with RF power and the additional dissipated power from this source is compensated by means of decreasing the total power supplied to the cavity heaters precisely maintaining the same thermodynamic conditions on the helium bath as for the reference (level, pressure). The dynamic heat load is given by the variance in total dissipated power by the cavity heaters.

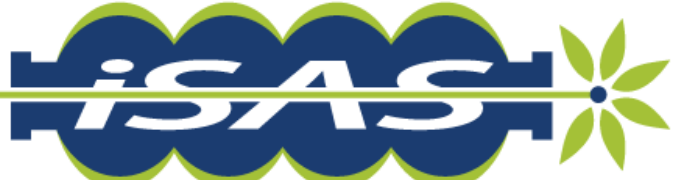

- Thermal Shield Measurements: For the 40/50 K, heat loads are typically evaluated by measuring the temperature rise between the inlet and outlet of the shield circuits, combined with the mass flow rate provided by the cryogenic plant.

Figure 12 shows a summary of the cryogenic performance during these measurements:

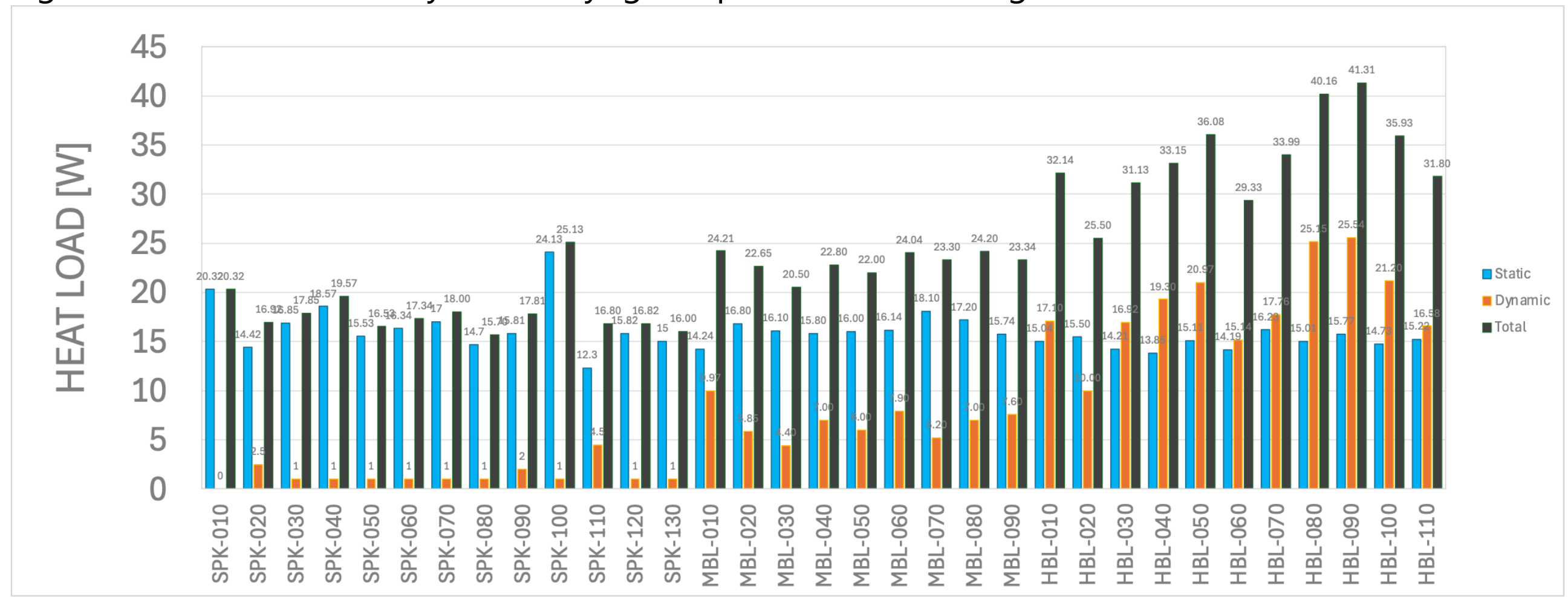

**Figure 12: Summary of Static, Dynamic and Total heat loads for all the cryomodules. (note that the measurements done on the spokes cryomodules were taken at another test stand facility FREIA).**

Here the averaged results for the each cryomodule (CM) family at ESS:

- 13 Spokes cryomodules 2 cavities each at 9MV/m):
  - static=16.7W, dynamic=1.4W, total=18.1W.
- 9 Medium-beta cryomodules (4 cavities each at 16.7MV/m)
  - static=16.2W, dynamic=6.8W, total=23 W
- 11 High-beta cryomodules (4 cavities each at 19.9MV/m):
  - static=15.0W, dynamic=18.7W, total=33.7 W

Regarding the heat loads to the thermal shield operated at ~35K:

- Spokes: (no result)
- Elliptical CM (MB , HB): 80W

The results show a good consistency with the design values, however the heat loads to the thermal shield are consistently higher. This was attributed to a wrong choice of material properties during the design phase.

A declining trend in SHL was observed as the assembly campaign progressed, indicating that manufacturing quality improved with experience and feedback from testing. All tested cryomodules demonstrated low dynamic heat loads compared to the design specifications. In the more recent tests, the specific DHL contribution of each cavity was measured at various gradients to act as a long-term performance gauge.

### 5.3. RF performance

The RF performance of the ESS elliptical cavities has been systematically validated through Vertical Tests (VT) at partner facilities (DESY and STFC) and Site Acceptance Tests (SAT) at Test Stand 2 (TS2) in Lund. The High beta cavities were qualified up to the nominal gradient of 19.9 MV/m. While most met the specification, a few cavities were accepted via non-conformity (NC) reports despite slight performance limitations (e.g., lower quench levels or early field emission), typically being allocated to positions in the linac with lower field requirements.

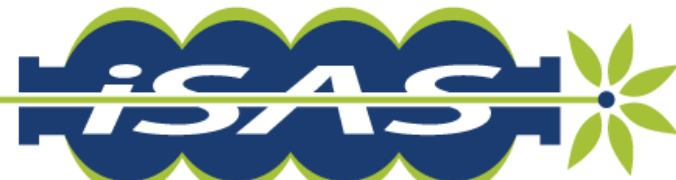

A strong 1:1 correlation was observed between vertical test results (naked cavities) and cryomodule performance (fully assembled). As can be seen in Figure 13, most cavities retained their performance within a 20% margin after assembly, proving the robustness of the cleanroom assembly and transportation processes.

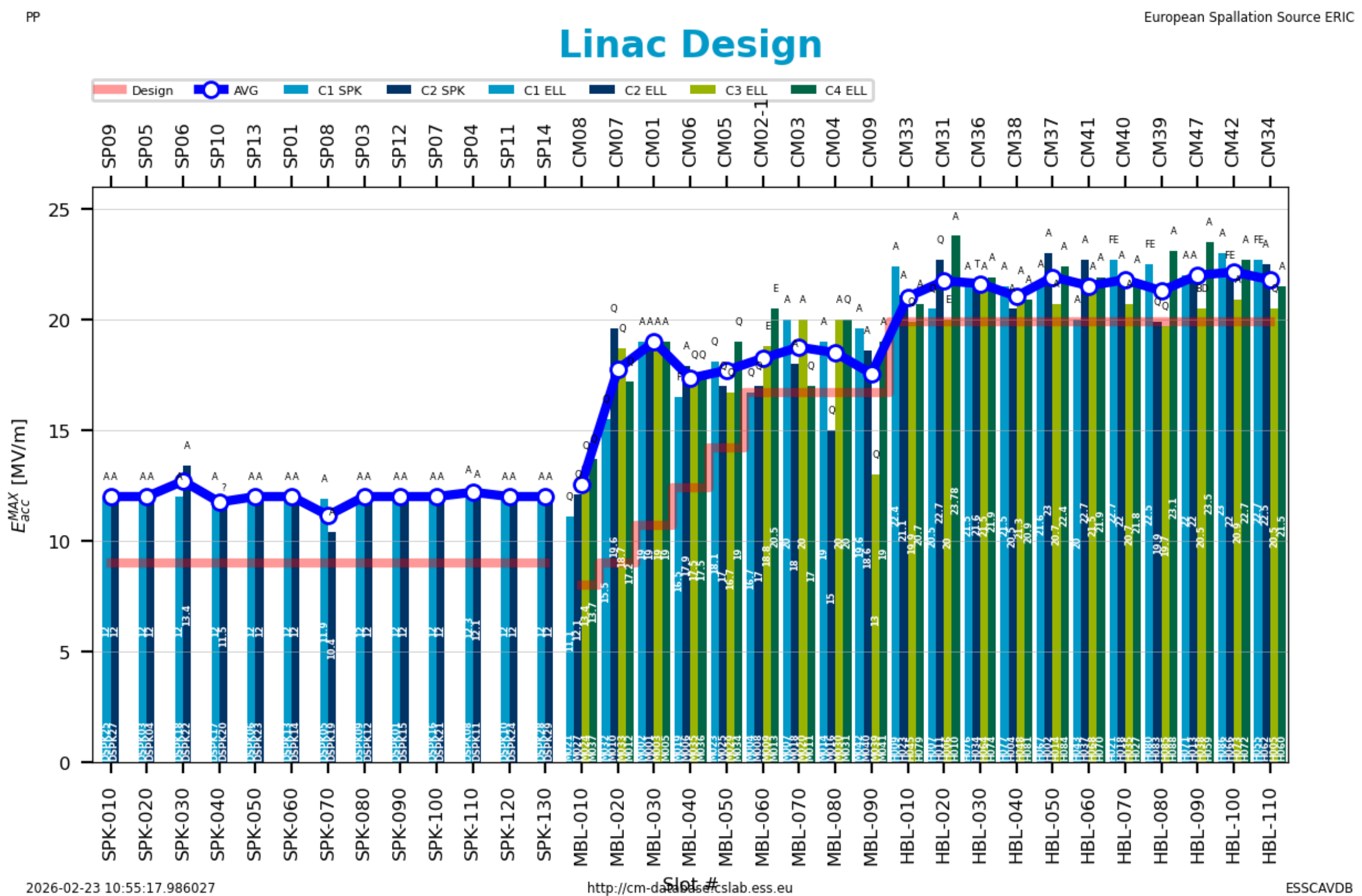

**Figure 13: Cryomodule assignment along the linac, with achieved gradient per cavity and indication of limiting mechanisms: (A is Admin limit, Q is fast quench, E is too high number of electros measured by the electron pick-up, T is thermal quench, FE is field emission).**

Also very important is to note that each cryomodule performance in the linac lattice achieved higher gradients than those required by beam physics, thus posing no constraints on beam delivery after installation.

## 5.4. Installation and Operational Challenges and "Lessons Learned"

### *5.4.1. Tunnel Installation*

The installation of the elliptical cryomodules within the ESS tunnel represented a complex integration phase that transitioned the project from individual component testing to full system synchronization. This period provided critical insights into the mechanical and logistical challenges of deploying large-scale superconducting infrastructure in a subterranean environment.

The physical installation of the 30 elliptical modules required meticulous coordination to ensure that the mechanical alignment achieved during assembly was preserved throughout the transition into the tunnel.

One of the primary lessons learned during this phase involved the vulnerability of the internal components to mechanical shocks and vacuum degradation during transport and handling. During the SAT phase, it was discovered that some internal piping could become displaced from its supporting hooks during transit, necessitating post-delivery inspections before tunnel integration.

Furthermore, the interconnection process to the LWU while maintaining very high beam vacuum quality required meticulous integration, necessitating the deployment of mobile clean rooms.

The interconnection of the cryomodule jumper connection process pipes to the Valve box, was considered a critical element, where the absence of leaks had to be maintained as well as the capability for pressure bearing. To make that step effectivelly, it was necessary to prepare the work beforehand by cutting all the required connections, that were used specifically for the SAT. Moreover, specific interconnecting pipe elements had to be prepared to length before the final welding step. The welding of the process pipes between a cryomodule and its valve box can be seen in Figure 14.

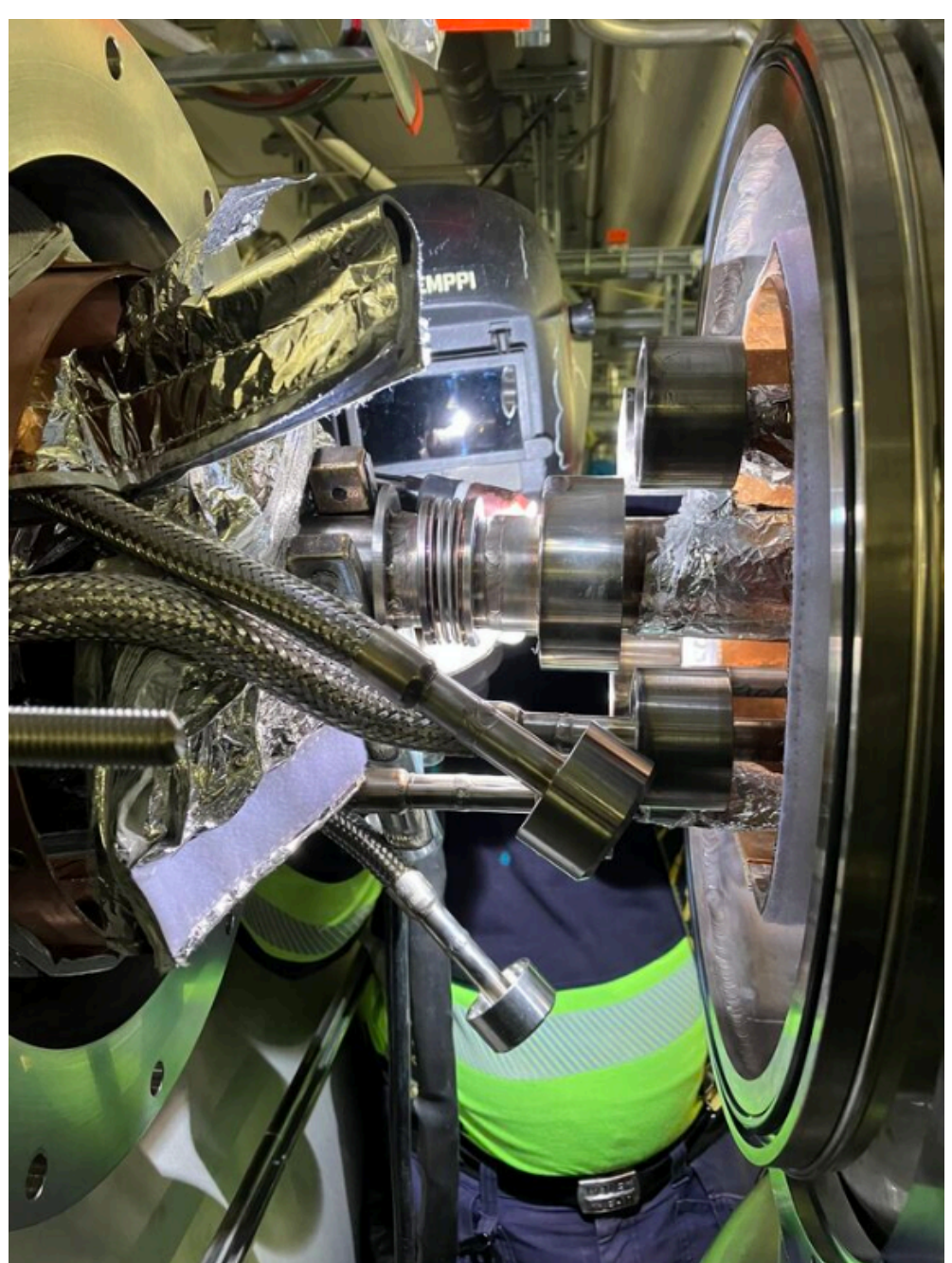

**Figure 14: Welding of process pipes between Cryomodule and Valve box.**

The installation phase for the cryomodules (CM) was characterized by a transition toward parallelized workflows and highly specific preparatory protocols. A significant portion of the efficiency gains resulted from performing time-consuming mechanical tasks, such as cutting process pipes and preparing water-cooling or compressed air piping, achieved during maintenance windows or within the gallery prior to the primary installation period. This preparation extended to radiological safety, where a dedicated Radiological Protection (RP) work environment collaboration meeting was established to create a step-by-step schedule based on projected radiation levels.

To ensure technical readiness, toolboxes were refined based on lessons from previous installations, including specialized tools for cutting valve box process pipes and general-purpose kits optimized for the specific size of the work crews

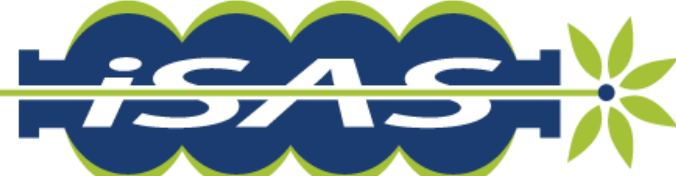

The execution of the installation sequence saw several strategic optimizations that reduced the overall timeline. For example, the decision to defer cable trays and wiring installation until the final stage prevented cable damage and afforded riggers and vacuum technicians' easier access to the underside of the CM.

Additionally, an example for the optimization in the management and improvement of interconnection to auxiliary systems was the water-cooling manifolds, where existing pipes were disconnected and lifted to the ceiling rather than fully dismantled. Through this change approximately two days per module were saved by eliminating the need for additional electrical support.

Furthermore, by transporting and aligning two cryomodules simultaneously, the team maximized tooling capabilities and reduced the time spent on instrument setup, effectively doubling the installation rate from one to two units per week.

To streamline future installations and the subsequent commissioning phases, several design modifications have been identified. Implementing cable patch panels would significantly improve the speed of electrical troubleshooting by allowing for the rapid identification of device-specific versus wiring-related faults.

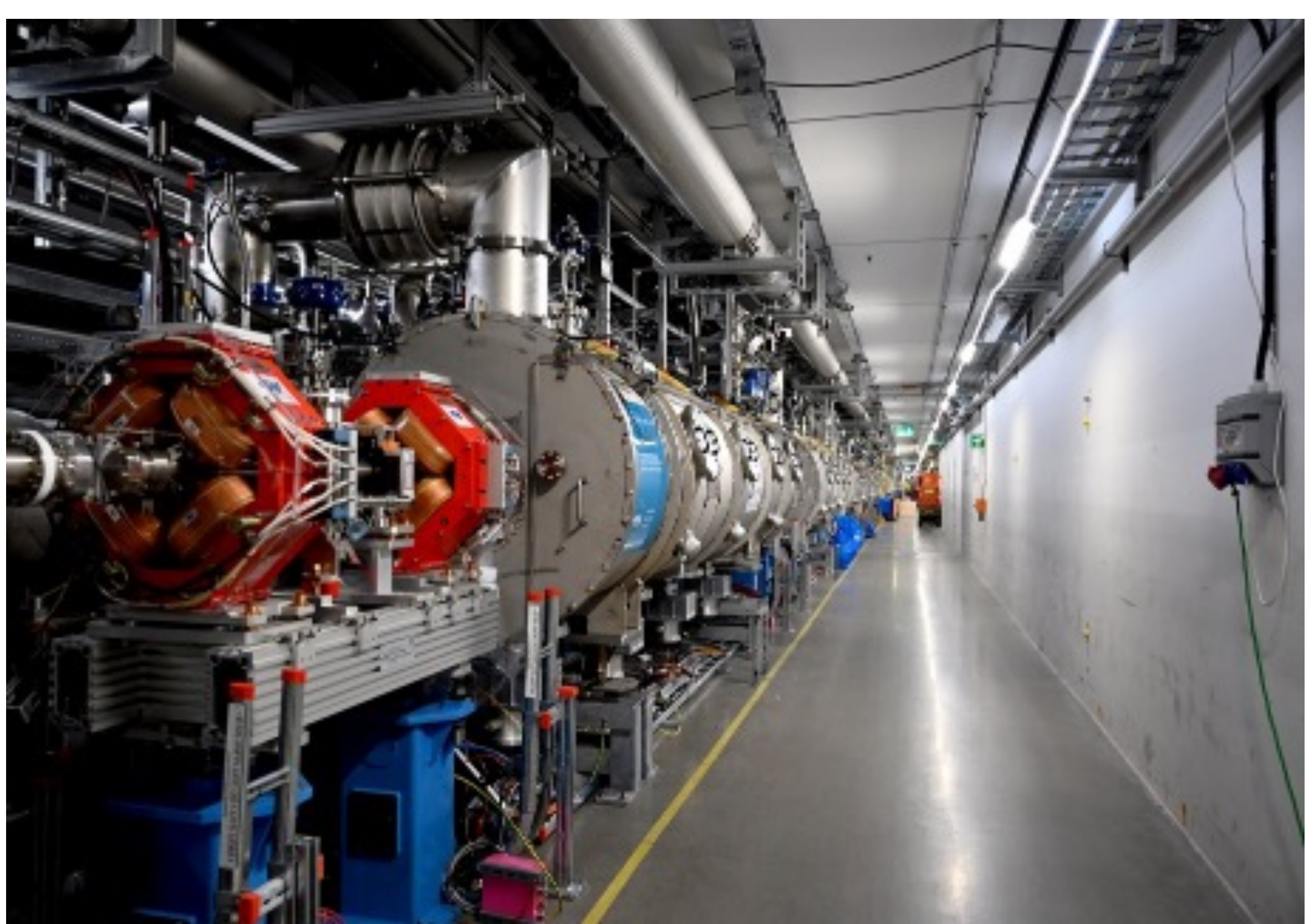

**Figure 15: Picture of the Elliptical Cryomodules installed in the tunnel.**

From a mechanical standpoint, increasing the "play" or tolerances within the design would facilitate the removal of components after cutting, especially in high-congestion areas of the tunnel. While flange connections remain useful for initial pressure testing, a hybrid approach using flanges for testing and welds for final installation is suggested to ensure long-term vacuum integrity while maintaining some modularity.

To reduce the reliance on manual labour and increase the consistency of joints, the adoption of automated cutting and welding tools is recommended. On the digital and organizational front, the commissioning phase would benefit from the early availability of critical signals through the Integrated Control System (ICS), which was a bottleneck in previous iterations.

Finally, the implementation of a centralized tool-tracking dashboard and a more robust spare parts availability would ensure that any repairs or adjustments during the installation window do not lead to cascading delays.

Figure 15 shows a photograph of elliptical cryomodules successfully installed into the tunnel.

#### *5.4.2. Operation and availability*

The ESS accelerator is currently in the technical commissioning phase, having already completed its initial beam commissioning stage. While real-time availability statistics have not yet been officially calculated, the project has established a clear operational goal of 95% availability, excluding periods for maintenance and scheduled stops. Current challenges are primarily tied to vertical system integration, though these issues are actively being addressed as the machine moves toward full functionality.

To ensure this high level of reliability, the system incorporates several layers of redundancy and failure compensation for most critical components. For example, each cryomodule is equipped with two redundant sensors for both helium bath pressure and level control.

Beyond hardware, there is ongoing development to use cavity heaters as a backup method for maintaining these process controls such as dynamic heat load compensation methods to minimise disruptions to the cryoplant. Furthermore, the system is designed to handle individual cavity trips with possibility to adjust the gradients of neighbouring cavities, a compensation strategy that is currently undergoing validation.

One of the most significant operational hurdles identified during commissioning involves recovery times following cryoplant trips. When these events occur, the process of re-establishing stable conditions at 2K is quite time-consuming, typically requiring between five and seven hours for stabilization and pump down.

Improving this recovery window is a major focus for future development, as it directly impacts the machine's ability to maintain its target availability during the operational phase.

The performance of the High Beta (HB) cavities within the installed cryomodules is primarily governed by administrative safeguards rather than immediate physical failures.

Currently, the most prevalent limiting factor is a set of administrative limits placed on the RF power driven through the power couplers, which affects approximately 75% of the cases. Despite these operational constraints, these specific cavities successfully reach performance levels that exceed their original nominal design values.

Beyond these power-related limits, a smaller subset of cavities face physical performance barriers related to thermal and electromagnetic limits:

- Approximately 14% of the cavities are limited by quenching—the sudden loss of superconductivity—with half occurring above the nominal gradient and the other half at the nominal gradient.

- A further 2% experience slow thermal quenches.
- Roughly 7% are restricted by field emission, where electrons are extracted from the cavity surface by high electric fields.

To maintain steady operations in the face of these limitations, the system relies on a combination of strict power management and the validation of gradient compensation strategies.

For the minority of cavities limited by quenches or field emission, the ongoing development of neighbouring cavity compensation allows the rest of the linac to adjust its performance, ensuring the overall beam requirements are met even when individual components hit their physical ceilings.

The maintenance strategy of cryomodules in the ESS tunnel is designed around a series of access hatches that allow for the replacement of critical internal components without necessarily requiring a full disassembly of the cryomodules. For example, the tuner systems, are positioned to be accessible and replaceable through these hatches in the event of a mechanical failure. Similarly, the level gauge assemblies are engineered for extraction and replacement through these same access points, although this specific procedure is recognized as being of high complexity and has not yet been performed in the tunnel environment.

Other critical repairs have already been demonstrated, most notably the successful replacement of a Fundamental Power Coupler (FPC) ceramic window within the tunnel, due to visible fracture of the ceramic window. This operation benefitted with technical experience gained during previous test-stand trial with support from CEA experts, proving that such sensitive vacuum components can be serviced in situ.

Furthermore, several external components, including pressure relief valves, fast-acting relief valves, and burst disks, are designed for straightforward maintenance and rapid replacement to ensure minimal downtime due to vacuum and pressure safety.

For the cryogenic and instrumentation systems, the approach balances accessibility with redundant design. The cryogenic valves are configured so that most parts can be serviced without opening the vacuum vessel hatches; however, a failure in the valve seat would require disconnecting the valve body from the inside. While the bath pressure gauges are doubled for redundancy, replacing a faulty unit is a more intensive process that requires cutting and re-welding.

To future-proof the system, spare instrumentation channels have been integrated into the design, providing the flexibility to add new sensors if needed, despite the existing thermometry setup already featuring multiple layers of redundancy.

Major failures of critical parts, such as the tuners or the Fundamental Power Coupler (FPC) windows, were explicitly foreseen during the initial design stage. Because of this foresight, it is possible to perform these significant repairs while the cryomodule remains in the tunnel, avoiding the logistical burden of removal for common mechanical or vacuum barrier issues.

In contrast, a cryomodule replacement is generally only required in the event of a catastrophic compromise of the beam vacuum. If the cavities are vented or become contaminated, the resulting loss of performance usually dictates that the entire unit be swapped out for a fresh one. This distinction ensures that while mechanical wear and tear are treated as field-serviceable, internal cavity degradation remains the primary trigger for a complete system exchange.

From a cryogenic safety perspective, the most severe failure mode considered involves a sudden loss of either the beam vacuum or the insulating vacuum. Such an event would cause a rapid increase in heat loads, leading to the fast pressurization of the helium bath and the subsequent activation of burst disks to vent the system.

While this scenario results in a significant loss of helium and requires the replacement of the burst disks, the hardware consequences are relatively minor compared to the performance loss associated with beam vacuum venting. Ultimately, if the integrity of the cavity environment itself is lost, the cryomodule is typically deemed unfit for continued operation and marked for replacement.

The ESS elliptical cryomodule design, in conjunction with its valve box, incorporates a high degree of thermal flexibility by providing each cryomodule with dedicated circuits, allowing for the individual warm-up and cool-down cycles independent of the rest of the linac.

To begin the warming process, the liquid helium supply is first cut off to empty the local inventory. The temperature then begins to rise naturally due to static heat loads, a process that can be accelerated and controlled by circulating warm helium from the high-pressure line into the recovery system. Operators can precisely manage the rate of this temperature increase by adjusting the supply flow from the high-pressure line.

A significant operational milestone has been the demonstration that a single cryomodule can be warmed up or cooled down while the remainder of the linac is maintained at its operating temperature of 2K. This represents an evolution from the initial design concept, which assumed the entire linac would need to be warmed to 4K for such interventions. While this capability is proven, it is not yet part of an automated sequence and currently requires the oversight of an experienced operator to manage the delicate thermal transitions.

While a single cryomodule unit can be warmed as its neighbours remain cold, a full replacement or any task requiring the disconnection of a cryomodule from its valve box is seen as a restricted operation, necessitating warming of the entire linac to room temperature.

In terms of hardware protection and system interlocks for cavity operation, the main logic is centralized by a "Cryomodule_OK" process variable, which serves as the primary gatekeeper for RF operations. This status is only granted when several critical parameters fall within strictly defined thresholds. Most notably, the helium level must be high enough to fully submerge the cavities, and the helium bath pressure must remain stable. Furthermore, the Fundamental Power Coupler (FPC) must maintain specific temperature and flow rates at the antenna to ensure thermal stability during high-power operation.

Beyond the standard operational gatekeeping, the system includes dedicated interlocks designed to prevent and mitigate helium level overfill and bath pressure overpressure scenarios. These protections are coupled with automated response mechanisms to stabilize the cryomodule before structural damage can occur. The tuner systems also feature specialized safeguards; they are designed to disengage during warm-up if the motor becomes immobile. To assist this process, localized heaters are strategically placed on the tuner mechanism, allowing operators to force a thermal disengage if the hardware becomes physically stuck.

Additional protective measures have been implemented to address environmental risks, such as the potential for water to freeze within the antenna cooling circuits. To mitigate this risk, the facility utilizes a variety of preventative strategies, including a dedicated water flushing mechanism. In situations where rapid intervention is required, this system can use compressed air to quickly purge the cooling lines, preventing ice expansion from damaging the sensitive antenna hardware.

# 6. SUSTAINABILITY ASPECTS

The design phase of any cryomodule is characterised by compromises coming either from design requirements, constraints or even alignment with project schedule and milestones.

Although energy savings from reduced heat loads are at the core of the cryomodule design engineering, there are additional compromises to be made that range between high up-front costs and lower running costs, compromise between robust and redundant instrumentation and static heat loads, compromise between bespoke, more efficient design solutions and the lower cost of known simplicity, compromise between simplicity and added functionalities or features, compromise between committing today and the possibility of the technological advances of tomorrow, and so on.

One further factor playing into the compromise of any decision is also weighing the various sustainability aspects of the chosen solution, be it in a direct way, from a longer-term sustainability perspective, or from a lifecycle perspective of the project.

## 6.1. Energy Savings

Minimal heat inleak by static heat loads plays an important role for example in the number of connections for instrumentation, where a fine balance between sufficient monitoring, sufficient redundancy, and the resultant heat losses need to coexist to ensure a low static heat load while maintaining a high operational availability (because of sufficient redundancy for critical equipment). The same approach happens with supporting structures that play an important role in the path for thermal conduction. Optimisation between the primary function and additional features as well as number of thermal barriers, complexity, etc., needs to be properly balanced in the design phase for best effect.

The proper architecture for heat extraction with the goal of further minimization of the cryoplant electrical power consumption and losses always points towards a strategy for removing heat loads at higher temperatures levels. The result is an architecture requiring thermal shield(s), or thermal intercepts, with consequences resulting from the design simplicity, access to components, number of components, space management, etc.

## 6.2. Materials and Component Circularity

Venting of helium gas during failure events is a subject that needs further analysis in the lifecycle analysis since helium is a finite resource.

Safety mitigations coming from pressure equipment Directives (PED) put requirements on the safety equipment required for failure mitigation, however very little attention is given to the amount of helium that can be lost into the environment.

The cryomodule design at ESS regarding PED follows below Category 1 of the code (the lowest category in terms of pressure risk). This was achieved by a conjunction of minimizing the product of liquid volume and maximum allowable working pressure before the ultimate action of the pressure relief devices.

The pressure relief devices primarily discharge into collectors or other circuits that allow for full recovery of helium, but in the case of catastrophic events resulting in extreme heat loads, the primary goals are first the safety of people, then equipment and finally, if possible, the recovery of helium.

At ESS measures are taken to recapture and contain the helium inventory in a closed loop under both normal and extreme situations, however there still exist critical scenarios where significant helium inventory could (and has) been vented to atmosphere (namely through the burst disks which form the last line of defence for over pressure of the CM).

Pure helium gas is also an expensive good. Minimizing the release to atmosphere through either recapture into sufficiently sized helium tanks, or through reducing the size of the inventory or likelihood of a critical venting event to occur, can all help with reducing helium inventory loss and increasing the sustainability related to the helium use.

### 6.3. Exergy analysis

Although considered to be at the boundary of the scope of the iSAS project, when looking for efficient cryogenic cooling in concepts, the design choices within a cryomodule should always be taken in conjunction with the cryoplant design.

Exergy analysis is essential in the design of accelerator cryoplants for particle accelerators because it identifies the true sources of energy destruction (irreversibility) in systems that operate at cryogenic temperatures. Cryomodules require very low temperatures to achieve superconductivity. Exergy analysis identifies major losses in components such as compressors, turbines, and throttling valves.

Long cryogenic transfer lines of hundreds of meters incur in significant pressure drops and heat leaks. Exergy analysis of the CDS identifies where pressure drops or heat leaks are most destructive to efficiency.

The final step to produce 2 K superfluid helium involves throttling, which causes a large, localized drop in exergy. Exergy methods help to optimize the valve design and pressure ratios to minimize this loss.

Modern accelerators like the European Spallation Source (ESS) produce significant dynamic heat loads from SRF cavities. Exergy analysis helps tailor the cooling capacity to these specific needs, enhancing sustainability.

### 6.4. Materials and Component Circularity

Materials and component circularity in cryomodule design is sometimes overlooked, however direct indicators such as availability, reusability, modularity, and repairability are critical areas for reducing costs, lowering high environmental impacts of material sourcing (e.g., Nb), and enhancing longevity in accelerators like ESS.

Circular design enables repairing or replacing of specific superconducting RF cavities rather than entire modules, maximizing availability and minimizing waste.

# 7. CONCLUSIONS

The development, production, and initial commissioning of the ESS elliptical cryomodules represent a significant milestone in superconducting radio frequency (SRF) technology.

This project has demonstrated that large-scale, high-performance accelerator infrastructures can be successfully realized through a distributed international collaboration model. The "lessons learned" captured in this document provide a blueprint for future sustainable accelerator systems, particularly within the framework of the iSAS (Innovate for Sustainable Accelerating Systems) program.

**Key Takeaways:**

- Standardization as a Sustainability Driver: The implementation of a "universal" cryomodule design for both Medium and High-Beta sections proved highly effective. This strategy simplified the industrial supply chain, reduced manufacturing waste through component commonality, and allowed for standardized assembly and testing procedures, directly supporting iSAS goals for industrial scalability and cost reduction.
- The In-Kind Contribution (IKC): The ESS experience validates the IKC model as a means to leverage "distributed intelligence" and effort across Europe’s leading research institutions. While introducing logistical complexities, the model enabled parallel development of complex subsystems, such as the coordination between INFN-LASA, STFC, and CEA for cavity production and cryomodule integration.
- Modular Cryogenic Architecture: Decoupling the Cryogenic Distribution System (CDS) from the cryomodule vacuum vessels provided essential flexibility for installation, maintenance, and future upgrades. This segmented approach allowed for sequential commissioning and provided a critical schedule mitigation strategy against unforeseen technical non-conformities.
- Mechanical supporting structure - the "Space-Frame": The adoption of the internal space-frame support system successfully decoupled the vacuum vessel's mechanical tolerances from the high-precision alignment required for the cavity string. This design choice facilitated easier assembly access and maintained structural stability during the thermal contraction of the cold mass to 2 K.
- Thermal Efficiency and Resource Management: The use of a nested temperature approach and advanced heat exchangers for pre-cooling the 4.5 K supply line significantly enhanced the efficiency of the Joule-Thomson expansion for 2 K operations.

**Closing Statement**

The ESS elliptical cryomodule project has transitioned from a "green field" concept to a fully operational superconducting linac. The technical successes in achieving stable 2 K operation and high-power RF performance, alongside the procedural lessons in industrialization and quality control, establish the ESS as a primary reference for the design of sustainable research infrastructures. The project serves as a definitive blueprint for the next generation of research infrastructures, proving that a unified, pan-European approach is an effective way to address the ambitious technical and sustainability goals of the global scientific community.

As the project moves into steady-state operations, these findings will continue to inform the parametric design studies and circularity goals essential for the future of particle physics.

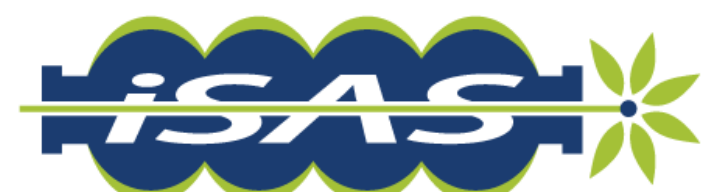

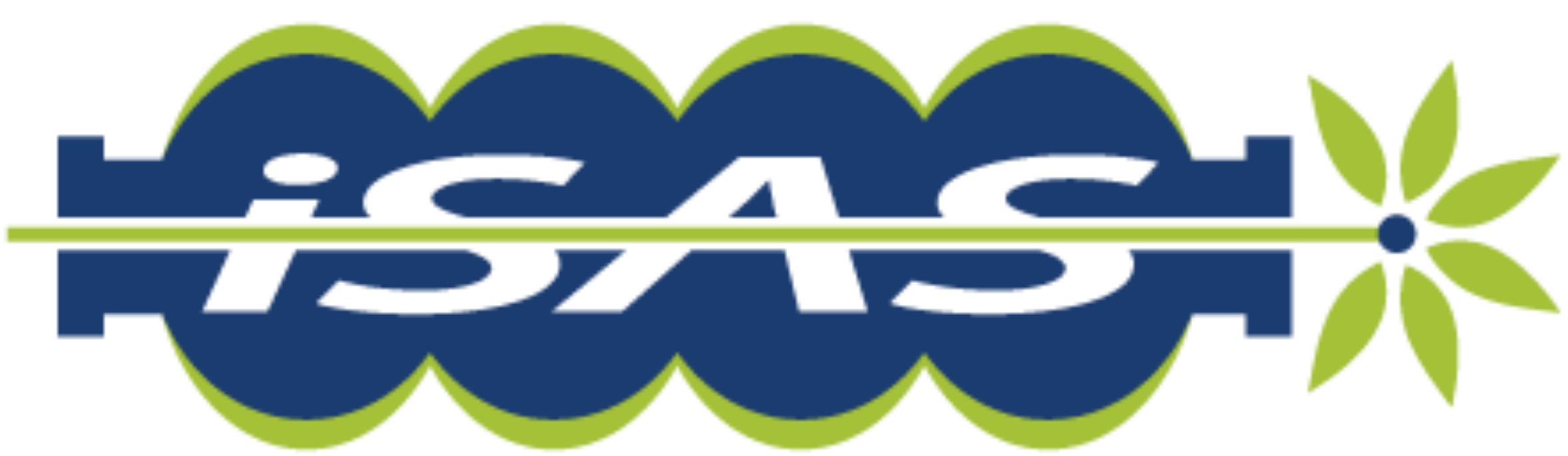

innovate for Sustainable Accelerating Systems

# ANNEX II
# Benchmarking Cryomodule Designs for iSAS WP5

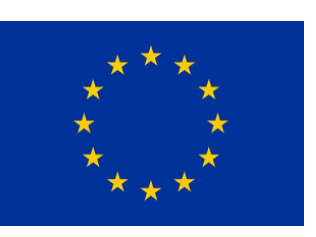

Funded by the European Union. Views and opinions expressed are however those of the authors only and do not necessarily reflect those of the European Union or the European Commission. Neither the European Union nor the granting authority can be held responsible for them.

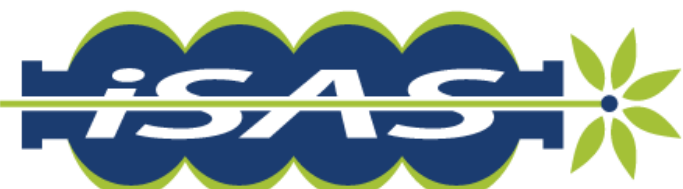

# DOCUMENT TRACKS DETAILS

| Deliverable Number | D 16 | Lead Beneficiary | ESS, CERN |
|---|---|---|---|
| Deliverable Name | Compile lessons learned from ESS and Benchmarking with Projects in implementation phase (**ANNEX II**) | | |
| Type | R — report | Dissemination Level | PU – Public, |
| Due Date | Project month 24 | Work Package No | WP5 |

## Deliverable Abstract

This report consolidates two complementary studies conducted within iSAS Work Package 5 (WP5) to support the development of a sustainable next-generation cryomodule.

The first study analyses the design, industrialization, commissioning, and early operation of the elliptical cryomodules developed for the European Spallation Source (ESS), capturing technical and procedural lessons learned across the full engineering lifecycle. It highlights the impact of standardization, distributed collaboration under the In-Kind Contribution model, cryogenic performance at 2 K, and industrial scalability as key enablers for long-term sustainability.

The second study benchmarks high-beta cryomodules across six leading international facilities—Spallation Neutron Source, Large Hadron Collider, PIP-II, European XFEL, Linac Coherent Light Source II, and ESS—through a structured comparative methodology covering beam and RF parameters, cryogenics, mechanical design, commissioning, and operational experience.

Together, these analyses establish a validated technical baseline and an international reference framework, forming the foundation for a WP5 roadmap toward a high-efficiency, sustainable "Cryomodule of the Future."

ANNEX I refers to the detailed report on compilation of lessons learned from ESS

ANNEX II refers to the detailed report on benchmarking of cryomodule designs for iSAS

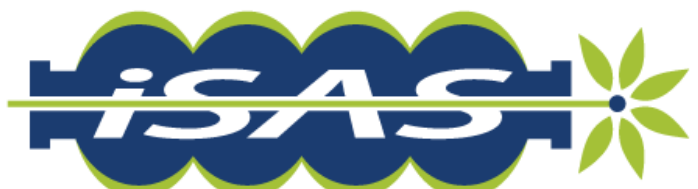

TABLE OF CONTENTS PAGE

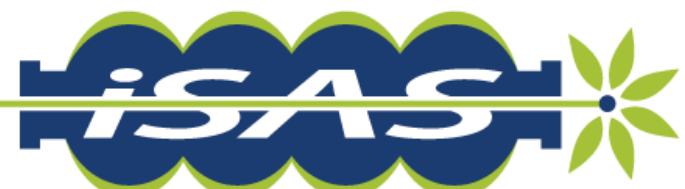

# 1. INTRODUCTION

The iSAS Work Package 5 (WP5) is dedicated to defining and documenting a comprehensive set of requirements and technical solutions for integrating emerging technologies into a next-generation sustainable cryomodule. To establish a robust, evidence-based foundation for these innovations, a global benchmarking study was conducted, evaluating high-beta cryomodules across six world-leading facilities: Spallation Neutron Source (SNS), Large Hadron Collider (LHC), Proton Improvement Plan II (PIP-II), European XFEL (XFEL), Linac Coherent Light Source II (LCLS-II), and ESS.

# 2. BENCHMARKING METHODOLOGY

Through a structured 17-step methodology, the benchmarking study analyses high-beta cryomodules across five key domains: beam and RF parameters, cryogenic systems and thermal management, cryomodule architecture and internal components, commissioning experience, and operational performance and availability. This comparative approach highlights how differing requirements, constraints, and strategic priorities shape design choices and lead to distinct technical configurations.

The main steps of the Benchmarking Methodology are reported below:

1. Definition of key technical Areas of Interest: This initial stage identifies the core domains required to capture the full cryomodule lifecycle, from initial beam physics requirements to long-term operational reliability.
2. Proposal of cryomodule design Concepts: The team selects the specific architectures that will serve as the technical basis for comparison, ensuring a diverse range of high-beta designs.
3. Identify Facilities: Six facilities were selected for the study: SNS, LHC, PIP-II, XFEL, LCLS-II, and ESS, representing a broad spectrum of missions and design philosophies.
4. Identify Experts: This step involves mapping out the technical leads and primary contact persons within each identified facility to ensure high-level technical engagement.
5. Defining Specific Points of Interest: For each primary area, detailed technical sub-topics are established to provide a focused framework for comparative analysis.
6. Draft Questionnaire: A comprehensive data collection instrument is created, structured into five sections: Energy and RF Parameters, Cooling Circuits and Cryogenics, Cryomodule Design, Commissioning Lessons, and Operational Availability.
7. Questionnaire Dry-Run: The tool is tested with two selected facilities to evaluate the clarity and effectiveness of the questions.
8. Dry-Run Feedback: Feedback from the initial tests is analysed to identify gaps or ambiguities in the query structure.
9. Final Improvements: The questionnaire is finalized based on the dry-run results to ensure it captures technical details and "as-operated" realities.
10. Literature Research: An exhaustive review is conducted using Technical Design Reports (TDRs), peer-reviewed scientific papers, conference presentations, and internal technical notes to establish a baseline of publicly available data.
11. Pre-filling Answers: Data from the literature research is used to populate the questionnaires. This "correction-based" approach optimizes expert time by allowing them to focus on correcting inaccuracies or providing context rather than providing basic parameters.
12. Initial Contact: The iSAS program and the specific scope of the benchmarking exercise are formally introduced to the facility representatives.

13. Distribution: The pre-filled questionnaires are sent to the facilities for an initial review and to allow experts to gather internal data.
14. Primary Face-to-Face Interview: Using video-conferencing to bridge geographic gaps, these deep-dive sessions transform the exercise from simple data gathering into a high-level technical exchange.
15. Follow-up Face-to-Face Interview: Subsequent sessions are held to conduct "deep-dives" into specific subsystems and capture information regarding commissioning hurdles and operational constraints.
16. Compilation of Answers: Validated data from all six facilities is consolidated into a structured comparison framework aligned with the five technical areas of interest (See Paragraph 3).
17. Completion of Report: The final phase involves drafting the report and the Benchmarking Synthesis, which identifies recurring design patterns and the performance trade-offs that impact cryomodule sustainability (See paragraph 4).

## 2.1. Detail of steps 2&3: Cryomodule design concepts and facility identification

The selection of the six facilities (SNS, LHC, PIP-II, XFEL, LCLS-II, and ESS) for the benchmarking study was based on several strategic and technical criteria, designed to create a robust, evidence-based comparison:

- Comparable Technical Characteristics (High Beta Elliptical Cavities):
  The study aimed to benchmark the ESS cryomodule, thereby similar “cryomodule types” were selected, focusing the study specifically on cryomodules with high-beta elliptical cavities with frequencies between 400MHz and 1.3GHz. This ensured that the comparison remained relevant and provided a validated technical baseline.

- Diverse Architecture and Machine parameters:
  The chosen facilities represent a broad spectrum of machine parameters - particles, current, beam power, frequency, duty cycle, etc. - and design philosophies – segmented / continuous layout, cavity supporting systems, cryogenic schemes, etc. In this way, a diverse range of high-beta designs was represented

- Different stages of advancement:
  The facilities represent a wide spectrum of maturity and lifecycle stages. This diversity allows for a comprehensive analysis of the transition from "as-designed" specifications to "as-operated" realities. By including facilities that range from those starting installation soon to those with years of proven performance, the benchmarking exercise can capture lessons learned across the full engineering lifecycle. This ensures that the proposed roadmap for a sustainable "Cryomodule of the Future" is informed by both cutting-edge design and the practical challenges of long-term maintainability and hardware degradation.

- World-Leading Facilities:
  The study specifically targeted world-leading facilities to ensure that the benchmarking was grounded in state-of-the-art international best practices and experts’ insights that are rarely captured in formal documentation.

A brief overview of the different cryomodules architectures is presented below, the detailed analysis can be found in the questionnaire outcome in Paragraph 3.

| | |
|---|---|
| *ESS* | The superconducting section in its nominal configuration comprises 43 cryomodules (13 Spoke, 9 Medium-beta, and 21 High-beta) bulk Niobium cavities operating at 2K. It operates in pulsed mode (4% duty cycle) at 704.42 MHz for elliptical modules, with a peak RF power of 1.1 MW.<br><br>Cryomodules have a segmented layout with stand-alone units connected via jumper connections to external cryogenic distribution line. The cooling scheme includes 4.5 K supercritical helium (3 bar), 2 K saturated helium (31 mbar), and a 35–50 K thermal shield.<br><br>The ESS cavities are supported and aligned with spaceframe at ambient temperature connected to the vacuum vessel. The FPCs are vertically oriented at the bottom, active helium cooling is used for the outer conductor and water cooling for the antenna. The tuning system includes a cold stepper motor and two fast piezo compensators. The design features 8 access hatches for in-tunnel maintenance of tuners, sensors, and valves.<br><br>Status: Technical and beam commissioning on going, aiming for "Beam-on-Target" milestone. |
| *SNS* | The linac uses 30 segmented cryomodules (11 Medium-beta, 19 High-beta) with bulk Niobium cavities operating at 2.1K. It operates in pulsed mode (7.8% duty cycle) at 805 MHz with a peak RF power of 550 kW.<br><br>Cryomodules have a segmented layout, the stand-alone units are connected via "end cans" and bayonet connections to the external cryogenic distribution line. The cooling scheme includes 4.5 K supercritical helium (3 bar), 2 K saturated helium (31 mbar), and a 35–55 K thermal shield.<br><br>Like ESS, the cavities are supported and aligned with a spaceframe, with the addition of a Nitronic rod for longitudinal anchoring. The fixed antenna FPCs are vertically oriented and use active helium cooling for the outer conductor. Cavities are equipped with a cold mechanical slow tuner, and fast piezo actuators, not used in standard operation.<br><br>It includes 10 access hatches to access tuning system, sensors, connectors and instrumentation. SNS has a standard procedure for replacing FPCs in the tunnel using a mobile local cleanroom<br><br>Status: in operation since 2006, now with the recently installed PPU cryomodules. |

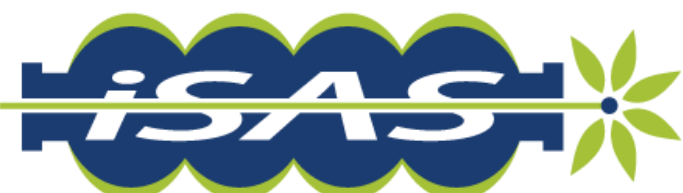

| | |
|---|---|
| *LHC* | The system consists of 4 cryomodules (4 cavities each) made of OFE Copper with a Niobium thin film coating operating at 4.5K. It operates in Continuous Wave (CW) at 400.8 MHz with a peak RF power of 300 kW.<br><br>Cryomodules have a segmented layout with stand-alone units connected via jumper connections to external cryogenic distribution line. The cooling scheme features the 4.5 K saturated He (1.3 bar) circuit, helium is spilled from the helium tank to actively cool feedthroughs. The thermal shield is floating, not active cooled, made of MLI blankets.<br><br>Cavities are supported by a frame anchored to the vacuum vessel through six tie rods. The FPCs feature a movable antenna (60 mm stroke), they are vertically oriented at the top, and the outer conductor is cooled with saturated 4.5 K helium. The cavities are equipped with a slow mechanical tuner, is driven by an external warm stepper motor. The design features removable lateral panels for access to internal components.<br><br>Status: in operation since 2010, Hi-Lumi upgrade installations / upgrade will occur in the Long Shutdown 3 (mid 2026). |
| *XFEL* | The main linac comprises 96 cryomodules with 8 cavities each in bulk Niobium operating at 2K. It operates in pulsed mode (1.37% duty cycle) at 1.3 GHz with peak RF power of 156 kW.<br><br>The accelerating modules have a continuous layout, thus are connected into single vacuum vessel. The cryogenic lines are embedded in the vacuum vessel, the cryogenic scheme includes two circuits for thermal interception (5-8K, 40-80K), and the 2.2 K supply line, with a dedicated JT valve and cooldown / warmup valve for cryo-strings of 12 modules.<br><br>Cavities are supported by the Helium Gas Return Pipe (GRP), which acts as the main structural backbone and an Ivar rod that provides longitudinal anchoring. The FPCs have a movable antenna and horizontal orientation. Cooling for FPC outer conductors is provided by fixed copper thermal intercepts. The tuning system uses cold stepper motors and fast piezo actuators. To minimize the heat loads to the cold mass the XFEL modules have two actively cooled thermal shields. The XFEL design does not include access hatches.<br><br>Status: XFEL has been in continuous operation since 2017 and recently completed its first full warm-up/cool-down cycle for maintenance. Feasibility study for operation at higher duty cycle on going. |
| *LCLS-II* | The main linac is composed by 35 accelerator modules with Nitrogen-doped bulk Niobium cavities, operating at 2K. It operates in Continuous Wave (CW) mode, at 1.3 GHz with a peak RF power of 7 kW.<br><br>The accelerating modules have a continuous layout, thus are connected into single vacuum vessel. The cryogenic lines are embedded in the vacuum vessel, the cryogenic scheme includes two circuits for thermal interception (5-8K, 35-55K), and the 2.2 K supply line, with a dedicated JT valve and cooldown / warmup valve for each module. The 2-phase line of each module is sectorized for independent level control given the tunnel slope (0.5%). Cavities require fast cooldown across the Nb transition temperature to sweep out the trapped magnetic flux and reach high $Q_0$.<br><br>As for the XFEL design, cavities are supported by the Helium Gas Return Pipe (GRP), which acts as the main structural backbone and an Ivar rod that provides longitudinal anchoring. The FPCs have a movable antenna and horizontal orientation. Cooling for FPC outer conductors is provided by fixed copper thermal intercepts. The tuning system uses cold stepper motors and fast piezo actuators. The LCLS-II design includes 7 access ports per module for maintenance.<br><br>Status: LCLS-II is undergoing a long shut-down for a major high-energy upgrade, adding 23 additional cryomodules to its existing infrastructure |
| *PIP-II* | The linac includes 23 SRF cryomodules of 5 different types. The focus of the questionnaire is the high beta 650 MHz cryomodules, with 6 Niobium cavities, operating at 2K, in Continuous Wave (CW), with a peak RF power of 43 kW.<br><br>Cryomodules have a segmented layout, the stand-alone units are connected with bayonet connections to the external cryogenic distribution line. The cooling scheme includes: 2 K saturated helium (30 mbar), and two circuits for thermal interception (4.5-15K, 35-50K). Cavities require fast cooldown across the Nb transition temperature to sweep out the trapped magnetic flux and reach high $Q_0$.<br><br>The cavities are supported by two support posts (one fixed, one sliding) mounted on a strongback support at ambient temperature. FPCs have a fixed antenna, horizontal orientation, and use fixed thermal intercepts at 5 K and 50 K for the outer conductor cooling. The FPC antenna is air-cooled. Cavities are equipped with cold mechanical slow tuners with stepper motor and fast piezo actuators. The design features access hatches on both sides for tuner and sensor maintenance.<br><br>Status: PIP-II is in the production phase, with series cryomodules expected to be delivered starting in the summer of 2026 and tunnel installation beginning late that same year. |

### 2.2. Detail of steps 10: Literature research

The literature research phase served as the analytical foundation of the benchmarking study. Rather than starting from a blank slate, the iSAS team reviewed existing documentation to "pre-fill" the benchmarking questionnaires. This approach serves three purposes:

- Optimized Discussion: Instead of spending time reviewing design parameters (e.g., operating frequency or cavity count), the discussion immediately dives into technical nuances and "lessons learned."
- Reduced Expert Burden: Allowing them to focus on correcting inaccuracies or providing context that isn't found in published papers.
- Higher Accuracy: This "correction-based" feedback loop often reveals discrepancies between the "as-designed" specifications and the "as-operated" reality.

To provide a comprehensive pre-filled baseline, data is collected from a wide spectrum of technical repositories:

- Technical Design Reports (TDRs): These serve as record of the initial machine architecture, cooling circuit logic, and design constraints.
- Scientific Papers & Published Articles: Peer-reviewed journals (e.g., Physical Review Accelerators and Beams) provide validated data on cavity performance and cryogenic efficiency.
- Conference Presentations (JACoW): Proceedings from IPAC and SRF workshops are valuable sources as they capture real-time technical challenges and machine availability data.
- Internal Technical Notes: Where available, internal facility documentation provided the internal details on sub-components like couplers and tuners.

It is important to note that the literature research does not provide an “homogeneous” set of information to be compared across facilities. Therefore, the interview process became essential component of the methodology, specifically for the part of commissioning and operation, where predefined questions alone cannot fully capture the range of scenarios encountered at each facility.

### 2.3. Detail of steps 14&15: Face-to-Face Interviews

The execution of the interview phase represented one of the most intensive periods of the benchmarking study.

Planning a series of technically rich meetings across the six world-leading facilities (SNS, LHC, PIP-II, XFEL, LCLS-II, and ESS) presented to be a complex endeavour, requiring coordinating the calendars of high-level experts across multiple time zones as the team had to adapt the availability of several key stakeholders simultaneously.

The use of video-conferencing tools for virtual meetings as a primary communication platform provided the necessary flexibility to bridge the geographic gap, allowing for a structured yet adaptable dialogue with each facility.

A critical factor for the success of these sessions was the pre-filled questionnaire. By presenting experts with data already harvested from scientific literature and their initial feedback, the iSAS team transformed the interview from a standard data-gathering exercise into a high-level conversation.

This approach was instrumental in driving the discussions forward, as it bypassed the need for experts to provide basic, publicly available information. Instead, it empowered them to focus on nuance, correction, and the provision of context that is rarely captured in formal documentation. This pre-filling strategy proved to be a highly effective asset, ensuring that every minute of the scheduled meetings was spent on high-value technical exchange.

The atmosphere of the interviews was defined by an exceptionally amicable, honest, and open spirit of collaboration. Despite the formal nature of the iSAS program—and the fact that many of the participating institutions are not official project members—the experts shared their insights with remarkable transparency.

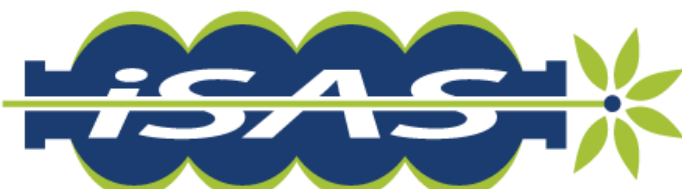

This openness yielded a "fruitful" dialogue that frequently moved beyond the initial scope, uncovering unexpected information regarding commissioning hurdles and operational constraints that were not originally anticipated.

Given the high volume of detailed information shared, the need for additional follow-up meetings became a necessity. These subsequent sessions allowed for thorough “deep-dives” into specific subsystems, ensuring that the final benchmarking report captures the operational reality of high-beta cryomodules at the selected facilities.

# 3. QUESTIONNAIRE OUTCOME

## 3.1. Section 1: Beam and RF Parameters

This area characterizes the fundamental machine requirements that dictate cryomodule architecture.

- Focus: Accelerating gradients, operating frequencies, and the choice between Continuous Wave (CW) and Pulsed modes.
- Goal: Understand how beam physics requirements drive the complexity and energy demands of the CM.

**Question 1.1**: Provide an indication of the machine size (Number of cavities and material / number of cryomodules).

| | |
|---|---|
| *ESS* | - The superconducting section in its nominal configuration is composed of 43 cryomodules: 13 Spokes, 9 Medium-beta, and 21 High-beta<br>- Each Spoke CM houses 2 superconducting cavities (26 in total), while MB and HB CM house 4 cavities (36 MB and 44 HB in total), all cavities are in bulk Niobium.<br>- The machine has a contingency space capable of housing 10 additional HB CM. |
| *SNS* | - The superconducting section is composed of 30 cryomodules: 11 Medium-beta CMs and 19 High-beta (7 of which recently installed as part of the Proton Power Upgrade)<br>- Each MB CM houses 3 superconducting cavities (33 in total), while the HB CM house 4 cavities (76 in total). Cavities are in bulk Niobium, electropolished, heat treated at 600C in vacuum for 6 hours for hydrogen degassing.<br>- After the PPU, all the penetration accesses from the klystron gallery are in use, thus there is no additional space for further cryomodules. |
| *LHC* | - The superconducting system had 4 cryomodules, 2 CM per each beam of the circular collider (M1B1 and M2B1 for the incoming beam (B1) and M1B2 and M2B2 for the outgoing beam (B2)).<br>- Each cryomodule houses 4 cavities (16 in total). The cavities are made in OFE Copper with Niobium thin film (1-2 um) coating by magnetron sputtering. |
| *XFEL* | - The main superconducting linac is composed by 96 1.3GHz accelerator modules, with 8 cavities each, for a total of 768 cavities (not considering the injector cryomodules).<br>- The cavities have 9-cells, made in bulk Niobium (deep-drawn / EB welding), chemically polished and HP rinsed, finally annealed at 800C in high vacuum. |
| *LCLS-II* | - The main Linac is composed by 35 1.3GHz accelerator modules, and 2 3.9GHz accelerator modules with 8 cavities each. Currently in the tunnel for the LCLS-II high energy (HE) upgrade 23 additional 1.3GHz modules are being installed, this will take all the remaining longitudinal space in the SLAC tunnel. The focus of the following questions will be on the 1.3GHz modules.<br>- Each module houses 8 cavities, for a total 280 1.3 GHz cavities and 16 3.9GHz cavities. The cavities have 9-cells, made in bulk Niobium with Nitrogen doping. |
| *PIP-II* | - The superconducting linac is composed by 23 SRF cryomodules of 5 different types HWR (162.5 MHz - 8 cavities in total), 2 SSR1 (325 MHz - 16 cavities in total) and 7 SSR2 (325 MHz - 35 cavities in total), 9 LB650 (650MHz - 36 cavities in total) and 4 HB650 (650MHz - 24 cavities in total). The focus of the following questions will be on the 650MHz cryomodules.<br>Cavities are in bulk Niobium. |

**Question 1.2**: What are the types of particles accelerated, the beam power and the beam energy?

| | |
|---|---|
| *ESS* | - Particles: Protons.<br>- Nominal Beam Energy: 2GeV (with the 43 CM), Average beam power: 5MW.<br>- During the first commissioning phase of the SRF linac only 27 cryomodules were installed which were able to reach 860MeV.<br>- Currently on its second commissioning phase, the SRF linac has 33 cryomodules installed capable of providing a Beam Energy:1.3GeV and Average beam power: 3MW (the RF sources for the additional 6 CM are under installation). |
| *SNS* | - Particles: Negative Hydrogen Ions (H-)<br>- Beam energy: 1.3GeV,<br>- Average beam power: 2MW.<br>Limit comes from the available power from the klystrons and the accumulator ring. |
| *LHC* | - Particles: Protons and ions.<br>- Beam energy: 6.8 TeV |
| *XFEL* | - Particles: Electrons.<br>- Beam energy: 17.5 GeV.<br>- Beam power: 600 kW |
| *LCLS-II* | - Particles: Electrons.<br>- Beam energy: 3.8-4 GeV.<br>- Beam power: 20 kW ramping up to < 250 kW |
| *PIP-II* | - Particles: Negative Hydrogen Ions (H-)<br>- Beam energy: 800MeV. Higher value than what is required in operation<br>- Average beam power on target: 1.2MW (6.7E12 particles per pulse at 20Hz) |

**Question 1.3**: What is the RF power, the current amplitude and the operating frequency?

| | |
|---|---|
| *ESS* | - Max RF power delivered by the FPC =1.1MW<br>- Current amplitude: 62.5mA<br>- Frequency: 704.42MHz MB and HB CM |
| *SNS* | - Max RF power delivered by the FPC: 550 kW<br>- Current amplitude: 35 mA<br>- Frequency: 805MHz |
| *LHC* | -Max RF power: 300 kW<br>-Current amplitude: 0.58 A (per beam)<br>-Frequency: 400.8 MHz for protons |
| *XFEL* | - Max RF power delivered by the FPC: 156 kW<br>- Current amplitude: 0.5 mA<br>- Frequency: 1.3 GHz |
| *LCLS-II* | - Max RF power delivered by the FPC: 7 kW (limit from the solid state amplifiers), 2kW in operation.<br>- Current amplitude: 0.062 mA<br>- Frequency: 1.3 GHz |
| *PIP-II* | - Max RF power delivered by the FPC: 43 kW, 32.5 kW in operation.<br>- Current amplitude: 2 mA<br>- Frequency: 650 MHz |

**Question 1.4**: Does the machine operate in continuous or pulsed mode? If pulsed, what is the duty factor?

| | |
|---|---|
| *ESS* | Pulsed operation with 4% duty cycle.<br>(Repetition rate: 14Hz and pulse width: 2.86 - 3.2 ms) |

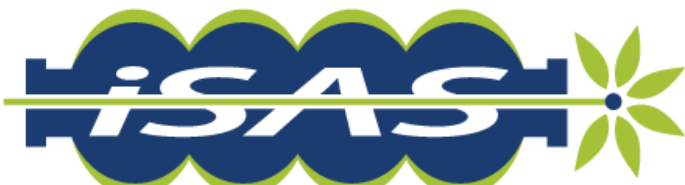

| | |
|---|---|
| *SNS* | Pulsed operation with 7.8% duty cycle.<br>(Repetition rate: 60Hz and pulse width: 1.3ms). |
| *LHC* | Continuous wave (CW) operation |
| *XFEL* | Pulsed operation with 1.37% duty cycle.<br>(Repetition rate 10Hz and pulse length 1.38ms) |
| *LCLS-II* | Continuous wave (CW) operation |
| *PIP-II* | Continuous wave (CW) operation |

**Question 1.5**: What is the cavities $Q_0$ and accelerating gradient [MV/m] in operation? What is the cavities operating temperature?

| | |
|---|---|
| *ESS* | - $Q_0$ in operation: Requirement for MB: >5E9 at 16.7 MV/m, HB: >5E9 at 19.9 MV/m. After testing cryomodules were sorted, to place cavities with lower performances in the transition regions of the linac, where the required gradient is lower for phase matching.<br>- Operating temperature: 2K (immersed in saturated He-II bath regulated at 31mbar) |
| *SNS* | - $Q_0$ in operation: 5E9. Accelerating gradient depending on the cavity to ensure stable operation of the klystrons, gradient averages between 14-15 MV/m, peak value of 16 MV/m for around 12 cavities.<br>- Operating temperature: 2.1K (immersed in saturated He-II bath regulated at 42 mbar) |
| *LHC* | - $Q_0$ in operation: 2E9 at 5.3 MV/m.<br>- Operating temperature: 4.5K (immersed in saturated He-I bath regulated at 1.3 bar) |
| *XFEL* | - $Q_0$ in operation: 1E10 at 21-23 MV/m.<br>- Operating temperature: 2K. |
| *LCLS-II* | - $Q_0$ in operation: 2.7E10 at 16 MV/m. 21MV/m required for cavity qualification in vertical testing. Nitrogen doping showed very good performances and at high gradients, reduced heat loads and cooling needs.<br>- Operating temperature: 2K. |
| *PIP-II* | - $Q_0$ in operation: 3.3E10 at 19 MV/m.<br>15% higher $Q_0$ and gradient required for the qualification of the jacketed cavities before installation in the CM.<br>- Operating temperature: 2K. |

**Question 1.6**: What are the main features of the FPC design?

Ex: Movable/fixed antenna, ceramic window (thickness, disk, cylindrical, single, double), impedance, Q loaded, orientation, maximum power in test bench, etc.

| | |
|---|---|
| *ESS* | - FPC has a fixed antenna, single ceramic alumina disk window(6mm), also works as vacuum barrier, 50 Ω impedance.<br>- Doorknob transition to allow RF matching to RF power source<br>- $Q_L$ for HB ~ 5.6 to 7.6 E5<br>- Coupler is positioned at the bottom of the CM: Vertical/ bottom orientation.<br>- Diagnostics are integrated on the FPC: 2 flanges for arc detection, 1 flange for electron pickup, 1 flange for beam vacuum monitoring (all used as diagnostics during conditioning). |
| *SNS* | - FPC with fixed antenna (9mm penetration in the beam pipe), single disk window, 50 Ω impedance.<br>- FPC design qualified for 1MW in travelling wave and 600kW in standing wave, limit on the maximum value of reflected power to the klystron. In operation, for a well-performing cavity the FPC power is limited to 700kW, maximum power that can be delivered by the klystrons.<br>- $Q_L$ ~ 7.0 E^5. The natural spread of $Q_L$ is around 10-15% of the design value, caused by waveguide circulator and klystrons. Values of $Q_L$ measured from the field probe data and phase scan.<br>- Coupler is positioned at the bottom of the CM: Vertical/ bottom orientation. |
| *LHC* | - FPC with movable antenna (60mm stroke), cylindrical/coaxial window, 75 Ω impedance. The different requirements at injection and collision with the heavy beam loading have imposed the use of a variable power coupler. The main challenge was to design a mobile RF power coupler without any sliding contacts for these very high power levels (300 kW).<br>- QL ~ 20000 @ injection (450 GeV), QL ~ 60000 @ flat-top (6.8 TeV)<br>- Coupler is positioned at the top of the CM: Vertical/ top orientation. |

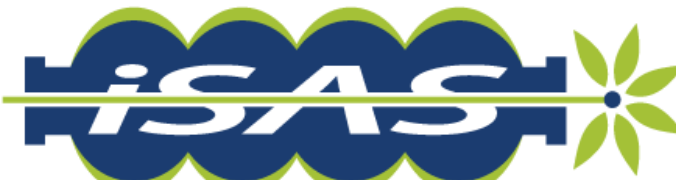

| | |
|---|---|
| *XFEL* | - FPC with movable antenna, 2 cylindrical coaxial windows (one warm and one cold), 50 Ω impedance.<br>- FPC design qualified for 1MW of pulsed power<br>- $Q_L$ ~ 4.6 10^6 (nominal) – 10^7 (to compensate for different beam loading conditions, facilitate power processing of the cavities against field emitters, upgrade to CW). The bellow in the inner conductor is compressed when the antenna is moved, it does not happen very often.<br>- Coupler has horizontal orientation, with bellows in the inner and outer conductor between the cold and the warm window. |
| *LCLS-II* | - The FPC design is the same as for XFEL cavities with movable antenna, 2 cylindrical coaxial windows (one warm and one cold), 50 Ω impedance.<br>- FPC design qualified for 7 kW in CW<br>- $Q_L$ ~ 4.6 10^6 (nominal) – 10^7<br>- Coupler has horizontal orientation, with bellows in the inner and outer conductor.<br>- To be noted: importance of cold testing FPCs (full assemblies, cold and warm parts of the antenna) at cold before installation. |
| *PIP-II* | - FPC with fixed antenna (fixed coupling in operation), single disk window, 50 Ω impedance.<br>- FPC design qualified for 50 kW in full reflection<br>- QL ~ 9.8 E^6 ± 20%. Can be fine-tuned during assembly, by rotation of the inner antenna with non-axisymmetric tip, this mechanism allows for the compensation of the manufacturing tolerances.<br>- The coupler has horizontal orientation and bellow on the outer conductor. |

**Question 1.7**: Do the cavities have High Order Modes extractors (HOM couplers)? What are the main design features? (Type, number, location)?

| | |
|---|---|
| *ESS* | The cavities do not have HOMs couplers. |
| *SNS* | Each cavity by design has 2 Niobium HOM couplers, on each side of the beam pipe, outside the helium tank (design based on the rescaling of the 1.3 GHz TESLA HOM couplers). Although they have found to be problematic (deterioration and leaks / failure at the RF feedthrough and ceramic window) and unnecessary and have been progressively removed from the cryomodule that got repaired through the years.<br>The new PPU cryomodules installed do not have HOMs. The HOM currently in place are picking between 1-8 W of RF power. |
| *LHC* | Each cavity has 4 Niobium HOM couplers.<br>2 narrowband hook type HOMs couplers (500 MHz and 536 MHz) and 2 broadband probe type HOMs couplers (760 MHz to 1300 MHz). |
| *XFEL* | Each cavity has 2 Niobium HOM extractors on the two cavity beam pipes (for the modes below the cut off frequency of the interconnecting tube), and there is one BLA between modules, every 8 cavities (for the propagating modes, 80% of the total HOM power).<br>The HOM extractors, based on the coaxial line technique, have a superconducting pick up probe and high conductivity feedthrough. The extractors are located on the two cavity beam pipes and do not extract a lot of RF power. |
| *LCLS-II* | As for the XFEL TESLA cavities, there are 2 Niobium HOM extractors on the two cavity beam pipes (for the modes below the cut off frequency of the interconnecting tube), and there is one BLA between modules, every 8 cavities (for the propagating modes). |
| *PIP-II* | The cavities do not have HOMs couplers. |

### 3.2. Section 2: Cryogenic Systems and Thermal Management

This section aims at characterizing the cryomodule cooling circuits, heat loads and the cooling capacity of the cryogenic plant.

- Focus: Machine architecture, Cooling circuit layouts, temperature and pressure levels, static and dynamic heat loads, and the specific cooling capacity of the associated cryogenic plant. Management of helium inventory and ODH risk
- Goal: Characterizing design choices for cryogenic systems.

**Question 2.1**: Characterize the machine architecture (segmented/continuous, cryogenic distribution line, quadrupoles).

| | |
|---|---|
| *ESS* | - Segmented architecture: cryomodules are stand-alone units.<br>- External cryogenic distribution line (CDS) comprising valve boxes.<br>- Cryomodules and CDS are interconnected via jumper connections at the valve boxes, both vacuum insulated, however CDS and cryomodules have different insulating vacuum systems, separated by a vacuum barrier placed on the Valve box side.<br>- Warm Focusing-Defocusing lattice with normal conducting quadrupole magnets and beam diagnostics are stand-alone systems (also called Linac Warm Units) and are placed in between cryomodules along the beam line, in the LWU there is also a particle free beam vacuum pumping unit, not part of the cryomodule.<br>- The beam line vacuum can be sectioned by all-metal gate valves placed at each extremity of the cryomodule<br>- Mobile local cleanroom (better than ISO 5) used for cryomodules assembly, installation and maintenance of the pumping groups, gate valves and actuators. After performing the connections, slow venting to monitor the vacuum and avoid unwanted contamination. |
| *SNS* | - Segmented architecture: cryomodules are stand-alone units.<br>- External cryogenic distribution line (CDS)<br>- Cryomodules and CDS are interconnected through end cans (2 per cryomodules), vacuum insulated. The vacuum cans contain the JT valves and sub-cooler HEX and connected through bayonets to the cryogenic lines in the CDS. The end cans / bayonets design was developed by Jefferson Lab for CEBAF and chosen for SNS given the familiarity with the design. End cans are made in industry, are less reliable for leaks and have worse thermal performances compared to welded connection but offer a lot of flexibility for maintenance or cryomodule removal maintaining the rest of the machine cold. Also, bayonets will not require cutting of pipes that may have been activated from radiation.<br>- Warm normal conducting quadrupole magnets and beam diagnostics are stand-alone systems and are placed in between cryomodules along the beam line<br>- The beam line vacuum can be sectioned by warm gate valves placed at each extremity of the cryomodule; these can be remotely controlled from the control room. Each cryomodule has also two manual cold gate valves, always in open position during operations, kept close only during assembly so that the vessel endplates and the warm valves could have been assembled outside the cleanroom.<br>- Each cryomodule has a dedicated pumping group, plus additional port for connection to auxiliary pumping station (SkyVac serves 12 CMs) in case of need. Usually, the pumping system is always on to allow operation of the cryomodules also in presence of helium leaks in the insulation vacuum. The leaks are mostly coming from the feedthroughs of the in-process thermal diodes in the end cans.<br>- Mobile local cleanroom was used for the assembly and maintenance of cryomodules and warm sections. After performing the connections, slow venting to monitor the vacuum and avoid unwanted contamination. |

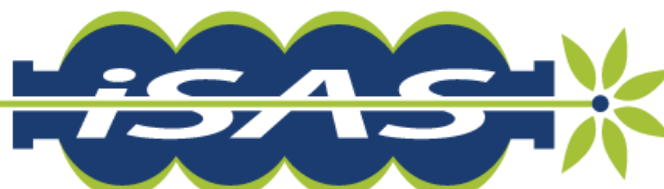

| | |
|---|---|
| *LHC* | - Segmented architecture: cryomodules are stand-alone units. Two cryomodules (M1B1 and M2B1) are accelerating the incoming beam (B1) and the other two (M1B2 and M2B2) are accelerating the outgoing beam (B2). Although each cryomodule embeds both beamlines, one going through the SRF accelerating cavities (Ø300mm) and the second (~ Ø70mm) simply crossing the vacuum vessel, given the limitation on the maximum spacing between the two beamlines.<br>- External cryogenic distribution line (QRL) comprising valve boxes.<br>- The cryomodules and the QRL are interconnected via jumper connections, cryomodules and QRL have different insulating vacuum systems, separated by a vacuum barrier placed in the jumper.<br>- Warm normal conducting magnets and beam diagnostics are stand-alone systems and are placed at the extremities of the RF system. The beam spacing (inter-axis) from the LHC sectors (194mm) is increased in the SRF section to 420mm by means of a separation/recombination dipoles system. Additionally, a undulator magnet is connected downstream the SRF cavities in each beam.<br>- The beam line vacuum can be sectioned by gate valves placed at each extremity of the cryomodule (on the beam line with SRF cavities).<br>- Each cryomodule has two dedicated pumping group beam line, placed before the gate valves on both sides (on the beam line with SRF cavities). |

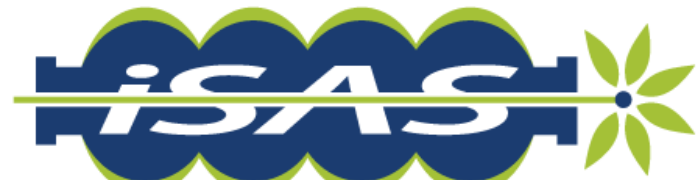

| | |
|---|---|
| *XFEL* | - Continuous architecture: the accelerating modules are integrated within a single vacuum vessel. Each module counts 8 cavities, a beam position monitor and one cold superconducting magnet with conduction cooled current leads (12.2m length), the beamline terminates at each end with an all-metal gate valve. The cold beam pipe interconnection region, between the accelerator modules gate valves, consists of a 300 mm long section that incorporates a beam line HOM-Absorber (BLA), a bellow and a vacuum pumping port.<br><br>- The cryogenic pipes are embedded in the vacuum vessel, there is one circuit for all 96 cryomodules for each of the thermal shields (5-8K, 40-80K). The 2.2 K supply line goes along the whole linac, but each string of 12 modules has a dedicated JT valve and cooldown / warmup valve (supply and level control for cryo-strings of 12 modules).<br><br>- Vacuum sectorization: every cryo-string of 12 modules (~144m), in the interconnection region, there are vacuum barriers that separates the insulating vacuum allowing independent pumping, and a connection for continuous pumping of the beam vacuum by sputter ion pumps and all-metal automatic beam valves.<br><br>- For the connection between the modules (850 mm), all the cryogenic pipes are connected by welding, to reduce the risk of leaks introduced by flanged connections. Each connection is made of 2 or 3 (for the gas return pipe and 2-phase pipe) welds, since a bellows is installed between the 2 pipe ends protruding from the modules (all welds are checked to be compliant with the EU PED regulations). For the beam line connection, a beam line absorber (BLA) is mounted on the beam line with flanged connections. A portable clean room is used for this installation, to avoid contamination. The two thermal shields are connected and one DN1100 interconnection bellow closes the interconnection region to guarantee the continuity of the insulation vacuum along the string. |

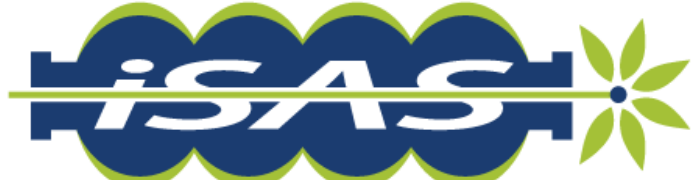

| LCLS-II | - The LCLS-II main linac cryomodule is based on the XFEL design, with modifications to accommodate operations in CW. The Linac is composed by 4 segments, with different numbers of cryomodules, separated by three warm beamline sections. Every segment has a continuous architecture: the accelerating modules form one continuous insulating vacuum space (modules' vacuum vessels bolted together with interconnect bellows).<br><br>- Each module counts 8 cavities, a quadrupole and corrector magnet package with conduction cooled current leads, and a beam position monitor (12.2m length), the beamline terminates at each end with an all-metal gate valve. The cold beam pipe interconnection region, between the accelerator modules gate valves, incorporates a beam line HOM-Absorber (BLA), a bellow and a vacuum pumping port.<br><br>- The cryogenic pipes are embedded in the vacuum vessel of each segment, there is one circuit for all the cryomodules in the segment, for each of the thermal shields (5-8K, 35-55K). The 2.2 K supply line goes along the entire segment, but every module has a dedicated JT valve and cooldown / warmup valve. A primary reason for not having an external transfer line like CEBAF or ESS is the space limitation of the SLAC tunnel, only a little over 3 meters x 3 meters cross-section.<br><br>- Vacuum sectorization: each of the linac strings, L0B (1CM), L1B (2CMs + 2x 3.9GHz CM), L2B (12CMs) and L3B (20CMs, divided into 10+10 by insulating vacuum break) has their individual insulating vacuum space, allowing independent pumping.<br><br>- The connection between the modules (850 mm), is done with an inter-module unit in-situ as for the XFEL modules. All the cryogenic pipes are connected by welding, to reduce the risk of leaks introduced by flanged connections. For the beam line connection, a beam line absorber (BLA) is mounted on the beam line with flanged connections. A portable clean room is used for this installation, to avoid contamination. The outer shell of the inter-module unit (DN1100 bellow) can slide over the vacuum vessel to allow space for making the internal joints, it is then closed to guarantee the continuity of the insulation vacuum. Bellows are used to mechanically decouple the modules and allow for thermal contraction or expansion during thermal cycling. |
|---|---|

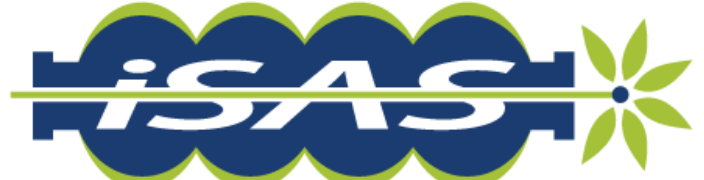

| *PIP-II* | - Segmented architecture: cryomodules are stand-alone units.<br>- External cryogenic distribution line (CDS).<br>- Cryomodules and CDS are interconnected through bayonets. Inside the vacuum vessel each cryomodule contains a localized sub-cooler heat exchanger, a JT valve and a cooldown valve.<br>The bayonets design was chosen given the familiarity with the design.<br>Bayonets are less reliable for leaks and have worse thermal performances compared to welded connection but offer a lot of flexibility for maintenance or cryomodule removal maintaining the rest of the machine cold. Even if the prototype showed higher heat loads compared to the design value, the decision of passing to welded connection was never discussed. Possibility of quick maintenance and reliability were the key drivers for the design choices.<br>The possibility of having a continuous cryostat with integrated cryogenic lines was discussed but given the challenging cavity parameters a segmented architecture was preferred to be able to maintain the linac cold and minimize the downtime in case of maintenance and/or replacement of a cryomodule. Thermal cycling could cause spread of contaminating particles for the warm interconnection regions (never particle free) to the cavities, degrading the performances over time.<br>- Warm normal conducting quadrupole magnets and beam diagnostics are stand-alone systems and are placed in between cryomodules along the beam line. In the warm interconnection region, there is also a particle free beam vacuum pumping unit.<br>- The beam line vacuum can be sectioned by warm gate valves placed at each extremity of the cryomodule. |
|---|---|

**Question 2.2**: List the temperature and pressure levels of the cryomodule (cold and warm) circuits and their stability levels.

| | |
|---|---|
| *ESS* | > **Interconnecting through the jumper connection**<<br>- Helium supply at 4.5K / 3 bar.<br>- Vapour Low Pressure (VLP) return line at ~2K / 31mbar. High pressure stability ensured with control valves in the VLP of each cryomodule (ex: quench pressure waves will not propagate). Cryoplant set to pump down to 24mbar, as design values: 1mbar Δp in the HEX and 1mbar Δp in the VLP line in the CDS.<br>- Thermal shield supply and return circuits operating between 35K-50K / 14-19 bar.<br>> **Internal Cryomodule branches**<<br>- FPC double wall cooling supply, as a branch of the helium supply line, with return at room temperature (one circuit per coupler) via a helium return line outside the CDS.<br>- Filling valve supply, as a branch of the helium supply line, used for the initial cooldown, supplies a JT valve (CV-02) which discharges to each individual cavity He tank from the bottom.<br>- JT valve supply, as a branch of the helium supply line, used for the nominal operation after cooldown, passing a sub-cooler heat exchanger (HEX), supplies a JT valve (CV-01) which discharges into the shared diphasic line, interconnecting to each cavity He Tank from the top.<br>>**Auxiliary circuits -** connect externally with the cryomodule, not through the jumper<<br>- Purge line, warm line for cleaning cycle ~1atm.<br>- Safety relief line, discharge line for pressure relief valves ~1atm. Helium is recovered at the cryoplant.<br>- Helium return line, return line for FPC cooling and return line for individual CM warmup (extinguish the inventory) ~1atm. Helium is recovered at the cryoplant.<br>- High pressure line, warm helium supply line for individual CM warmup (pressure up to 12 bar).<br>- Safety helium collector, discharge line for pressure relief ~1 atm. Helium is not recovered |
| *SNS* | - Helium supply at 4.5K / 3 bar line that feeds two loops in parallel: one to the sub-cooler heat exchanger (HEX) and JT valve, to supply 2K liquid to the cavities, and the other one to the FPCs outer conductor and Cold Warm Transitions (CWT).<br>- Vapor Low Pressure (VLP) return line at ~2.1 K / 41mbar.<br>- Thermal shield supply and return circuits operating between 35-55K / 4 bar. |

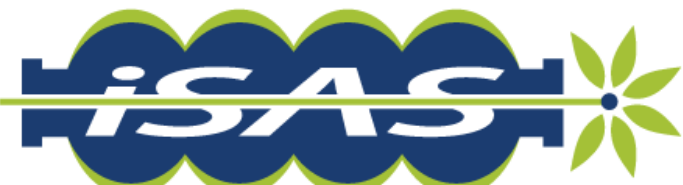

| | |
|---|---|
| *LHC* | - Liquid helium supply at 4.5K / 1.5 bar.<br>- Saturated helium vapour return at ~ 4.5 / 1.3 bar.<br>- From the top part of the helium tank (vapor box) saturated helium vapour at 4.5K is spilled with a capillary for cooling the FPC Double Walled Tube (DWT) outer conductor and the beam pipe CWT. The helium flow warms up to room temperature and is collected in the helium warm return line (WRL outside the cryogenic distribution line).<br>- From the helium tank LHe at 4.5K is spilled with capillaries for the isothermal cooling of the inner antenna of the HOM couplers (both hook and probe type), the helium vapor at 4.5K return to the helium tank vapor box. |
| *XFEL* | - Helium supply at 2.2K / 1.4 bar. The 2.2K supply comes from a centralized heat exchanger for the whole cryogenic loop of 96 cryomodules, for each cryo-string (12 modules) the 2-phase circuit and the cool down / warm up line are filled with helium through a Joule-Thompson (JT) valve and a cool down / warm up valve from the 2.2 K line.<br>- Gas Return Line (GRL – Ø300mm) return line at ~2 K / 30mbar. The gas return pipe is connected to the 2-phase pipe at each module connection and flows through the whole linac. The pressure stability of this line is better than 1%.<br>- Low temperature heat intercept supply and return circuits operating at 5-8K<br>- High temperature heat intercept supply and return circuits operating at 40-80K |
| *LCLS-II* | The cryogenic circuits have been modified from the XFEL cryomodule to account:<br>- for higher dynamic heat loads from operation in CW (80-90W per module): bigger diameter for the connection between the helium tank and the 2-phase line, bigger diameter of the 2-phase line<br>- higher slope of the SLAC tunnel (0.5%): sectorization of the 2-phase line of each module with valves for independent level control in each module.<br>- need of fast cooldown to provide high thermal gradient through 9.2K transition temperature to minimize flux trapping: dedicated cooldown / warmup circuit per module with enlarged diameter.<br>For what concerns the cryogenic circuits in the cryomodule:<br>- Helium supply at 2.2K / 1.4 bar. The 2.2K supply comes from two centralized heat exchangers, one for sector 1-2 and one for sector 3, for a total mass flowrate of ~200g/s, for each module the 2-phase circuit and the cool down / warm up line are filled with helium through a Joule-Thompson (JT) valve and a cool down / warm up valve from the 2.2 K line.<br>- Gas Return Line (GRL – Ø300mm) return line at ~2 K / 30mbar. The gas return pipe is connected to the 2-phase pipe at each module connection and flows through the whole linac.<br>- Low temperature heat intercept supply and return circuits operating at 5-8K<br>- High temperature heat intercept supply and return circuits operating at 35-55K |
| *PIP-II* | - Helium supply at 4.5K / 3 bar. It divides in three streams:<br>(I) Cooldown valve supply for the initial cooldown, which discharges to each individual cavity He tank from the bottom.<br>(II) JT valve supply used for the nominal operation after cooldown, after passing sub-cooler heat exchanger (HEX), supplies a JT which discharges to 2-phase line, interconnecting the cavity He Tank from the top.<br>(III) Low temperature heat intercept, for the thermalization of the feedthrough, returning at <15K<br>- Pumping line return line at ~2K / 20mbar. High pressure stability requirement: 0.5mbar<br>- High temperature heat intercept supply and return circuits operating between 35K-50K / 14-19 bar. |

**Question 2.3:** What is the cooling strategy for the FPC (Antenna and outer conductor)? What are the reasons for these design choices?

| | |
|---|---|
| *ESS* | - The FPC outer conductor is a Double Walled Tube (DWT) cooled with ~5K supercritical helium at 3bar. The helium is supplied to the FPCs with a dedicated line connected to the 4.5K / 3 bar helium supply line.<br>The mass flowrate is fixed to 23.1mg/s (1.38 g/min) in operation, controlled with metering valves and flow switches. Active cooling with helium seemed a better choice due to the high average RF power, compared to copper braids. Copper braids, with their thermal inertia, seems to be less reliable in accommodating dynamic changes.<br>Heaters are placed at the base of the coupler double wall cooling, and allow for the external surfaces of the coupler to be maintained above dew point<br>- The FPC antenna is water cooled until the tip, fundamental to ensure a constant coupling factor with fixed antenna, avoiding dimensional changes coming from thermal deformation that cannot be compensated.<br>The typical water flow is set to 2lpm. An additional system is set to act in emergency case where water could freeze, allowing the antenna to be vented with compressed air.<br>- The ceramic window is not actively cooled<br>The helium mass flowrate is extracting ~30W, exiting at 235K. To avoid condensation in the connection with the warm flanges of the DWT there are electrical heaters, with maximum heating power of 30W. |

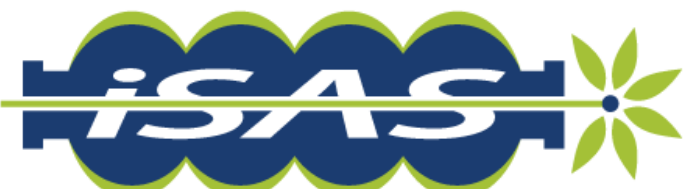

| | |
|---|---|
| *SNS* | - The FPC outer conductor is a DWT actively cooled with ~5K supercritical helium at 3bar and set mass flowrate (around 40mg/s) manually regulated with needle valves at the outlet (not accessible in operation). There is a tank inside the cryomodule, feeding all FPCs in parallel, the temperature and pressure of the helium in the tank is regulated according to the temperature measurements in the FPCs outer conductors (no independent regulation for each cavity). The maximum target heat load from each FPC to the helium bath is 2-3W. The same tank feeds in parallel as well the cooling loops for the CWT.<br><br>- The FPC antenna is water cooled on the air side, no cooling of the antenna in the vacuum side. The water cooling system is designed to remove a maximum of 200 W per FPC with a 4 g/min minimum flow<br><br>- The ceramic window is not actively cooled<br><br>The helium mass flowrate is extracting ~30W, exiting at 210K, to avoid condensation in correspondence of the warm flanges of the DWT there are electrical heaters, managed by P&ID controls. For each cryomodule the average total power is ~200W, split among the four cavities, and not always on. |
| *LHC* | The FPC outer conductor is a DWT cooled with saturated helium vapor at 4.5K, spilled with a capillary from the helium tank vapor box. The mass flowrate (~20mg/s) is not actively controlled in operation, and it is determined by:<br><br>- the pressure difference between the helium tank (1.3 bar) and the warm helium return line (1.1 bar)<br><br>- the geometry of the DWT heat exchanger and the upstream and downstream piping.<br><br>- the opening of the downstream valve in the warm helium return line. Being only one valve, it is not possible to have independent regulation for each FPC and for the CWT cooling circuits.<br><br>The FPC antenna and the ceramic window are cooled with forced air flow. |
| *XFEL* | - The FPC outer conductor is cooled with two thermal intercepts @5-8K and @40-80K. The expected static heat load from each FPC are @2K: 0.06W, @5-8K: 0.15-0.37W, @40-80K: 2.3-1.98W.<br><br>- The FPC antenna and ceramic cylindrical windows are not actively cooled.<br><br>This cooling system with thermal intercepts is realised with copper braids clamped on one side to the FPC outer conductor and on the other side to the cryogenic lines (minimizing the thermal resistance as much as possible). However, because of the fixed temperature, thickness, length and connection of the copper braids, there are some operational limits when increasing the duty cycle (compared to the nominal value of 1.37%), since the temperature in the outer conductor increases between the thermalization points. |
| *LCLS-II* | - The FPC outer conductor is cooled with two thermal intercepts @5K and @55K.<br>- The FPC antenna and ceramic cylindrical windows are not actively cooled. The heat dissipated in the antenna is partially evacuated through the ceramic.<br>No overheating experienced in the inner or outer conductor with operation in CW, since the delivered RF power is much lower compared to the maximum design value. |
| *PIP-II* | - The FPC outer conductor is cooled with two thermal intercepts @5K and @50K. The expected static/dynamic heat load from each FPC are @2K: 0.54W / 0.56 @5K: 2.2/0.04, @40-80K: 7.5/1.1W.<br>Cooling of the outer conductor with similar principle as XFEL and LCLS-II for design familiarity, the possibility of using active cooling with helium was not considered, although a comfortable margin for heat extraction through the thermalization was considered, to ensure reliable operation in CW. The position of the thermalization was studied to minimize the heat loads to the helium bath.<br>No overheating experienced in the inner or outer conductor with operation in CW, since the delivered RF power is much lower compared to the maximum design value.<br>- The FPC antenna is air cooled (5g/s).<br>- 2 x 10W electric heaters are placed on the ceramic window to prevent condensation. |

**Question 2.4:** What is the cooling scheme for the HOM couplers? (Port, antenna and jacket if applicable) and the reasons for these design choices.

| | |
|---|---|
| *ESS* | The cavities do not have HOMs couplers. |
| *SNS* | The two HOM extractors are not actively cooled with helium.<br>Heat is evacuated by conduction to the bath and through the feedthrough thermally intercepted at 2.1K and 50K. |
| *LHC* | The HOM cavity port is integrated in the helium tank, thus cooled by liquid helium at 4.5K. The HOM antennas (hook and probe types) are hollowed and isothermally cooled with helium at 4.5K. The liquid helium fills the antenna, through a capillary connected to the helium tank, it evaporates and the helium vapor (with lower density compared to the liquid) evacuates through another capillary connected to the top part of the helium tank (vapor box) filled with vapor only. The mass flowrate is not actively controlled in operation and cannot be regulated.<br>According to the design documentation, this cooling system has the capacity to evacuate the thermal losses generated by 800W / 1000W of extracted RF power, which is much more than the nominal extracted power in operation. Although, since 2023, the overheating of one probe HOM coupler caused several hours of beam downtime, the cause of overheating is still unknown since it occurs regardless of beam parameters. |
| *XFEL* | The HOM extractors are located outside the helium tank, thus not actively cooled with helium. The feedthrough is connected, with a copper braid, to the 2K-30mbar circuit, to ensure proper heat evacuation so that the Niobium probe remains in superconductive state. The BLA (positioned between two consecutive cryomodules) is connected to the 40K forward line with a copper braid, the cooling circuit has been dimensioned considering an absorbed power value of 100W, for operation mode in CW. |
| *LCLS-II* | The HOM extractors are located outside the helium tank, thus not actively cooled with helium. The feedthrough is connected, with a copper braid, to the 2K circuit, to ensure proper heat evacuation so that the Niobium probe remains in superconductive state. The BLA (positioned between two consecutive cryomodules) is connected to the 35K forward line with a copper braid.<br><br>During the acceptance test, some HOM antenna showed overheating caused by issues in the thermalization assembly. In operation, in a few cases HOM antenna overheating is experienced in correspondence of leaks in the insulation vacuum, operation is going on with additional pumping, and the cavity is operated at lower gradient. |
| *PIP-II* | The cavities do not have HOMs couplers. |

**Question 2.5**: Does the cryomodule have a thermal shield? At what temperature?

| | |
|---|---|
| *ESS* | The cryomodule has an active thermal shield at operating 30-45K / 19bar.<br><br>The thermal shield is made of aluminium alloy sheet covered by MLI, the thermal shield supply line is divided in two cryogenic circuits, welded onto the shield for a quasi-homogeneous temperature distribution (Aluminium – stainless steel bimetal weld).<br><br>The Cold/Warm Transitions, the tie-rods supporting the cavities, and the instrumentation wires are thermally anchored to the thermal shield. Jumper thermal shield connected to the thermal shield return line. |
| *SNS* | The cryomodule has an active thermal shield at 35-55K / 4bar.<br><br>The 35 K stream which first cools the supply transfer line (TL) shield, then the CM shield and thermalizations, and finally the return transfer line shield before returning to the refrigerator at 55 K. The tie-rods supporting the cavities, the cables and instrumentation wires are thermally anchored to the thermal shield.<br><br>There are two layers of MLI, one between the vacuum shell and the cold mass and the other is around the cavities and beampipe. |
| *LHC* | The cryomodule has a floating thermal shield made of several layers of MLI blankets. Since it is not actively cooled the temperature depends on the thermal equilibrium in the vacuum vessel during operation. |
| *XFEL* | The XFEL modules have two aluminium thermal shields, one at 5-8K and one at 40-80K. For both circuits, the supply line goes through the linac, without connection to the thermal shield used for the thermalization of the internal components, the shield temperature is stabilized with the return line. All cables, the FPC, the support posts and the current leads are thermalized. |
| *LCLS-II* | The LCLS-II modules have one thermal shield at 45K.<br><br>For both circuits, the supply line is used for the thermalization of the internal components, the shield temperature is stabilized with the return line. All cables, the FPC, the support posts and the current leads are thermalized. |
| *PIP-II* | The cryomodule has one active thermal shield at 40K. Even if there are two temperature levels for thermalizations: Low Temperature Thermal Source (LTTS) for intercepts at 5K-10K and High Temperature Thermal Shield (HTTS) for thermal shielding and intercepts at 40-80K.<br><br>The possibility of introducing a second thermal shield was considered but the effective reduction of the heat load to the 2K was not considered enough to justify the cost and space occupation of a second shield.<br><br>The HTTS circuit is used for thermalization and for the stabilization of the thermal shield temperature (supply split in two) while the LTTS is used for thermalizations only. Every feedthrough from ambient temperature is intercepted twice (FPCs, support posts, RF cables, tuner motor, etc.), For the HB650 cryomodules the heat load allocation is 22.9W for the LTTS and 205W for the HTTS (static + dynamic). |

**Question 2.6**: What are the maximum dynamic and static heat loads, at the different temperature levels, for which the cryomodule has been designed? And what are the current operational values?

| | |
|---|---|
| *ESS* | Design values / Average values measured in Test Stand 2 at nominal gradient:<br><br>- Static heat load @2K: 13.5W / 15W<br>- Dynamic heat load @2K: 24W / 19W<br>- Thermal Shield heat loads: 50W / 80W<br><br>Discrepancy in the thermal shield load caused by a different real emissivity of the MLI compared to the one considered at design stage. |
| *SNS* | Design values per cryomodule:<br><br>- Static heat load @2K: 28W<br><br>- Dynamic heat load @2K: 10W (MB) / 33W (HB)<br><br>- Static heat load @50K: 200W<br><br>Values measured for the entire superconductive linac: static load~540W and dynamic heat load ~400W, considering that there are 30 CM in the linac the operational heat loads are smaller than the design values. |

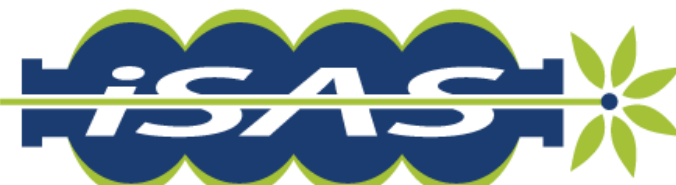

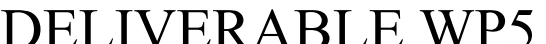

<table>
<tr><td>LHC</td><td>Design values couldn’t be retrieved, the heat loads measured in operations are:<br><br>- Static heat load @4.5K: 120W.<br>- Dynamic heat load @4.5K: 23.4W.</td></tr>
<tr><td>XFEL</td><td>Design values per cryomodule are:<br><br>- Static / Dynamic heat load @2K: 4.8W / 8.6W.<br>- Static / Dynamic heat load @8K: 13W / 2.3W.<br>- Static / Dynamic heat load @40K: 83W / 40W.<br><br>The measured XFEL static and dynamic heat loads for all the three circuits are in line with the XFEL design values. The deviations could be explained remembering that the measured values at the XFEL linac are conservative values, since together with the cryomodules value, they include the contribution of the transfer lines, feed and end caps, and string connection boxes.<br><br>- Static / Dynamic heat load @2K: 6.1W / 7.2W.<br>- Static / Dynamic heat load @8K: 8W / 2W.<br>- Static / Dynamic heat load @40K: 102W / 32W</td></tr>
<tr><td>LCLS-II</td><td>The measured values are below, and are in accordance with the predictions<br><br>- Static / Dynamic heat load @2K: 9W / 80W.<br>- Static + dynamic heat load @8K: 17W.<br>- Static + dynamic heat load @40K: 140W.<br><br>Some cryomodules experienced leaks in the insulation vacuum (source not easy to detect in such a long string) strongly increasing the static heat loads to the 2K, using up the cooling capacity margin on the 2K circuit. The GRP, with Ø300mm, has a big surface impacted by the leaks in the insulation vacuum, strongly increasing the heat load on the return line to the cryoplant.<br><br>If the temperature of the return line is too high, the sub-cooler heat exchanger is bypassed, with a resulting higher temperature in the supply line (2.3-2.5K and higher vapor fraction after expansion in the JT valve. This is compensated by supplying at higher flowrate, possible given the overhead on the cryogenic capacity of the plant at 2K.<br><br>It is planned to address the leaks to the insulation vacuum after the warmup for the long shutdown (2026) foreseen for the installation of 23 additional cryomodules for LCLS-II-HE. In this way the margin on the cooling capacity can be recovered, not to be limited on the beam energy.</td></tr>
</table>

| *PIP-II* | The final design values for the HB650 series are:<br>- Static / Dynamic heat load @2K: 19.1W (6.5W non-isothermal) / 133.6W.<br>- Static / Dynamic heat load @LTTS: 20.1W / 2.8W.<br>- Static/ Dynamic heat load @ HTTS: 190W / 15W.<br>The initial estimates were more optimistic, assuming heat loads approximately 30% lower than the current figures. With the test of the first prototype emerged that the measured values were considerably higher than the initial estimates, therefore some design modifications were integrated for the series production of all cryomodules, and the heat load budget was increased to the current reported values.<br>The first cryomodule from the series production (SSR2) will be delivered at FNAL and tested in 2026, to assess impact of the design modifications and the current heat load estimates.<br>The non-isothermal heat load at 2K represents the contribution from cryogenic distribution system (pipes in the CDS/bayonets), heat sources external from the CM. |
|---|---|

**Question 2.7**: What devices are installed to protect the equipment from overpressure (burst disks, pressure relief valves)? Where is the helium discharged? Is there a system for helium recovery?

| *ESS* | - The helium tanks circuits are classified by the European Pressure Equipment Directive, as article 4.3, corresponding to Sound Engineering Practice, with a Maximum Allowable Working Pressure of 1.04barg set pressure for the safety devices.<br>The ultimate safety devices selected were burst disks (2 units with same burst pressure setting), releasing either in the tunnel or in the Safety helium collector, when the tunnel can be accessed by the personnel.<br>-Additionally, two other pressure levels are used in order to minimise actions and venting through the rupture disks: (I) A pressure relief spring loaded valve (SV90), with an opening pressure of 0.67 bar(g) (II) A cryogenic fast acting control valve, with a set opening pressure at 0.5bar(g); both devices are releasing to the safety relief line.<br>-The insulation vacuum has a valve with a set pressure of 0.1bar(g). |
|---|---|

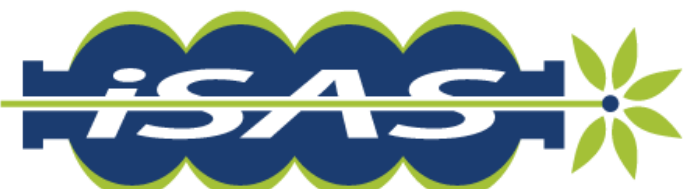

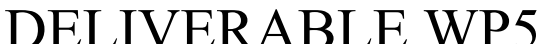

| | |
|---|---|
| *SNS* | Each cryomodule has multiple devices for equipment protection:<br>- The helium tanks circuit has a parallel relief plate (45psi) for large He discharge events venting in the tunnel, and self-sealing relief pipe (17 psi) for small He release, recovered back to the cryoplant. The relief plate is dimensioned for the worst case scenario of loss of beam vacuum, in case of rupture of the bellows in the warm section (even if protected, human error must be considered).<br>The relief plate discharging to the tunnel has an elastomeric O-ring seal and helium guard to prevent air inleaks to helium circuit at sub-atmospheric pressure in case of non-perfect tightness of the seal.<br>- The vacuum vessel has two parallel plates (or one parallel plate and a burst disk for the PPU design).<br>- The beamline has a burst disk installed just outside the cryomodule.<br>- In case of power shut down and temperature drift, SNS has storage space for half of the helium inventory, in case power is not recovered, the rest of the inventory get discharged in the tunnel through the pressure relief devices. |
| *LHC* | Each helium tank has a safety exhaust pipe with:<br>- operational depressurization device that can be activated through the control system in case of emergency.<br>- control valve with set opening pressure at 1.5 bar(a).<br>- passive safety valve with set opening pressure at 1.7 bar(a) .<br>- safety rupture disc with set opening pressure at 2.1 bar(a).<br>All safety devices discharge helium in the tunnel, and there is no system for helium recovery.<br>- The vacuum vessel has two relief plates.<br>The burst disk maintenance done in 2022 revealed that the burst disks in place had opening pressure at 2.3 bar(a), so they were replaced with new burst disk at 2.1 bar(a), corresponding to the pressure test of the cryomodules after production. After this intervention, several burst disks ruptured prematurely, thus a controlled fast-depressurizing valve with set opening pressure at 1.5 bar(a) was introduced to mitigate the pressure build up during nominal operations. Active controlled proved to be necessary, given the tight pressure range between operating pressure and MAWP. |
| *XFEL* | The helium circuits are protected from overpressure by pressure relief valves, opening at 3 bar(g) for the 2K circuit, 19 bar(g) for the 5/8K and 40/80K circuits. The safety valves in the tunnel discharge helium into warm gas collector.<br>The insulation vacuum vessel has relief plates venting in the tunnel. |
| *LCLS-II* | The safety valves are sized for the loss of insulation vacuum, while the burst disks are sized to handle the worst case scenario of loss of beam vacuum.<br>All process reliefs and rupture disks are outside of the linac tunnel and ducted outside of buildings. The insulation vacuum vessel has relief plates venting in the tunnel.<br>There is no system for helium recovery from relief valves and rupture disks.<br>Power outages have been a recurrent scenario causing helium release (6 over the lifetime of the machine); the stop of the helium supply results in pressure build up in the circuits and venting at 2bar(a). The amount of inventory released depends on the time needed to recover the power, although because of the leaks in the insulation vacuum, the warmup is quite rapid, leading to losses up to 2/3 of the total inventory on some occasions.<br>Helium lost during a power cut cannot be recompressed for storage, the cost of storing space for uncompressed helium in case of emergency is higher than the cost of purchasing additional helium according to the amount released.<br>It is currently developed a system to ensure faster connection to the power back up lines for some systems, to minimize the loss of inventory. |
| *PIP-II* | The cryomodule has one relief valve (2.05 bar(a) - 15psig) and one burst disk (4.10 bar(a) - 40psig). The burst disks are sized for the worst case scenario of beam vacuum break (heat flux of 2W/cm2).<br>The helium is released to atmosphere through a dedicated line, to prevent ODH risk in the tunnel. The maximum pressure allowed in the helium release line is 121% of the MAWP: 4.96 bar(a). There is no system for helium recovery. |

**Question 2.8**: What is the installed cooling capacity of the cryoplant at the different temperature levels (if applicable)? What is the installed liquefaction capacity?

| | |
|---|---|
| *ESS* | The installed cryogenic capacity is 3.0kW @2K, and 11.4kW @40-50K, and 9g/s liquefaction (FPC). This installed capacity allows staged operation and covers already the needs for the future upgrade of the linac.<br>The installed cryogenic capacity includes safety margins on static heat loads 1.5, and operational safety margin 1.15.<br>The cryogenic capacity includes heat loads with origin on the CDS (~27%), SPK CM (13%), MB CM (13%) and HB CM (47%). |
| *SNS* | The original SNS cryogenic system was designed with a 100% capacity margin at 2K, and a 35% margin at 35-55K. This was done to account for the uncertainty of the design, to mitigate risk to the project, and to ensure capacity for upgrade. The installed cryogenic capacity is 2400W at 2K and 8300 at 35-55K. 2K flow capacity ~125g/s and liquefaction capacity: 15 g/s.<br>The extra installed cryogenic capacity also helps in operation to decrease the sensitivity to leaks in the insulation vacuum. |
| *LHC* | The LHC SRF system (in P4) does not have a dedicated cooling plant.<br><br>The cryomodules are cooled with the cryogenic plants for to the adjacent superconducting magnet sectors (Sector 3-4 and Sector 4-5) with a dedicated cryo extension of about 10m on each side, equipped with an 80K thermal screen.<br><br>Each magnet sector cryoplant (8 in total) have cryogenic capacity of 18 kW at 4.5K and 2.4 kW at 1.8K, about 10% of the helium mass flow provided by the cryoplants of the adjacent sector is used for the SRF cavities. |
| *XFEL* | The requirements for installed cooling capacity are 1900W at 2K, 3600W at 5-8K and 24000W at 40-80K. |
| *LCLS-II* | The LCLS-II is supported by two Cryoplants, each capable of delivering 18 kW at 4.5 K or 4.0 kW at 2.0 K.<br><br>Currently, Cryoplant #1 is in operation and connected to the LINAC. Cryoplant #2 is off and will only return to operation to support an additional 23 cryomodules, which will be added to support the LCLS-II High Energy (HE) upgrade. The installed cryogenic capacity at 2K gives a considerable margin in operation, enabling operation also with leaks in the insulation vacuum.<br><br>With nitrogen doped cavities, fast cooldown is necessary across the Nb transition temperature to sweep out the trapped magnetic flux and reach the required $Q_0$ (2.7E10) at high gradients. The cryoplant is dimensioned to deliver the necessary flowrate to cooldown all the modules together (modules cannot be cooled independently).<br><br>The automatic control system has been fine-tuned to properly balance the flow and ensure a homogeneous cooldown along the sectors. Monitoring is performed using temperature sensors located on cavities 1 and 5 of each module, as well as on the HOM coupler feedthroughs, FPCs, and stepper motors. Prototype cryomodules, equipped with more extensive instrumentation, were used to guide the tuning of the control process. |
| *PIP-II* | The installed cryogenic capacity is:<br><br>- 2500W at 2K and 475 at 2K non-isothermal.<br><br>- 1680W at 4.5K and 10680W at 40-50K.<br><br>The initial estimate for the cooling capacity was smaller, based on the more optimistic estimates on the heat loads, thus compared to the initial installation, an upgrade in the cooling capacity was necessary (entirely FNAL scope).<br><br>Although, even with the upgraded cryoplant the margins are still a bit tight: 20-23% effective margin at 2K. The cryoplant is dimensioned to deliver the necessary flowrate for the fast cooldown of the cavity, to cross the Nb transition temperature with a high temperature gradient and ensure that no magnetic field is trapped in the cavities (20K/min between 45 and 4.5K – required flowrate ~100g/s per CM). |

### 3.3. Section 3: Cryomodule Design

This section aims at summarizing the main characteristics of the cryomodule internal components, and the main constraints leading to the described design solutions.

- Focus: Characterize Cryomodule sub-systems features, Cavity string assemblies, power couplers, tuners, and magnetic shielding.
- Goal: Map the constraints, design choices and integration aspects regarding the cryomodule design and its assembly sequence

**Question 3.1**: What are the main features of the helium tank?

| | |
|---|---|
| *ESS* | -The helium tank is made in titanium, with a simple tubular geometry. The tank has an extension and embeds the FPC cavity port. The tanks are connected at the bottom by the supply line and at the top by the diphasic line, partially filled with liquid helium. The di-phasic line is in titanium, with titanium bellow to avoid transitional welds on pressure equipment, titanium welds must be performed in a glovebox with protective inert gas. The cavity supply line is connected through a bimetallic weld stainless steel – titanium.<br>- The helium level is measured in a parallel circuit to the tanks, ensuring that the top of the cavities is covered by liquid and that the liquid is reaching the diphasic line, acting as a phase separator and communicating channel.<br>- The helium tank also acts as supporting structure to the cold tuning system of each cavity. |
| *SNS* | - The helium tank is made in titanium with Ti/Nb dished heads (transition alloy to enable the weld to the Nb cavity ports).<br>The tank has simple tubular geometry, 150L volume, and does not embed the cavity ports for FPC and HOMs couplers.<br>The tanks are connected at the bottom by the supply line and at the top by the bi-phasic line, partially filled with liquid helium.<br>The connections from the titanium tank to the cryogenic lines in stainless steel is made by explosion bonding.<br>- The helium level is measured by two level gauges in the furthest helium tank from the primary supply JT valve, one is used, the second is a spare.<br>- The helium tank also acts as supporting structure to the cold tuning system of each cavity. |
| *LHC* | - Helium tank made of 2mm thick stainless steel, with a complex geometry following the cavity shape. Its cross-section is cylindrical around the cavity cell and octagonal at the location of the ports, the tank has a helium vapor box / collector connected on the top part, containing only gaseous helium in nominal operation. This shape minimizes the helium inventory and cleanroom assembly time, although increasing the manufacturing complexity.<br><br>- The four helium tanks within a cryomodule are interconnected through the supply line at the bottom and the vapor boxes / collectors on top of the tanks are connected through the helium vapor return line (there is no bi-phasic line in favour of compactness).<br><br>- There is a level gauge in each helium tank with 250mm length, corresponding to 1/3 of the helium bath depth. The liquid level is regulated to fully cover the cavity (60% reading on the level gauge) while not filling the helium collector. The level control logic is very complex to deal with the level fluctuation without a biphasic line or liquid inventory buffer. |
| *XFEL* | - The helium tank is made in titanium, with bellow, so that the length can be adjusted by the mechanical tuner. All the eight helium tanks of one module are connected to the 2-phase pipe, the 2-phase pipe of each module is connected to the helium return pipe on top.<br><br>- The JT-valve reacts on any changes of the 2-K cryogenic load within the cryo-string (12 modules) by keeping the liquid level constant. In addition, electrical heaters, at the extremities of the cryo-string (in the feed / end caps and connection boxes), can be used to compensate for fast changes of the dynamic load. The electrical heaters will buffer the fast load changes and will give the refrigerator (in particular the cold compressors) time to adapt.<br><br>- At the end of each string of 12 cryomodules, the 2-phase tube is connected to a Helium reservoir, equipped with a liquid level probe and a pressure sensor. During steady state operation, the two-phase tube is filled with liquid to about 1/3 of the diameter. |
| *LCLS-II* | - The helium tank is made in titanium, with bellow, so that the length can be adjusted by the mechanical tuner. All the eight helium tanks of one module are connected to the 2-phase pipe, through a "chimney" sized for helium-II heat transport without bubbling with worst-case cavity heat load. The 2-phase pipe runs the length of each module and terminates with liquid level sensors at each end. The liquid helium level is controlled for each cryomodule in the 2-phase pipe, separate liquid level control for each cryomodule is required due to a 0.5% tunnel slope (so 6 cm per 12 meter CM).<br><br>- The JT-valve reacts on any changes of the 2-K cryogenic load within the cryo-string by keeping the liquid level constant. In addition, electrical heaters fixed outside the tank (to avoid feedthrough) can be used to compensate for fast changes of the dynamic load (ex. RF trips). The electrical heaters will buffer the fast load changes, to stabilize the inlet parameters at the cold compressor (used for the feedback loop).<br><br>- Each tank has two connections to the cooldown supply line at the bottom, in favour of a homogeneous temperature sweep from the bottom to the top for flux expulsion. |

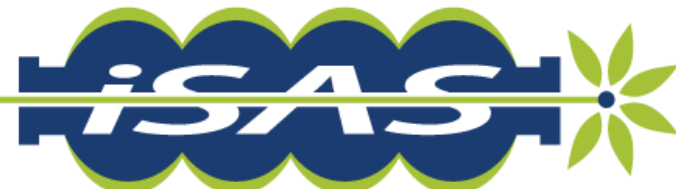

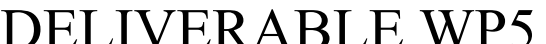

| | |
|---|---|
| *PIP-II* | - The helium tanks are made in titanium, with a simple tubular geometry with bellow, so that the length can be adjusted by the mechanical tuner.<br>- The tanks are connected at the bottom by the supply line, each tank has two connections to the cooldown supply line at the bottom, in favour of a homogeneous temperature sweep for flux expulsion during fast cooldown. The design of the diameters of the different branches of the supply line to the tanks have been studied to balance the mass flowrate in the different cavities.<br>At the top the tanks are connected by the bi-phasic line, partially filled with liquid helium. |

**Question 3.2**: How are the cavities supported and aligned in the cryomodule? What is the fixed longitudinal reference point during thermal contraction? How are vacuum forces compensated?

| | |
|---|---|
| *ESS* | - The spaceframe serves as the hanging and alignment structure for the cavities, supported with tie-rods.<br>- The spaceframe internally anchored to the vacuum vessel, shares cryomodule insulating vacuum and operates at ambient temperature.<br>- The FPC is the fixed point longitudinally and the intercavity bellows compensate the longitudinal thermal contraction.<br>- The FPC DWT thermal contraction is accounted for in the preloading of the springs of the barometric compensation system connected to the DWT flange (system accounts as well for the compensation of the vacuum forces).<br>- The alignment of each cavity is performed at room temperature and atmospheric pressure, before pumping the beam vacuum, taking into consideration all the following steps of the cavity assembly and operation. |
| *SNS* | - The spaceframe with a Nitronic rod serves as hanging and alignment structure for the cavities, which are supported with tie-rods to the space frame and fixed longitudinally on the FPC side.<br><br>- The spaceframe internally anchored to the vacuum vessel, shares cryomodule insulating vacuum and operates at ambient temperature.<br><br>- The FPC is the fixed point longitudinally, anchored to the spaceframe and Nitronic rod and the intercavity bellows compensate the cavity longitudinal thermal contraction.<br>- The FPC DWT thermal contraction is allowed by the bellow in correspondence of the warm flange and the vacuum. No need for vacuum forces compensation since they are transferred to the rigid frame.<br><br>- The alignment of each cavity is performed at room temperature, trough threaded fastener at the tie rods extremities, after that, a cap is welded on top, to stop any further movement and ensure leak tightness.<br><br>In case of cavity string disassembly, it can be done in loco, since a space was set up for CM refurbishment, the caps can be slow milled (better than grinding on activated material).<br><br>Special supports for shipping: cryomodule is shipped in vacuum monitored with vacuum gauges. It was inserted an additional bracket support on the FPC to block inner and outer conductor, avoiding stresses on the ceramic window, dimensioned for a worst-case scenario acceleration of 10g. The warm (air side) part of the FPC is not in place for shipping, it was assembled upon arrival at SNS. |
| *LHC* | Each cavity is supported by a frame anchored to the vacuum vessel through six tie rods, three on each side of the frame, the frame has two thick endplates joined by four longitudinal rods. One side of the frame is fixed to the cavity flange on the FPC side, making that the fixed longitudinal reference. While the other side is connected to the cavity through sliding blades that compensates the cavity thermal contraction and allows for the mechanical tuning.<br><br>The tie rods are accessible from the side opening of the helium vessel for the cavity alignment. |

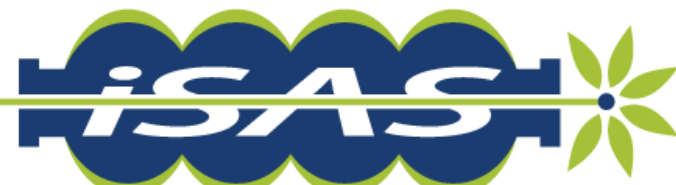

| | |
|---|---|
| *XFEL* | Each cavity is supported in two points, on the helium gas return pipe (GRP), and it is anchored in one point to an Invar rod bar providing fixed longitudinal reference on the side of the FPC port. The connections, between the cavity tank and the supports to the GRP, are made with four C-shaped clamps with needle bearings. Rolling needles drastically reduce the longitudinal friction, lateral and vertical positions are defined by reference screws. The cavities are independent from the elongation and contraction of the He GRP.<br><br>The maximum longitudinal displacement of the FPC is 2mm.<br><br>The cavities are connected by copper coated bellows without RF shielding and can be aligned individually.<br><br>The GRP is supported by three connections to the vacuum vessel. The middle post is fixed to the vacuum vessel and the invar rod, and the other two can slide in the longitudinal direction to compensate for the contraction of the GRP and the cavity string itself. |
| *LCLS-II* | Each cavity is supported in two points on the helium gas return pipe (GRP), and it is anchored in one point to an Invar rod bar providing anchoring in the beam direction (TESLA design) on the side of the FPC port. The connection, between the helium tank and the supports to the GRP, are made with four C-shaped clamps with needle bearings. The bearings drastically reduce the longitudinal friction and support the cavity vertically and laterally. The cavities are independent from the elongation and contraction of the He GRP.<br><br>The maximum longitudinal displacement of the FPC is 2mm.<br><br>The cavities are connected by copper coated bellows without RF shielding and can be aligned individually.<br><br>The GRP is supported by three connections to the vacuum vessel. The middle post is fixed to the vacuum vessel and the invar rod, and the other two can slide in the longitudinal direction to compensate for the contraction of the GRP and the cavity string itself. |
| *PIP-II* | Each cavity is supported by two support posts (one fixed and one sliding) positioned on the strong back support at ambient temperature. The support posts are made of low thermal conductivity material to limit the thermal load between the room temperature strongback and the helium tanks at 2K. The cavities installed on C-clamp assemblies with adjustable bearings that allow motion in the horizontal plane. The C-clamp closest to the FPC does not allow longitudinal motion, to fix the position of the cavity along the beam axis, while on the other clamp the longitudinal displacement is free to compensate for the cavity thermal contraction.<br><br>The cavity alignment starts in cleanroom, blocking the relative rotation between cavities (not to generate shear stresses on the bellows), final alignment of the cavities mounted in the strongback is possible before closing the thermal shield, through the adjustable bearings. HBCAMs are used for monitoring the alignments of the cavities during assembly, cooldown and operation. |

**Question 3.3**: What are the main features of the cavity tuning system?

| | |
|---|---|
| *ESS* | - Each cavity has a cold tuning system comprising a mechanical slow-tuner with internal (cold) stepper motor and a fast tuner with two piezo compensators.<br>- The CTS is placed inside the cryomodule and operates at a temperature ~20K. Each CTS weighs about 30kg and is mostly made of stainless steel.<br>- The slow-tuner combines lever arms and excentric shafts actuated by a motor gearbox and main screw assembly to provide the slow mechanical tuning action with a range of +/-3mm used for cavity elongation (pull force) corresponding to a tuning range of 650kHz and 590kHz for MB and HB cavities. However, the cavities are pre-tuned requiring about 200kHz final adjustment after cooldown. The tuner is always engaged: the preloading ensures very good linearity in all the range of motion. The tuning speed is ~200Hz/s.<br>- The Fast tuner comprised of two piezo elements act with LLRF to provide the final tuning during the pulse and beam compensation max adjustment 1kHz. |
| *SNS* | Each cavity has a mechanical slow tuner with internal (cold) stepper motor with range of ±200kHz. The fast tuners are not used, and many have been removed, the Lorentz Force Detuning (LFD) is a problem only for some cavities and is controlled by LLRF, with a feedback loop on the slow tuner. The slow tuner is kept off resonance, so that summing up the LFD the right frequency is reached. The tuning resolution is 500 steps per revolution of the stepper motor, and 1 revolution is ~1.5 mm, or ~170 Hz.<br>The tuner does not have an encoder, it is driven by the motor when RF is turned on, RF reading is needed to derive the position of the tuner, the feedback loop is regulated with RF resonance data. |
| *LHC* | Each cavity had a mechanical tuner driven by an external (warm) stepper motor.<br>The tuning range is 180kHz, with a tuning speed of 9kHz/s. |
| *XFEL* | - Each cavity has a cold tuning system comprising a mechanical slow-tuner with internal (cold) stepper motor and a fast tuner with two piezo compensators.<br>- The mechanical tuner has a tuning range of 400kHz, with a resolution of 0.47 Hz/step. The fast reactive tuner has a tuning range of 1KHz over a voltage range of ±85-100 V/DC. |
| *LCLS-II* | - Each cavity has a cold tuning system comprising a mechanical slow-tuner with internal (cold) stepper motor and a fast tuner with two piezo compensators.<br>- The mechanical tuner has a tuning range of 450kHz, with a resolution of 1-2 Hz/step. The fast reactive tuner has a tuning range of 1-3kHz, with resolution of 1Hz. |
| *PIP-II* | - Each cavity has a cold tuning system comprising a mechanical slow-tuner with internal (cold) stepper motor and a fast tuner with two piezo compensators.<br>- The mechanical has a range of 200kHz, with resolution of 2Hz/step. The fast tuner has a range of 1.2 kHz, with a resolution 0.5Hz. |

**Question 3.4**: Does the cavity have magnetic shielding? If yes, how many layers? Cold or warm? Where is it /are they located?

| | |
|---|---|
| *ESS* | The ESS cavities are shielded, with a cold magnetic shield of Cryophy outside the helium tank, with thickness of 1.5 mm covered by MLI. The aim is to limit the external static magnetic field at the Nb surface to 14mG. |
| *SNS* | The cryomodule has a warm global magnetic shield, plus there is a cold magnetic shield on each helium tank. |
| *LHC* | The LHC cavities are not sensitive to magnetic field, so no magnetic shielding is present. |
| *XFEL* | The cavities are shielded, with a cold magnetic shield outside the helium tank, with thickness of 1 mm covered by MLI. The vacuum vessel in low carbon steel did not require systematic degaussing to remove residual magnetization. |
| *LCLS-II* | The nitrogen doped cavities require magnetic shielding to reach high gradient.<br>Every cavity has two layers of cold magnetic shield outside the helium tank. The magnetic shields are made of Cryoperm 10, they are conduction / radiation cooled.<br>The aim is to limit the external static magnetic field at the Nb surface to 10mG.<br>Some modules required degaussing to improve the Q0. |
| *PIP-II* | The cavities are shielded, with a cold local magnetic shield outside the helium tank.<br>Additionally, the cryomodule has a warm global magnetic shield, anchored to the vacuum vessel. The target magnet field at the cavity surface is <5 mG, and to minimize the residual magnetic fields the magnetic permeability of the components between the local and global shields should be below 1.05 (μ <1.05). |

**Question 3.5**: Does the cryomodule have access hatches? What components can be accessed?

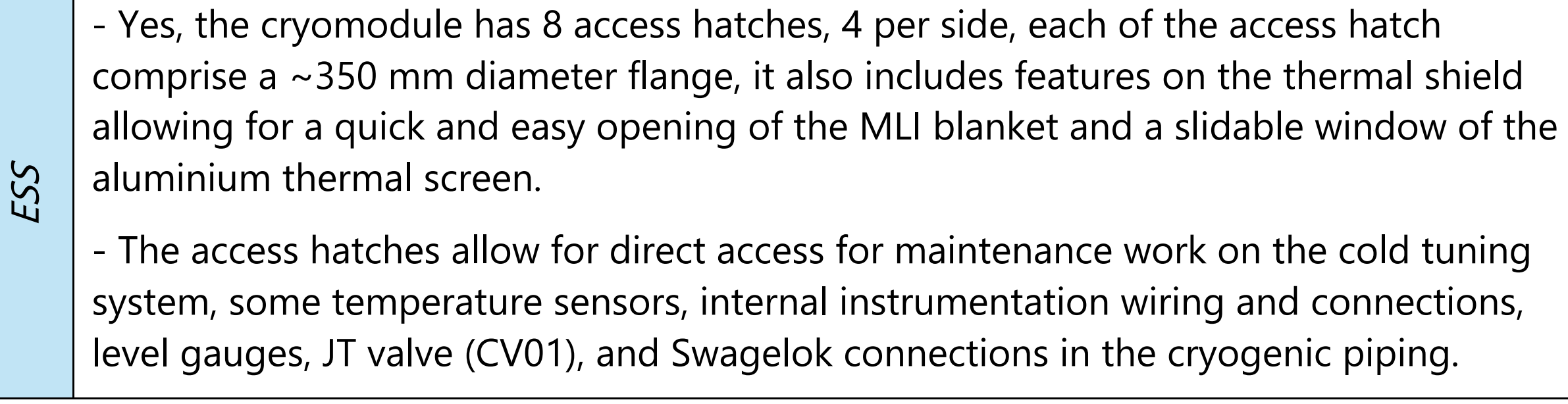

| | |
|---|---|
| *ESS* | - Yes, the cryomodule has 8 access hatches, 4 per side, each of the access hatch comprise a ~350 mm diameter flange, it also includes features on the thermal shield allowing for a quick and easy opening of the MLI blanket and a slidable window of the aluminium thermal screen.<br>- The access hatches allow for direct access for maintenance work on the cold tuning system, some temperature sensors, internal instrumentation wiring and connections, level gauges, JT valve (CV01), and Swagelok connections in the cryogenic piping. |
| *SNS* | - Yes, the cryomodule has 10 access hatches, 5 per side, sliding opening are present in the global warm magnetic shield, thermal shield and MLI blankets.<br>- The access hatches allow for direct maintenance work on the cold tuning system, some temperature sensors, internal instrumentation wiring and connections. The accessibility to the cavities and tuners at the extremities is reduced, compared to the intermediate ones, since the thermal shield and connected cryogenic circuit is wrapped around those areas. Hatches is now a standard practice for cryomodules, but the ergonomics of the repair operation shall be studied in detail (space required for tooling and relative motion) considering the space restrictions in the tunnel. |
| *LHC* | -Yes, the cryomodule has removable lateral panels all along the length.<br>A modular construction was adopted for the vacuum tanks of the cryomodule, each tank is a stainless steel cylinder, without any welds, with large lateral openings to permit easy access to the cryomodules inner components |
| *XFEL* | There are no access hatches |
| *LCLS-II* | Yes, each module has 7 access ports that are adjacent to each tuner for access to tuner motors and piezo actuators. Access ports were vital for inserting and removing shipping constraints on the power couplers to eliminate bellows fatigue failures during transport.<br>The cryomodule has the waveguides on the passage side (accessible for FPC adjustment or maintenance), while the hatches are on the wall side, thus access for maintenance work is possible but tight, given the space limitations of the tunnel (3m x 3m cross section).<br>There is also an additional plate for access to the JT and cooldown valve, that can be useful for troubleshooting and/or leak detection. |
| *PIP-II* | Yes, the cryomodule has access hatches on both sides, to access the tuner motors and piezo-actuators, temperature sensors and level/vacuum gauges.<br>There is also an additional port to access the area with bayonets connections.<br>Access hatches were fundamental for inserting and removing the extra structures / supports required for transportation, example: for the FPC and to eliminate bellows fatigue failures during transport.<br>Also, the cryomodule endcaps can be opened in case of need. |

**Question 3.6**: Are there any design compromises implemented in favour of assembly/alignment/cost which impacted the performances of the cryomodule?

| | |
|---|---|
| *ESS* | - The jumper connection to the valve box is placed at one end of the cryomodule, resulting in pressure drops imbalances in the circuits leading to/from the cavities. This design choice was preferred over a central position of the jumper, to allow the final assembly steps around the heat exchanger to be performed by removing the cryostat endplates (assembly).<br>- The two supply JT valves (CV01 and CV02) are at opposing ends of the CM, due to the lack of space near the jumper connection, and the need for finalizing the valves' assembly in the cryostat just by opening the cryomodules endcaps. The JT valve after the HEX has been placed on the other end of the cryomodule, which has consequences in final JT performance and liquid production, but allows the assembly in the cryostat when endcaps are open (space, assembly).<br>-Schedule pressure drove many decisions to go ahead without final validation / verification and possible improvements (schedule).<br>- Size of the diphasic line should have been bigger to allow for higher liquid inventory and easier liquid management during trips (cost, space)<br>- The cold tuning system should have been better thermally anchored to reduce the cooldown time (cost).<br>- The thermalizations were realized without indium foil on the copper braids, to simplify the design compromising on the resulting static heat loads (cost).<br>- The cryoplant was one of the first systems delivered at ESS, before the heat load could have been validated with experimental measurement, therefore large safety margins were included. The prototyping phase showed good agreement with design values, thus the installed cryoplant has a very large / comfortable margin for operation (schedule).<br>- PED classification played a big role in the design of the pressure safety devices. The design goal was to stay in the SEP categorization to reduce the effort in documentation, inspection and maintenance. Although this resulted in a very low MAWP and setting for the safety devices, causing strict requirements in the cryogenic controls and pressure stability (schedule). |
| *SNS* | - The He circuit thermocouples were installed inside the cryogenic piping, for accurate measurements of the helium streams. Their feedthroughs tend to fail creating helium to vacuum leaks, since the ceramic part of the feedthroughs was frequently damaged by overheating in the assembly / welding process. Due to their position in the end cans, fixes are not easy (space constraints)<br>The PPU cryomodules are equipped with surface mounted diodes or Cernox sensors.<br>- Design of end cans inherited by CEBAF (schedule, design familiarity).<br>The design was then simplified and optimized for the PPU cryomodules: simplification of piping routing, displacement of JT valves in the vacuum vessel, simplification of the support and connection to vacuum vessel.<br>- Warm beampipe gate valves were often cycled for sectorizing the beam vacuum during cryomodule maintenance to avoid the risk of cavity contamination. Because of radiation damage few gate valves failed mechanically, and some started leaking since the O-rings lost their elastic properties (Industrial components not suitable for high radiation level).<br>- High voltage bias power to prevent multipacting cannot be controlled for the individual cavities, set value for the entire cryomodule (cost).<br>- Cavity beam pipe and ports are made of Niobium with low RRR (~RRR70) resulting in normal conductive hotspots in the FPC port not immersed in helium (cost).<br>The cavities in the PPU cryomodules are fully made with high RRR niobium. |
| *LHC* | - The FPC assembly, with the motorization for the antenna, is assembled in cleanroom after inserting the cavity string in the vacuum vessel. Since the cryomodule needs to enter the cleanroom for the FPC insertion, the design was kept as minimalistic as possible, choosing a floating thermal shield made of MLI blankets, positioned outside the cleanroom, just before closing the vessel lateral openings (assembly).<br>The lack of an actively cooled thermal shield results in high static heat loads to the cold mass and fast temperature drift in case of problems with the cryogenic supply.<br>- The absence of documentation or strong discrepancy between engineering documentation and real equipment installed in the cryomodule makes it difficult in operation when problems appear (schedule).<br>- The pressure drops in the helium return line is too high and the helium supply valves are oversized (supply lines shared with SC magnets), consequently the cavity cooldown must be done slowly and carefully to avoid overpressure in the helium tank and opening of the pressure relief devices.<br>- The low MAWP of the jacketed cavity imposes strict pressure control of the cryogenic circuits and slow cooldown transients.<br>- Passive cooling system for the HOM couplers cannot compensate the overheating registered in the last years of operations causing beam downtime (cost, assembly). |
| *XFEL* | - The absence of access hatches implies that if any of the tuner cold motor fails the entire module must be removed from the tunnel. It happened during testing that one cold motor got stuck, being on one of the cavities at the extremities of the string, it was fixed from the side and the cryomodule was qualified for installation in the tunnel. |
| *LCLS-II* | - The removal of the 5-8 K thermal shield. With heat loads of about 10 W per cavity, so about 90 W per cryomodule, we decided that the cost and complication of the shield was not worthwhile.<br><br>- Air inleaks in the insulation vacuum (concentrated in the downstream half of string L3) likely caused by O-rings.<br><br>In the CM design there are Ø1.1m for the interconnection regions and Ø300mm for the 7 access ports, but the main issue seems to be one O-ring under an instrumentation flange which may have been frozen by thermo-acoustic oscillations. Small damper volumes are currently being designed to block that oscillation in the next cold run. Additionally, if needed, double seals may be introduced to improve the situation and avoid other long terms vacuum problems.<br><br>- Some cavities had to be lowered in operational gradient because of microphonics, caused by issues with the cool-down helium supply valve. The valve showed leakage, which cause hasn't been determined yet, could be the valve preload, seat not well fixed, or thermal chapping of the bellows inducing vibrations in the cavity. Depending on the location of the issue, it may be reparable in the tunnel during the long shutdown. |
| *PIP-II* | - Choice of a segmented layout that surely generates higher heat loads in favour of easier maintenance / replacements of the modules, higher machine reliability and lower risk of cavity contamination.<br><br>- Since the cryomodules are assembled at CEA and have to be transported overseas, a rigid supporting structure for the cavities was selected, to bare the unexpected loads and accelerations during the different transport phases, impacting on the other side the thermal performances. The more critical steps for the loads and acceleration are the handling operations, more than the actual transport. |

### 3.4. Section 4: Lessons learned in the commissioning phase

This section aims at collecting the insights and the knowledge gained during the commissioning phase of the facility.

- Focus: Knowledge gained during the assembly, testing, and initial installation phases.
- Goal: Document technical hurdles and breakthroughs that occurred during the commissioning phase, providing useful feedback for future sustainable designs.

**Question 4.1:** What are the main differences between the design values and the cryomodule performances measured in the test stand? What was the identified cause of these differences?

Ex.: Calculated and measured heat loads, overheating of certain components, leaks, etc.

| | |
|---|---|
| *ESS* | - The measured static heat loads are in line with the design estimates at 2K.<br>- The static heat loads on the thermal shield, however, were estimated to be on the order of 50W and measurements show results on the order of 80W. The discrepancy in these values was attributed to wrongly estimated parameters of the MLI during design.<br>- The FPC DWT requires electric heaters on the coupler flange to avoid water condensation onto the external surfaces. The required heating power is close to the limit of the electrical heater (~30W) which limits the possibility of increasing the helium flowrate in case of need.<br>- The HEX design assumes a temperature at the low pressure inlet side equal to the bath saturation temperature, experience shows slightly higher values decreasing the performance.<br>- The cold tuning system thermal anchorage to the cold mass is poor and requires about 7 days for full thermal equilibrium (~20K). |
| *SNS* | Most cryomodules though perform as expected, apart from a very small number of outliers where a cavity may be performing slightly worse than the average. Occasionally some temperature signals do not match what is expected, and the cause has been identified in miswiring during the fabrication phase.<br>- In terms of cavity performances, it is difficult to make a comparison since the test stand does not have a power limit for the RF source while in the tunnel the klystrons are limited to 500 kW (700 kW nominal but not stable).<br>- Dynamic heat loads are not directly measured in the tunnel, the opening of the cryomodule JT valve is monitored to identify the cavities that are dissipating more. It is known that local hotspots are presents in the cavity ports not immersed in helium, anyways, given the big overhead in cryogenic capacity, it is not a primary concern.<br>Note: it was experienced a progressive slipping the JT valve position, corrected with active control and monitored with linear voltage transducer.<br>- During testing the FPC $Q_L$ couldn't be accurately calculated, it is monitored in operation to ensure good matching with the design value and ensure small reflection to the circulator.<br>- Only few medium beta cryomodules were tested, so they were not sorted in the tunnel according to their performances. The sorting according to the cavity gradient was done in a second stage, knowing the cavities performances in operation, and replacing / moving around cryomodules to place high performing cryomodules in the right areas to maximize gain. |
| *LHC* | - There is limited information about design values for the heat loads to compare the values measured in operations.<br>- The target gradient is 5.3 MV/m and is achieved in each cavity even if they have been operating for more than 18 years (no issues with under-performing cavities that needs to be operated at lower gradient). $Q_0$ cannot be calculated precisely since there is no dedicated instrumentation in the cryomodule, and the high static losses are shadowing any variation in the dynamic heat loads of the single cavities.<br>It is important, at the design stage, to include the instrumentation needed to retrieve to measure precisely the quantities relevant for the control logic and the monitoring of the performances.<br>- In the last 5 years of operation no significant changes have been registered in heat loads or radiation level. The measurements taken in the past had different goals and different methods so cannot be used for comparison with current data.<br>- Documentation available only for the performance in the test stand (SM18) for a cryomodule removed from the tunnel in 2019 – long shutdown 2 (LS2) - showing worst performances than the cryomodules in operation. |

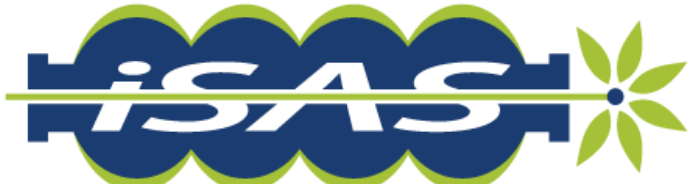

| | |
|---|---|
| *XFEL* | - It is difficult to compare the performances of the cavities in the test stand with the performances in the tunnel, measured with the evaporation method, on the full string of modules which is not very precise in monitoring the performances of each single cavity. Compared to the nominal $Q_0$ in operation of 1E10, the measures show that the modules have a $Q_{0\,f}$ 1.3-1.8E10.<br><br>- In the test stand cavities have been tested for the acceptance tests but not to the limit, and in the tunnel with 1 amplifier for 32 cavities the performances are limited by the lowest performing cavities, there is no way to measure the limits of a single cavity once it is installed in the tunnel.<br><br>- In total 105 CM were built, of which 3 for the pre series and 1-2 prototypes. Out of the 100 in the series production, only 4 were out of specs (not reprocessed for installation), so 96 were installed in the tunnel. |
| *LCLS-II* | - The main difference between the performances in the test stand and in the tunnel has been the static heat load at 2 K, for the modules in the L3 segment of the linac due to several small helium-to-insulating vacuum leaks. Additional pumping has improved but not eliminated the problem. Cryomodule replacement and/or repair may eventually provide more information and resolution. These leaks were not observed in the test stand, they appeared after installation in the tunnel, all the issues detected in the test stand (cavities, FPCs, vacuum leaks) were repaired and retested before installation. |
| *PIP-II* | We have observed discrepancies between calculated values and experimental results in terms of heat loads for the prototypes. This was a fundamental step to adjust the design and implement the lessons learned from the prototype units, before starting the production of the series. Main modification focused on:<br><br>- Improve thermal performances, reduction of heat inleaks<br><br>- Space required for the tooling around bolts and nuts<br><br>- Routing and supporting of cryogenic pipes<br><br>- Replacement of leaky instrumentation flanged feedthroughs with welded connections<br><br>- Changing in the procurement strategy, grouping subassemblies, when possible, to have fewer procurement orders and interfaces to verify (interfaces verified by the vendor – outsourcing of several quality control steps).<br><br>- No radical changes in the assembly sequence or handling, only more quality control steps were added and small details were improved.<br><br>The major issue concerned the measured static heat loads, with respects to the initial estimates, the modifications introduced were propagated in all cryomodule types.<br><br>Having a prototype phase to validate the design choices before the series production was very beneficial, also useful for the staff to gain hands-on experience (ex. Critical aspects to check, positioning the MLI, etc.). |

**Question 4.2:** Describe the key points of the experience gained in the installation phase?

examples: Mismatching component interfaces, alignment, problems in accessibility / tunnel integration / space limitations, wrong staging in the installation of the different services, collaboration with partners and external contractors, etc.

| | |
|---|---|
| *ESS* | **Preparation work before installation window in the tunnel:**<br><br>- Preparation and cutting of the cryogenic pipes (given the different jumper configuration between test stand and tunnel)<br><br>-Time consuming work such as ancillaries, water and compressed air piping can be connected before lowering the cryomodule in the tunnel.<br><br>- Check mismatching components.<br><br>- Organizational aspects: general installation plan involving different groups should be communicated, checking with RP potential access limitation, prepare toolboxes with required special tools and spare parts for simple mistakes during the work, identify periods of limited availability or of extra support needs, flexibility speeds up the sequence.<br><br>- Tool / Dashboard to track the progress.<br><br>- Ensure availability of spare parts.<br><br>- Ensure readiness of the control system for checks during commissioning.<br><br>**Optimisation in the installation sequence**<br><br>- Start by dismounting the existing set up (dummy beam pipe, LWU).<br><br>-RF waveguides system should be correctly positioned and certified before installation of cryomodules.<br><br>- Cable routing and location of the connection points should be studied to ease installation, for example by using patch panels.<br><br>- Cable trays and cable installation should be done last: limit cable damages and reduce need of retesting cables, ensure easier access while working around the CM.<br><br>**Parallel work and multiple CM installation in parallel**.<br><br>Preparation work (positioning, alignment) can be done once for multiple cryomodules. |
| *SNS* | The only issue that was encountered was related to connections (thermocouples, HV connectors to ion pumps, CCGs connectors) which were sorted out during installation and commissioning. All in all, the collaboration between different teams and groups was great, as much as important to get right.<br><br>At the beginning of operation several issues were detected in the cryomodules in the transition region between medium and high beta, requiring removal of HOM probes, maintenance of ion pumps and thermal diodes wiring. This resulted in cavity contamination and cryomodules replacements to place the contaminated cavities operating at lower gradient in non-critical locations. |
| *LHC* | - Even if the cryomodules have lateral access panel all along the vacuum vessel, there is very limited access for installation, intervention and maintenance to critical components behind the cryomodules – easy access only on the passage side.<br>- The cable routing should be smart, organized and accessible for exchanging cables or connectors that may have been degraded by vibrations or radiation damage.<br>- Important to choose the right materials to be used in a radiation environment to minimize the need of replacement.<br>- Important to define, at the design and tunnel integration phases, what are the accessibility paths and space requirements to all the components that require maintenance (ex. Burst disks, spring valves, waveguides, etc.). Adding small leaders, platforms, etc. can prevent equipment damage or safety hazards for the operators, plus, the space required to perform the maintenance should be considered.<br>- Several teams will be working on and around the cryomodules, and not everyone has full knowledge of the more sensitive components (temperature sensors, small cooling pipes, ceramic windows, bellows, etc.), thus some protections should be foreseen at design stage. |
| *XFEL* | In the tunnel the CMs are suspended from the ceiling, and all the electronics and services are installed underneath.<br>The installation sequence was carefully planned with all the teams involved and no major issues were encountered. The following points were mentioned:<br>- Starting the module installation from the end of the linac showed a strong optimization of the installation flow.<br>- In-house expertise combined with a limited number of in-kind components facilitates smooth integration and interfaces between the various teams and systems.<br>- Assembly of the FLASH accelerator was a useful practice to build the knowledge for the installation of XFEL. |
| *LCLS-II* | The installation sequence was carefully planned with all the teams involved and no major issues were encountered. The following points were mentioned:<br>- Up to two cryomodules a day were installed on their stand in the tunnel, then the mobile cleanroom was put in place to perform the particle-free operations, before welding the pipes (to avoid cutting welds in case of problems in the particle-free operations).<br>- During installation work, tasks were done in parallel when possible, having different parallel activities can facilitate the organization of tasks that cannot be performed simultaneously in the same location. For example: welding is not compatible with other work in the vicinity, or while the parts in the portable cleanroom area were getting clean, that part of the tunnel gets isolated.<br>- All the cables were pre-installed, to minimize the work to be done after the cryomodule was put in place, given the space limitations of the tunnel (3m x 3m cross section) the back of the cryomodule is not easily accessible.<br>- The interconnection regions are crowded area, so the tooling was carefully designed and accounted for at the cryomodule design stage. For example: enough space between pipes for the orbital welding machine, use of a fiberglass collars to ensure the alignment of the two pipe extremities to be welded together, use of a mock-up to test the tooling and train the staff.<br>- It was very helpful to have an internal team of specialized welders, and technicians for the particle-free installations, to ensure flexibility in the installation sequence. |
| *PIP-II* | The installation of the cryomodules in the tunnel has not started yet. The prototype units have been assembled and tested and in terms of installation of the cryomodule services no strong feedback was reported from the installation team. From the initial design phase, the interfaces were discussed with inputs from the different teams to ensure smooth installation. |

**Question 4.3:** What design modifications could have facilitated the installation and commissioning phases?

| | |
|---|---|
| *ESS* | - The set up of the cryomodule infrastructure hasn't been considered at the cryomodule design stage. Services designed independently had to be matched during installation.<br>- Integration of simple details in the cryomodule design that could ease / prepare the installation with the services (ex. Cables patch panels, etc).<br>- Engineering of the interfaces to account for the tooling (example: orbital welding machine).<br>- The CM connection to the jumper was much more complicated than expected: in the test stand the four cryogenic circuits are connected using 3 compression fittings and 1 flanged connection. This solution was selected to facilitate the multiple connections in the Test Stand valve box. However, in the tunnel welded connections are used, requiring cut and weld of new fittings. Flanged connections present inside the cryomodule (not always accessible) could be damaged in the procedure of subsequent welds and cuts.<br>- More flexible waveguide routing could simplify the alignment with the FPC doorknob.<br>- Introduction of an additional pumping group in the CM between gate valves, high range pressure gauge, and additional beam vacuum ports (pumping, flushing, measurements).<br>- More attention to the cryomodule integration in the tunnel: platforms required for installation / maintenance work, space requirements for assembly with local movable cleanroom, access to waveguide system for installation / maintenance, etc. |
| *SNS* | Surface mounted thermal diodes instead of in-process diodes, it slightly penalizes the accuracy but avoid risks of helium leaks to vacuum insulation. Already 7 cryomodules had their end cans opened because of failure of in-process diodes. This modification was done for the PPU cryomodules. |
| *LHC* | - For the LHC cryomodules the FPC assembly, waveguides and burst disks on top are very tight. Burst disks need to be changed regularly, in case of opening or every 10 years for regular maintenance, and waveguides need to be maintained every two years (dust removal to decrease the risk of arching on the air side), all these operations are very frequent but yet very critical and time consuming because of the space limitations (tight assembly, cable trays on top).<br><br>- HOM couplers are critical components requiring regular monitoring / diagnostics (especially in the last years because of overheating) but the one behind the cryomodules are very critical to reach for maintenance of for measurement. HOM couplers should be accessible and easy to monitor, a decoupled patch panel with the connectors could help at least for performing the RF measurements for diagnostics.<br><br>Still related to the HOM couplers, since many factors can compromise their functionality, instrumentation and access to raw signals is key. With the current setup, it is difficult to extract the temperature pattern, raw RF signals, frequency of the RF power extracted, so the understanding of the root cause behind the overheating is still not complete.<br><br>- Having an integrated design process where the cryomodule and the dedicated services in the tunnel are developed together, otherwise several layers of complexity will add up compromising functionality and accessibility of components (risk of damaging one system while fixing the neighbouring one). |
| *XFEL* | During installation the main difficulty was related to the welding of the modules at the interconnection regions. Stainless steel AISI 316 or 304 from different companies where not always easy to weld together.<br><br>Using German specifications from AD2000 leaflets (1.4541, 14571 for example) might help. |
| *LCLS-II* | There was no negative feedback on the cryomodule design from installation, confirmed also from the fact that the additional 23 CM, for the HE upgrade, were not modified. |
| *PIP-II* | The lessons learned, during the assembly and installation in the test stand of the prototypes, were then incorporated into improvements for the production units. |

**Question 4.4:** Describe the procedure for in situ warm/cold RF conditioning? How much time does it take? How could it be made faster (automatized?) to reduce commissioning and machine down-time?

| | |
|---|---|
| *ESS* | - The FPCs come pre-conditioned before installation in the CM. The power sweep (0 – max. value) is performed several times progressively increasing the pulse length and repetition rate. The FPC conditioning procedure is run with automated scripts and on average it takes 20h at warm and 13h at cold. Knowledgeable SRF operators can sometimes relax thresholds to allow for a faster conditioning.<br>- Cavity conditioning is done manually to react in case of exceptional behaviour, depending on the cavity it can take anything from ~1h-10h. |
| *SNS* | - After warming up a cryomodule (for maintenance) to room temperature, conditioning of FPC and cavities happens at 2K, the warm conditioning or non-resonant conditioning of the FPCs alone is not performed. The conditioning is done at low repetition rate (10Hz) while increasing the gradient while monitoring different parameters (vacuum activity in the beam line, liquid He level, all temperatures). The repetition rate is then subsequently increased in steps (20 / 30 / 60Hz) while keeping the gradient on target.<br>- The startup of the cavities after a transition from 4K to 2K takes about 4 hours (3 people powering up different sections of the SCL LINAC). An algorithm would probably speed up the process but may not be capable of dealing with unexpected cavity behaviour (multipacting, field emission, quenches). Some cavities are more difficult to bring up to the nominal gradient, it requires experience to know how to deal with the vacuum spikes, how much time to wait and when is it safe to increase the power without arching.<br>- The bias voltage power is set in the conditioning process, and it does not depend by the beam intensity |
| *LHC* | - Warm conditioning or non-resonant conditioning of the FPCs alone is not performed<br>- The conditioning is done only at cold, FPC and cavities together, the FPC has a movable antenna, changing position between beam injection and flat top, so the conditioning is done at for the two different positions of the antenna. The conditioning time depends on the maximum temperature reached and the exposure time. If the temperature rises above 40K, it takes about 6 hours of time needed for regeneration, if up to 300K, it takes 2.5weeks for 4 cryomodules.<br>The process is semi-automatized with loops of set parameters, the four CMs are conditioned in parallel. If one interlock gets triggered, the automatic process stops and an SRF expert decides if it is safe to restart the system immediately or if it better to wait longer / address the problem differently.<br>The interlocks are well adjusted (not too conservative), the process could be done faster with machine learning, to train the model with different events to verify / adjust the thresholds for interlocks, increase the resolution, understand if it is safe to restart RF immediately after or flag for the check of an operator. |
| *XFEL* | - Warm conditioning of the FPCs is performed in the test stand<br>- In the tunnel, the cold conditioning of FPC is fully automated, sequence is performed gradually increasing the power and the pulse length monitoring vacuum and electron activity. On average it takes 2weeks for the cold conditioning of all the 96 modules. |
| *LCLS-II* | - Warm conditioning or cold non-resonant conditioning of the FPCs alone is not performed, given that the maximum RF power is 7kW, in CW and full reflection<br>- In the tunnel, every cavity was conditioned separately to characterize the behaviour, the conditioning can be done in parallel on different modules, usually it takes few hours per cavity. The conditioning is performed manually with automatic interlocks, so that to ensure the adequate reaction in each scenario.<br>- For the HE upgrade, the gradient will be pushed above 16MV/m, which is the current threshold were multipacting appears, so it will take a more time for conditioning, processing the multipacting barrier. |
| *PIP-II* | - Warm conditioning or cold non-resonant conditioning of the FPCs alone is not performed,<br>- FPC and cavities are conditioned together at cold. The HB650 cavities showed soft multipacting barriers, that can be processed in the test stand and be avoided in the tunnel. |

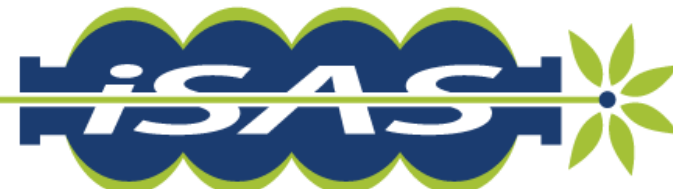

### 3.5. Section 5: Status of the machine in operation and Current Availability.

This section aims at collecting the experience in operation and the main factors limiting the machine availability.

- Focus: Factors limiting cryomodule performance, mitigation strategies, machine availability, capability for maintenance, repair and replacement.
- Goal: Use real-world operational data to ensure that energy-saving innovations do not compromise the facility's scientific output.

**Question 5.1:** What is the target availability of the machine? What are the redundant components/systems or overhead installed to compensate in case of failures? What is the real availability achieved in the last years?

Ex.: RF sources, extra capacity in the cavities accelerating gradient, overhead in the cryogenic capacity, etc.

| | |
|---|---|
| *ESS* | - The facility is undergoing the technical commissioning phase, the goal availability in operation is 95% (excluding maintenance and scheduled stops).<br>- During the commissioning phase we experienced trips of the cryoplant, which almost always require warm up 3.5K (~400mbar) or 4.3K. The process to re-establish stable conditions at 2K is time consuming: 2-4hours for cryoplant stabilization, 2 hours for pump down to 2K and 1 hour for linac stabilization. A goal for the future is to reduce the cryoplant trips and improve the recovery time.<br>- Some of the compensation measures in case of failure or trips still needs to be validated: e.g. cavity trip and gradient compensation on neighbouring cavities.<br>- In terms of instrumentation: each CM has 2 redundant helium bath pressure sensors and 2 redundant helium bath level sensors (used for process control).<br>- Additionally it has been proven that both pressure control and level control can be achieved with the cavity heaters (under development).<br>- The real operational availability is still not calculated. |
| *SNS* | - The facility is required by the DOE to maintain 90% annual availability. Significant effort is dedicated to proactive equipment monitoring and maintenance to enable early detection of potential issues and prevent their progression.<br>- The frequency of electrical trips / power shut down is twice a year. Which impact the availability according to the duration of the event and the temperature reached in the SCL (impacting the conditioning time).<br>- For the SCL, the extra installed RF capacity is equivalent to 5 cavities. Moreover, there is 1 spare per cryomodule type (1 medium beta, 1 high beta, 1 PPU).<br>- The cryoplant has additional capacity to address additional heat loads. In terms of hardware, there is a spare warm compressor, given its long delivery time in case of repair (52 weeks).<br>- In terms of instrumentation: each CM has one spare level sensors, and each helium tank has one spare internal heater installed (although there were not many failures of the heaters).<br>- All the cryomodule have vacuum recovery systems connected to preserve the insulating vacuum. |
| *LHC* | - The availability of LHC is 99% (excluding maintenance and scheduled stops).<br>- The faults related to the RF hardware in 2025 were 17, causing a total of ~17 hours of beam downtime. The main causes were overheating of one HOM connector (5 events) and full cryomodule quenches (12 events) with helium release in the tunnel through the control pressure release valve (no opening of the safety spring-loaded valves), all experienced by the same cryomodule (M1B2) foreseen to be replaced during LS3.<br>In case of HOM connector overheating, it takes 1-5 minutes to return to the nominal temperature, and the beam can restart in almost 15 minutes. For the case of the quench, it takes 30 minutes to recover the cryogenic conditions in the cryomodule, and 2-3 hours are required to set up the beam.<br>- There is one spare cavity and one spare cryomodule, the one removed from the tunnel in 2019, and helium processed in 2021 to recover the RF performances.<br>- In terms of instrumentation: each cavity has two field antennas, one is spare, since the signal is required for the control loop. It happened that one antenna cable broke, and the spare one prevented from warming up the cryomodule. There are redundant temperature and pressure sensors in the helium tanks. |

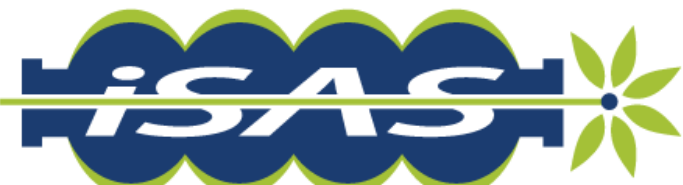

| | |
|---|---|
| *XFEL* | - The availability of the RF system is above 99% (excluding maintenance and scheduled stops). XFEL operates at different energies, and availability improve at low energy.<br>- Monday is the day for maintenance and testing. Maintenance addresses mostly the replacement of electronics that got radiation damaged.<br>In terms of installed overhead:<br>- 1-2 RF stations are available ready to automatically compensate for trips (less than 2 minutes) in case of need, even if normally klystrons are operating at lower power compared to their nominal performances to avoid trips.<br>- An Electronics Development Program is in place to qualify equipment suitable for runs longer than 10 years.<br>- Each cavity has a spare piezo actuator for the fast tuning system.<br>- The design of the XFEL-refrigerator includes redundant screw compressors for the low and high pressure stages plus an additional redundant cryoplant. Turbines as well as cold compressor cartridges can be exchanged easily in case of failure to keep the linac cold (no RF operation). |
| *LCLS-II* | - The facility has a ~90% availability. Availability not only impacted by the technical systems but also by power cuts and malfunctioning in the safety system interlocks.<br>In terms of installed overhead:<br>- Comfortable margin on the cryoplant cooling capacity at 2K.<br>- Installed energy overhead to compensate for cavities that need to be operated at lower gradient or detuned. LCLS-II operates with the Reverse Phase Operation scheme (RPO), each cavity has a fixed voltage amplitude in operation, but the phase vectors are not aligned. The total RF voltage amplitude of the linac can be increased by reducing the dephasing between the phase vectors of the cavities, this regulation can be very quick, allowing to switch off problematic cavities without resetting the beam. This regulation solution also ensure stability in the cooling capacity required, only dependent on the voltage amplitude.<br>- Each cavity has a spare piezo actuator for the fast tuning system. |
| *PIP-II* | The facility will start soon the technical commissioning phase, the goal availability in operation is 90% (excluding maintenance and scheduled stops), the real availability is still not calculated.<br>In terms of installed / planned overhead:<br>- One spare CM of each type, and a series of spare components.<br>- 20% margin on the cryoplant cooling capacity.<br>- Margin on the installed cavity gradient, although limited since the performances of the cavities are already pushed. |

**Question 5.2:** What are the mechanisms limiting the performances of the cavities? How were they addressed to continue the operations?

Ex. of limiting mechanisms: Multipacting, field emission, hotspots, contaminations, etc.

Ex. for non-nominal cavity operation: Reduced gradient, gradient compensation with other cavities, etc.

| | |
|---|---|
| *ESS* | For HB cavities in installed CMs:<br>- The main limiting mechanisms during operation are administrative limits on RF power driven through the coupler (75%), in these cases all cavities reach above nominal design values.<br>- Additionally a small number of cavities are limited by quench above nominal gradient (7%), quench at nominal gradient (7%) and slow thermal quench (~2%), field emission (~7%). |
| *SNS* | - The strongest limiting factor comes from the saturation of the klystrons, the field in some cavities is limited to 2MV/m below nominal gradient to avoid trips of the RF sources.<br>- There is a bias system for multipacting, field emission is controlled by reducing gradient. Operation into the field emission regime avoiding trips comes with experience.<br>- Small RF trips triggered by reflected power interlock, chatter-faulting, etc. can be quickly recovered within some minutes (few events per day), in case of drop in the vacuum level or arc detection in the FPC the pause is longer. For the future it is planned to develop an automatic scheme that rephase the machine automatically after a cavity trip, without dumping the beam, and slowly adapts while the tripped cavity is ramped up to its operational gradient.<br>- Arc detection fibres and photomultipliers are tested with LED every day, no aging / damage from radiation has been noticed for now.<br>- Hot spots have been dealt by warming up cryomodules during maintenance periods.<br>- Importance of logging data to monitor the evolution of the performances in time |
| *LHC* | - In the last years the main limiting mechanisms were emitted radiations or HOM overheating. The events are independent by the bunch number, beam intensity or beam type, it is recorded a temperature fluctuation and a higher RF signal picked up, probably because of the thermal expansion causing the HOM to pick up the fundamental mode.<br>- The strongest limiting factor is the maximum power available in the klystrons, especially during beam injection there is no margin on the deliverable RF power.<br>It is important at the design stage to include a margin in the RF sources to be able to deliver additional RF power in case compensation is needed. |
| *XFEL* | - During commissioning multipacting was experienced in 80% of the cavities but successfully conditioned.<br>- Field emission spotted during test and few cases of degradation in operation<br>- 23 detuned cavities, 15 since beginning of operations because of problems with quench, field emission, waveguide distribution system.<br>- The waveguide distribution is fixed, in a way that the gradient of a cavity cannot be individually decreased, so the low performing cavities are detuned, so the remaining ones can be pushed at high gradients, keeping the same total energy. For now, plasma processing is not an option, curing by processing can cause secondary emission and contaminate other cavities Adjust the waveguide distribution system can be the solution. |
| *LCLS-II* | - The main limiting mechanism is field emissions causing radiation levels above the acceptable limits. Degradation of the performances of some cavities were observed over time without any clear indication of contamination since the linac is a continuous system that always stayed cold. There is not a clear pattern for the cavities experiencing degradation, not necessarily in proximity of the interconnection regions, quicker degradation was observed only in the injector, closer to the warm part of the linac.<br>The cavity radiation is not continuously monitored (radiation sensor for the safety protocols are not design for the cavity signals), so the evolution of the degradation cannot be identified. The movable survey radiation sensors are the same used for the cryomodules acceptance test so that the measures can be compared.<br>Because of field emission some cavities are detuned and some others are operated at very low gradient ~5MV/m.<br>- Problem with stepper motor, cavity was detuned to safe position to continue operations and avoid problems in case of emergency warmup. The issue will be addressed when the linac is warm for the long shutdown.<br>- Quenches are very rare, the RF interlocks act very quick to avoid any perturbations in the cryogenic lines and cold compressor and to avoid helium release. |
| *PIP-II* | - Field emission is definitely the phenomena that has mostly limited performance of the cavities causing several reprocessing (additional costs and time). Reliability of the cleanroom facilities and procedures is fundamental. For example, the high pressure rinsing (HPR) procedure should be finely tuned to fully clean the inner complex shape of the cavity.<br>- Risk of contamination from warm interconnection regions. |

**Question 5.3:** What components can be accessed and maintained in the tunnel? (FPC, tuner, etc.)

| | |
|---|---|
| *ESS* | - All the components accessible through the access hatches: tuner motor and gear box, Swagelok connections in the cryogenic piping, instrumentation, helium bath level gauges (complex operation never exercised), helium bath pressure gauges (requires cut and rewelding).<br>- One FPC ceramic window was successfully replaced in the tunnel, profiting from the experience gained for the FPC window replacement in the test stand. Both replaced windows were located on the side of the cryomodule, in case of maintenance required by the FPC in the middle the procedure / local cleanroom shall be adapted to more strict space limitations.<br>- Some external components (pressure relief valves, fast acting pressure relief valves, burst disks) are easily maintainable / replaceable.<br>- For the cryogenic valves (both JT valves) most components can be replaced through the hatches, however if issues appear in the valve seat (of CV02), then the valve body need to be disconnected from inside the vacuum vessel, opening the endcaps. |
| *SNS* | - The components that can be maintained in the tunnel are: FPC (see next question for more details), tuner motors, cold cathode gauges (CCGs), ion pumps, valves, burst disks (on the beam line) and all the cryomodule relief valves can and have been maintained in situ in the tunnel.<br>- Importance of having platforms and protections to prevent damage to sensitive components during maintenance work.<br>- In case of access for maintenance work in the tunnel the RP team checks for hardware contamination, no air sampling or fixed time interval before starting the work on cryomodules is imposed. Only one cryomodule showed activation, first one after the warm section of the accelerator, the warm magnets are usually the more critical part. |
| *LHC* | - The FPC ceramic window cannot be replaced in the tunnel in case of failure. Although it is accessible for visual inspection and cleaning on the air side. The FPC antenna motor can be replaced (it proved to be necessary because of aging and radiation damage).<br>- The tuner has an external motor that can be replaced in case of failure without opening the vacuum vessel, only the liquid inventory is emptied for safety reasons, but no need to warm up. All the mechanical parts, cables and mechanisms inside the vacuum vessel are accessible through the lateral panels, after cryomodule warm up.<br>- By opening the lateral panels, the following elements can be accessed and replaced: the HOM and field antenna RF cables (radiation damage) and some sensors. In some locations the MLI has some Velcro openings, but for some places, for example the helium tank sensors, it would be required to cut through the MLI layer. |
| *XFEL* | None |
| *LCLS-II* | - All the components accessible through the access hatches: tuner motor and gear box, some connections in the cryogenic piping, instrumentation.<br>- The FPC can be repaired in the tunnel with a local cleanroom.<br>- For the cryogenic valves the JT and cooldown valves can be accessed through a dedicated bolted plate in the back of the cryomodule.<br>To perform any of these tasks the full linac needs to be warmed up and insulating vacuum vented. |
| *PIP-II* | All the components accessible through the access ports: tuner motor and gear box, some connections in the cryogenic piping, instrumentation.<br>The cryomodule has access ports on both sides, the ones on the passage side are meant to be used after installation in the tunnel for maintenance, in principle, there will be no need to access the ports on the wall side. |

**Question 5.4:** If any, what were the cryomodule failure modes? Were they foreseen during the design stage? Could the cryomodule be repaired in the tunnel or was a replacement needed?

| | |
|---|---|
| *ESS* | - Taking into consideration the intended long-term operation of the cryomodules, it is not to be excluded that total failure of important components such as tuners or FPC windows could occur. In both cases it is possible to perform repair of the ESS CM with the CM in the tunnel. These have been foreseen in the design.<br>- The ESS cryomodule would need replacement only in the case of beam vacuum break and cavity venting / contamination. For the case of loss of insulation vacuum, it is expected that the helium bath pressurization coming from the heat load will result in the breakage and venting through burst disks and loss of helium, without consequences on the cavity performances.<br>- No plan for preventive maintenance is foreseen at the moment, no plan for waveguide maintenance if arching problems do not appear. |

| SNS | - The cryomodule would need replacement in the case of beam vacuum break and cavity venting / contamination. Because of leaks in the cryogenic circuits, excessive leaks from the feedthroughs of the thermal diodes, several cryomodules were removed and repaired outside the tunnel, at the beginning of operation. In the last years not many cryomodules have been removed, only few for training the staff, focus was more on repairing components in the tunnel. On average it takes one month to remove and replace a cryomodule, of which one week is needed for warm up, in the meantime a dummy beampipe is installed to continue operation.<br><br>- FPCs are currently maintained in the tunnel with a standard in situ procedure that doesn't require cryomodule removal and replacement. 15 RF couplers have been replaced over the last 20 years, most of failures were not visible, detected from the vacuum data, localized in the brazing region between the ceramic window and the outer conductor. Two FPCs showed radial cracks in the ceramic window (one in the test stand and one in the tunnel) because wrong operation caused arching, one more FPC had to be removed because of overheating caused by discharges on the air side.<br><br>A mobile local cleanroom ensures laminar flow under the cryomodule for the FPC removal and replacement, lifting tools are used to help the operator, which also enters the cleanroom. This is the same procedure used for maintenance of the gate valves and components in the warm sections. Several local cleanrooms have been designed to fit in the different parts of the machine, the procedure is now standard at SNS, and it is mastered in a relative quick time interval: the cryomodule warm-up is done in parallel with the setup of the local cleanroom, once the particle count in the local cleanroom drops the FPC replacement is performed. The gasket on the cavity port cannot be damaged, and it is not removed, thus the CF flanges are closed with the same gasket in place, that proved to be leak tight even after opening and reclosing<br><br>The FPCs with very small leaks from air to vacuum can be kept in operation, lowering the cavity gradient to 1MV/m below the nominal value, avoiding discharges, with active pumping. The FPC are then replaced when the size of the leak increases.<br><br>- A few cryomodules show helium to insulating vacuum leaks, but they can be dealt with additional pumping. These leaks will be fixed if the cryomodule is warmed up for more important maintenance work. |
|---|---|
| *LHC* | - In case of failure of both pick-up antennas the cryomodules cannot be operated, but the cryomodule can be controlled with just the spare one. During Run 2 one cable of the pick-up antenna failed, operation continued until the end of the run and the cable was fixed in LS2, to restore the second signal which is useful to compare measures and monitor the quality of the cables.<br><br>- In case of failure of the HOM antennas cooling, not much can be done without removing the cryomodule and opening of the beam vacuum in cleanroom.<br><br>- During Run 1, one cavity could not work reliably above 1.2 MV/m, so that cryomodule was replaced during LS2, performances were recovered with high RF pulse processing (2015) and since then no voltage limitation was observed in any of the cavities.<br><br>- Accidental venting and contamination of one cryomodule that had to be replaced in LS2. The contamination in the removed cryomodule was solved with He processing (2021) and now it is considered a valid spare.<br><br>- Unavailability caused by rupture of the burst disks, managed through intervention in the tunnel. (2022/2023)<br><br>- Overheating of HOM coupler and full cryomodule quenches still unknown / unresolved (2023-2026). It is foreseen for LS3 to remove the cryomodule interested by the overheating and quenches, to be replaced with the valid spare, currently being prepared for testing in early 2026. |
| *XFEL* | Cryomodules cannot be repaired in the tunnel, the whole string of 96 modules needs to be warmed up and the cryomodule removed to be exchanged or repaired.<br><br>Cryomodule failure modes that would require removal, but never happened up to now, are: accidental venting/contamination of the beam line, rupture of the FPC windows, stuck slow tuner stepper motor in a position that prevents cavity detuning, too many degraded cavities that the user cannot be satisfied (16.3GeV max requested).<br><br>The cryomodule removal would cause long downtime, estimated to be several months,<br><br>given the architecture of the linac and that the process was never performed / studied in detail, since was never necessary. |
| *LCLS-II* | For the moment because of the installed overhead in energy and cooling capacity, it was possible to operate detuning the degraded cavities, and with leaks in the insulation vacuum. To ensure smooth operations for the HE upgrade, with the warm linac the issues will be addressed in the long shutdown.<br><br>Cryomodule failure modes that would require removal are: accidental venting/contamination of the beam line, rupture of the FPC windows, stuck tuner stepper motor in a position that prevents cavity detuning, too many degraded cavities, excessive losses to the insulation vacuum that will saturate the cooling capacity and limit the increase of beam energy. |
| *PIP-II* | The cryomodule failure modes that would require removal are: accidental venting/contamination of the beam vacuum, rupture of the FPC windows, excessive degradation in the cavity performances because of contamination.<br>For what concerns the FPC window cracking, if the leak in the insulation vacuum is small and the performance of the cavity are not degraded the replacement doesn't need to be immediate. Additionally, if the cavity is not needed, it can be fully detuned, and a plug with active pumping can be inserted on the air side of the FPC ceramic window to stop the leak and prevent contamination.<br>In principle the replacement of a FPC ceramic window in the tunnel could be done by a robot in a local cleanroom, to minimize the possibility of human error, but a dedicated procedure must be implemented, for the time being the possibility has been studied, adding features in the cryomodule that can help (ex. Port for nitrogen purging of the system to create an overpressure in the beam line before opening the FPC port). |

**Question 5.5:** In case of maintenance in the tunnel, is it possible to warm-up/cool-down a single cryomodule? Is the LHe inventory emptied in the adjacent cryomodules? If yes, how is the temperature rise controlled?

| | |
|---|---|
| *ESS* | - Each cryomodule has a dedicated circuit that allows for individual warm-up/cool-down. The liquid inventory is first emptied (no supply) and then the cryomodule is warmed up by static heat loads and by circulating warm helium from high pressure line and returning to the helium recovery line. The temperature rise can be controlled by adjusting the HP line supply flow.<br>- It has been demonstrated that a CM can be warmed up and cooled down while keeping the Linac at 2K (initial design considered that the linac had to be warmed up to 4K), although this is still not part of an automatic sequence, it requires intervention of an experienced operator.<br>- Interventions on a single cryomodule can be performed by first proceeding with warming a single cryomodule with the Linac kept cold and with full liquid inventory on adjacent cryomodules. However, for safety reasons it is not allowed to replace a cryomodule with the linac cold, since it requires disconnection from the valve box leaving a single valve separating the cold CDS from the ambient. To remove a cryomodule it is required a thermal cycle for the entire machine.<br>- During intervention the cryomodule gate valves are closed to limit the contamination in case of accidental beam vacuum venting. |
| *SNS* | Each cryomodule has a dedicated circuit that allows for individual warm-up/cool-down for maintenance. In case of cryomodule replacement, also the two adjacent cryomodules are warmed up to reduce contamination risks and to have two gate valves separating the cold and warm portions of the beam pipe. The liquid helium inventory in the cold cryomodules stays around 87%.<br>The transport tools are small two wheels carts with trailers, electrically controlled and moves slowly not to accidentally damage the equipment. |
| *LHC* | - For safety reasons, in case of work on or next to the LHC cryomodules, all the cryomodules are emptied from the liquid inventory (2 hours) and the cryogenic system is locked (cryogenic lock-out) while the maintenance work is performed.<br>- If the maintenance work does not require opening of the vacuum vessel (ex. Replacement of the motor of the FPC antenna), the temperature is controlled not to get above 40K to avoid reconditioning of the cavities. The cryomodules warm up to 40K in about 30 minutes, therefore, if the intervention is longer than 30 minutes, the work is performed in several stages, after about 30 minutes the work is stopped and the cryomodules are cooled again, then a lock-out is performed again and the work continues, and so on until the end.<br>- If the maintenance work requires opening of the vacuum vessel, it is possible to warm up only the cryomodule interested by the maintenance work. While for the others the liquid inventory is emptied and the temperature is controlled as mentioned above. All the beamline gate valves are closed to ensure minimum two valves separating warm and cold parts of the beam pipe.<br>- When the liquid inventory is emptied the helium is recovered and stored for the following cooldown. |
| *XFEL* | No, the whole string of 96 modules needs to be warmed up. |
| *LCLS-II* | No, the whole linac must be warmed up. |
| *PIP-II* | Yes, it is possible to warmup a single cryomodule.<br>For the warmup procedure the beam line gate valves are closed to always have a minimum of two gate valves separating the warm and cold parts or the beam pipe. |

**Question 5.6:** What is the instrumentation installed for hardware (cavities, FPC and helium tank) protection? What values are used as interlocks?

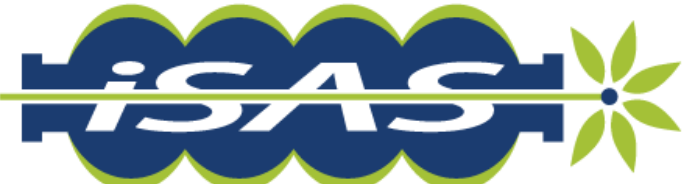

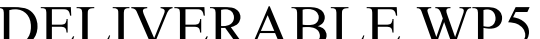

| | |
|---|---|
| *ESS* | - RF is allowed if "Cryomodule_OK" is enabled. "Cryomodule_OK" depends on:<br>- Helium level within good thresholds (covering the cavities).<br>- Helium bath pressure stable (a threshold is applied to ensure pressure stability and allow for the compensation of transient heat loads and/or modifications of the cavity gradient).<br>- Good temperatures and water flow on the FPC antenna.<br>The complexity of "Cryomodule_OK' logic might evolve with experience.<br>There are interlocks and protection on "Overfill and Overpressure" with adequate response mechanisms to mitigate the issue.<br>The tuner system is designed to disengage on warm up in case there is no capability to move the tuner motor, additionally heaters placed on the tuner mechanism allow for a localised warm-up of the mechanism to force the disengage.<br>One considered risk is the possibility of the water in the antenna to freeze. If a rapid removal of the water is required a water flushing mechanism with compressed air is in place. |
| *SNS* | Each cavity has 3 temperature sensors. The FPC has 2 temperature sensors. The cryogenic circuits have temperature and pressure sensors. The cryomodules have two LHe level sensors (one used, one spare).<br>The RF is interlocked with:<br>- the LHe level and pressure,<br>- the FPC ceramic window temperature, the water cooling flow, and arc detection system, RF signals from HOM, electron probes, klystrons and circulator<br>- the beamline vacuum.<br>The interlocks act in the milliseconds range. In the LLRF control system some interlocks can be bypassed, (ex. allowing temperatures to be a bit higher in case a cavity shows hotspots known to be non-problematic), fault counters per component to monitor the behaviour over time.<br>After the interlock of the RF system, the ramp-up procedure is defined by an RF specialist according to the cause of the trip (no automatic program).<br>The tuners cannot be automatically released, to avoid problems with the cold motor, but there are limit switches and hard mechanical stops. Cavities are preloaded (compressed) at warm temperature to reach a fixed frequency, in case of emergency warmup, the cavities go back to their preloaded condition at warm, not inducing detrimental stresses. |
| *LHC* | Each cavity has: four temperature sensors on the helium tank (2 on top and 2 on bottom – no temperature sensors on the cavity even if it would be very beneficial), one pressure sensor and one level gauge.<br><br>RF interlocks with:<br><br>- the LHe level and pressure in the tank (in case of emergency pressure readings are changing faster than temperature readings on the helium tank)<br><br>- the FPC and HOM temperature interlocks<br><br>- the cavity vacuum<br><br>When RF is switched off the beam is dumped.<br><br>The tuner can be manually released in case of fast thermal transient (feasible with 16 cavities, for bigger machines should be automatic), to make sure that there is no mechanical constraint to the cavity expansion in case of sudden transient or loss of the cryogenic conditions to avoid risk of plasticisation. |
| *XFEL* | RF is allowed if "Cryo OK" is enabled. Cryo OK depends on 2-K Helium liquid level and the 2-K vapour pressure.<br><br>In case of emergency warm up the detuning of the cavities is not automatic, the script for releasing the tuner must be launched from an operator. The cavity is detuned to safe position, loose enough to avoid plastic deformation, the constraint in the free deformation stays within the elastic limit. The tuner is not fully disengaged to save motor lifetime (~15 thermal cycles) |
| *LCLS-II* | RF is allowed if "Cryo OK" is enabled. Cryo OK depends on 2-K Helium liquid level and the 2-K vapour pressure. The current in the superconducting magnet is interlocked if its temperature goes above 2K.<br><br>RF is interlocked in the full module by cryo control, usually it takes 20-30 minutes for reaching again a stable condition to enable RF.<br><br>In case of emergency warm up, for example in the case of power loss, there is a system powered by batteries that can be used to release the cavity tuners from the klystron galleries. In nominal conditions with power on, the tuner can be easily controlled / released according to the situation. |
| *PIP-II* | RF interlocks with:<br><br>- the LHe level and pressure in the tank<br><br>- temperature sensor in the cavities<br><br>- the FPC temperature sensors and air flow in the FPC antenna<br><br>- the beamline vacuum (no vacuum gauges in the FPCs).<br><br>No automatic procedure is in place yet for tuner release in case of emergency warmup. |

**Question 5.7:** What is the current status of the facility? What is the foreseen next upgrade and what are the main challenges associated? What is the driver for the upgrade?

| | |
|---|---|
| *ESS* | - ESS accelerator is still under technical commissioning, although beam commissioning phase-1 has already taken place.<br>- The availability issues encountered during technical commissioning, are mostly related to vertical system integration. And are being solved.<br>The ESS facility is currently in technical commissioning, where the next important milestone is "Beam-on-Target" where for the first time the protons accelerated in the LINAC will hit the target and produce the first neutrons from spallation. To reach that milestone the current phase is Beam Commissioning to the proton beam dump.<br>In the current configuration the LINAC has an installed cryomodule capacity for producing 3MW proton beams (however only 2MW installed capacity for RF sources), therefore the next upgrade will be to complete the RF installation to allow for 3MW beams.<br>The following upgrades will depend on availability and demonstration of the Target systems, nevertheless the qualification of cryomodules to reach 5MW is continuing, in view of taking advantage of any possibility and opportunity in the schedule to increase the linac capability beyond 3MW. |
| *SNS* | The facility is up and running with recently installed PPU cryomodules, the linac was tested already for a beam power of 2.3 MW and currently proceeding to 2.8 MW for the Second Target Station project. Some adjustments on the hardware and software are foreseen (ex. Faster recovery after RF trips, easier monitoring)<br>The next upgrade to the facility will be the new 2K cold box, estimated in the next 5 years. The current 2K cold box is an older design and some issues that cannot be easily addressed, and newer 2K box designs would cut our downtime in the case of a trip considerably. |
| *LHC* | The LHC is currently operational, with no drop of performances observed in the performances of the cavities even if ageing issues may appear when pushing the performances. 18 more years of operations are presently foreseen with the Higher-Luminosity (Hi-Lumi) upgrade installed in LS3.<br>For the Hi-Lumi runs, the LHC cryomodules will see the following upgrades:<br>- high-efficiency klystrons are required for the higher power needs<br>- Change in the control logic (ex. Positioning of the tuner (pre detuning) in injection and flat top)<br>- Replacement of the FPC antenna and tuner motors, obsolete after long time in operation in radiation environment<br>- Reconditioning of the FPC up to 500kW in CW (maximum tested value in test box but never conditioned above 300 kW in the cryomodule), currently delivering 300 kW of RF power, will have to increase to 350 kW, hopefully without issues with arching or cooling of the DWT. Currently the cooling is not used at his full capacity, there is still margin to increase the mass flowrate by manually adjusting the return valve onto the helium warm recovery line. |

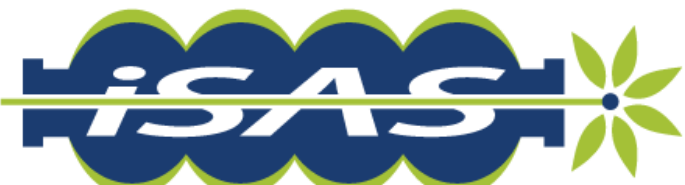

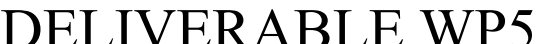

| | |
|---|---|
| *XFEL* | The XFEL facility has been operating no stop since 2017 and has 10 more years ahead. The facility just underwent its first warm up and cool down cycle for mandatory maintenance and inspection of the helium valves and to double the safety valves to avoid future warmups due to maintenance and inspection.<br>It is also being study the feasibility of increasing the duty cycle, to deliver more bunches to the users, for that, the current cryomodule designed must be reassessed to verify its capability to deal with higher dynamic heat loads. In parallel, new CW capable cryomodules (based on the XFEL design) are being developed and tested and will replace the first part of the linac (L1 and L2, i.e. 16 cryomodules), while the existing 16 cryomodules will be moved at the end of the existing L3 linac.<br>The feasibility study will be finalized in 2026, for a CDR in 2027.<br>Operation in CW will imply a maximum beam energy of 8GeV, but if this is not interesting for the users, and hybrid approach will be studied to increase the duty cycle (ex. 50%), thus having longer pulses with a maximum beam energy of 12-14 GeV.<br>Some of the design aspects limiting the increase of the duty cycle are: diameters of the cryogenic lines, connection between the helium tank and the 2-phase pipe, FPC outer conductor temperatures, possibility of cavity quench, and others. |
| *LCLS-II* | The linac is currently warm, the facility is going through a long shutdown period for the installation of 23 additional cryomodules for the LCLS-II-HE upgrade, to achieve a beam energy of 8GeV. |
| *PIP-II* | The cryomodules from the series production will be delivered at FNAL starting from summer 2026, the tunnel should be ready to start with the cryomodule installation by the end of 2026. |

# 4. SUMMARY OF THE BENCHMARKING ANALYSIS

Compiling the same questionnaire for each facility enables a consistent and structured collection of data, allowing the transition from facility-specific information to a coherent cross-facility synthesis. This dataset supports a more comprehensive and systematic analysis, making it possible to identify common trends, key differences, and best practices across installations.
Such an approach provides a solid foundation for developing a robust comparison framework focused on the technical areas of interest. Ultimately, this comparative understanding can be used to define informed guidelines and a strategic roadmap for the design and development of next-generation, more sustainable cryomodule systems.
This section highlights recurring design patterns, key divergences, and performance trade-offs that directly affect cryomodule sustainability, drawing on representative examples from the data collected in the previous stage.
By synthesizing inputs from multiple facilities, it emphasizes common engineering approaches as well as areas where design choices diverge due to differing constraints, operational priorities, or technological maturity. Particular attention is given to the compromises inherent in these choices—such as between performance, efficiency, reliability, and maintainability—and how they influence long-term sustainability.
The intent is to provide a high-level, integrative overview with the main insights without reproducing the full level of detail. For a rigorous, point-by-point comparison of specific technical aspects and quantitative parameters, the reader is referred to the previous section, where the complete set of questionnaire responses and supporting data is presented.

## 4.1. Beam & RF Parameters

The benchmarking revealed that beam structure requirements and RF parameters, including choice of frequency, fundamentally dictate the SRF and cryomodule architecture.

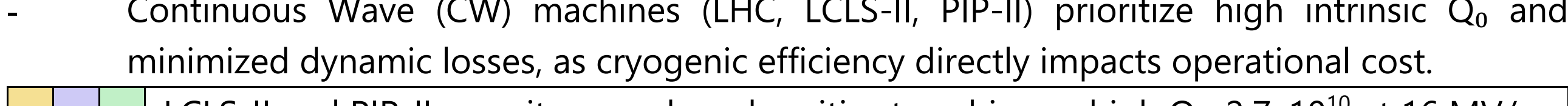

- Continuous Wave (CW) machines (LHC, LCLS-II, PIP-II) prioritize high intrinsic $Q_0$ and minimized dynamic losses, as cryogenic efficiency directly impacts operational cost.

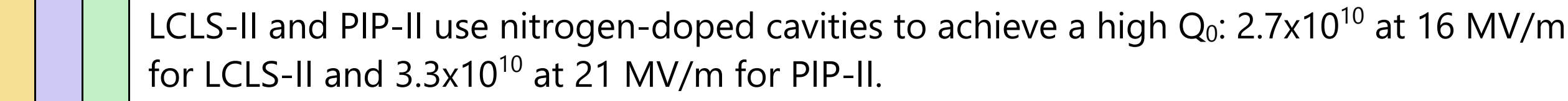

LCLS-II and PIP-II use nitrogen-doped cavities to achieve a high $Q_0$: $2.7 \times 10^{10}$ at 16 MV/m for LCLS-II and $3.3 \times 10^{10}$ at 21 MV/m for PIP-II.

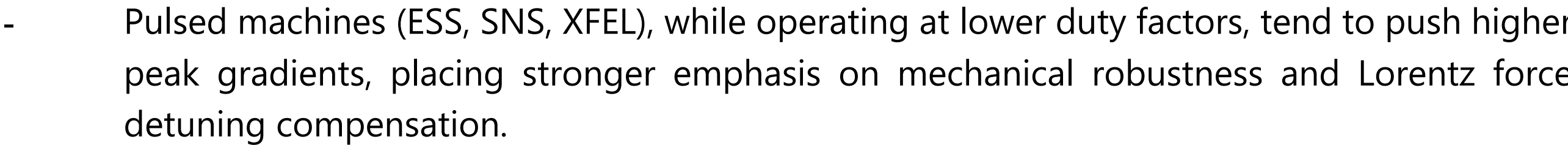

- Pulsed machines (ESS, SNS, XFEL), while operating at lower duty factors, tend to push higher peak gradients, placing stronger emphasis on mechanical robustness and Lorentz force detuning compensation.

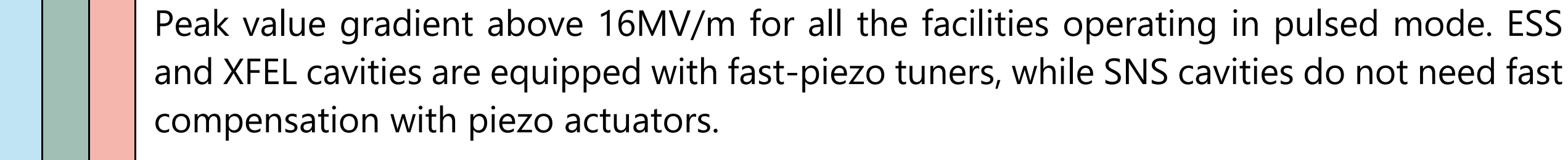

Peak value gradient above 16MV/m for all the facilities operating in pulsed mode. ESS and XFEL cavities are equipped with fast-piezo tuners, while SNS cavities do not need fast compensation with piezo actuators.

- The type of machine (circular multi-pass, linear single-pass) and particles to be accelerated also dictate different requirement (alignment, losses, beam stability thresholds, etc.)

- It was observed a very wide range in delivered FPC power amongst the various facilities. Moreover, within the limit of the six facilities considered, two main families were identified, when it comes to power coupler orientation within the cryomodule, either vertical or horizontal.

| | | | |
|---|---|---|---|
| | | | XFEL, LCLS-II, and PIP-II operates with average RF power of 2.15kW, 7kW and 35kW respectively, with FPCs designed for a peak power of 156 kW, 7 kW and 43kW respectively, at the nominal duty cycle. The FPCs have horizontal orientation and bellows in the transmission line ensure sufficient flexibility to compensate for the cold mass thermal contraction. |
| | | | ESS, SNS and LHC operates with average RF power of 43 kW, 44kW and 300 kW respectively, with FPCs designed for a peak power of 1.1 MW, 550kW and 350 kW respectively, at the nominal duty cycle. The FPCs have a vertical - bottom or top for the LHC- orientation, these designs do not have bellows in the transmission line. |

- In proton linear accelerators, the beam exhibits lower sensitivity to trapped HOM perturbations; consequently, HOM couplers are non-essential at low currents and may be omitted to reduce design complexity. This does not apply to circular colliders, where factors such as current amplitude, beam emittance, beam stability limits, and bunch intensity must be considered.

| | | | |
|---|---|---|---|
| | | | ESS and PIP-II cavities do not have HOM couplers. SNS originally included HOM couplers, however operational experience indicated they were not required to achieve goal performances, and given occasional leaks and failures, they have been progressively removed during cryomodule repairs. |
| | | | Circular colliders like the LHC, and electron machines like XFEL and LCLS-II require the integration of HOM couplers in the cavity design. |

## 4.2. Cryogenic Architecture and Thermal Management

Significant diversity exists in cooling circuit segmentation, temperature levels, and thermal shield configurations. However, common trends emerged:

- A multi-level temperature architecture (2K / 4.5/ 5-8/ 10K / 35–80 K) is standard across all facilities.

| | | | | | |
|---|---|---|---|---|---|
| | | | | | A three-level architecture (typically 2 K, 4.5–10 K, and 35–80 K) is the standard for optimizing thermal efficiency across all facilities. |
| | Due to assembly compromises, the LHC uses a single temperature level (4.5 K) and a floating thermal shield made of MLI blankets, which significantly increases static heat loads to the helium bath | | | | |

- Static heat loads represent a substantial fraction of total cryogenic consumption, for cavities operating at low duty cycles. For cavities operating in CW instead, the cryogenic losses are driven by the dynamic heat loads. Except for LHC where the total cryogenic consumption is dominated by the static losses, given the low number of cryomodules (4) stringent optimization of the thermal design was not prioritized. Minimizing heat loads becomes critical for high energy machines with high number of cryomodules.

| | |
|---|---|
| | The percentage of static heat load, compared to the total heat load to the cavity bath, is indicated per facility: ESS 44.1%, SNS 73.7%, LHC 83.7%, XFEL 45.8% |
| | The percentage of static heat load, compared to the total heat load to the cavity bath, is small for facilities operating in CW, LCLS-II 10%, PIP-II 12.5%. |

- FPC inner conductor cooling plays a major role in designs with fixed antenna, to prevent thermal deformation from changing the coupling.
  For the outer conductor, it was observed a preference for the active cooling with helium, for operation with high average RF power (>40/50 kW). Active cooling is considered more flexible, than fixed thermal intercepts, as it allows adjustable heat removal via the mass flow rate (within design limits) and provides more uniform cooling, reducing the risk of hotspots. On the other side, thermal intercept cooling - realised with clamped copper braids – may show some limits in the maximum heat that can be extracted, given the fixed temperature, thickness, length and connection of the copper braids.

| | |
|---|---|
| | Active helium cooling is considered more flexible for high average RF power operations or for designs qualified for higher values of peak power, ESS, SNS and LHC use ~5 K helium to cool the FPC outer conductors. |
| | Facilities with lower average power, such as XFEL, LCLS-II, and PIP-II, the outer conductor is cooled with fixed temperature thermal intercepts (copper braids) connected to the cryogenic lines. |

- Cryogenic distribution strategy (continuous vs segmented) strongly affects maintainability and failure recovery capability.
- Facilities with more modular cryogenic segmentation demonstrate improved operational flexibility, particularly for maintenance scenarios. For the case of segmented architecture, the cryomodule connection to the external cryogenic distribution line – welded or bayonets – impacts the possibility of removing a cryomodule without warming up the entire linac.

| | |
|---|---|
| | Accelerating modules are integrated into a single vacuum vessel string. In case of maintenance the entire linac must be warmed up. Depending on the component failing, for the case of LCLS-II, the maintenance work may be performed through the access hatches. For XFEL instead, the modules do not have access hatches and must be removed from the tunnel for maintenance. |
| | Cryomodules are stand-alone units, with access hatches, connected to the external distribution line via bayonets. Depending on the component failing, the maintenance work may be performed through the access hatches, otherwise the cryomodule can be removed maintaining the rest of the machine cold, considered a significant operational advantage for long term sustainability. |

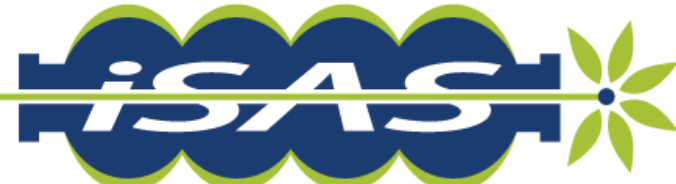

| | | |
|---|---|---|
| | | Cryomodules are stand-alone units, with access hatches, connected to the external distribution line via jumper with welded connections. Depending on the component failing, the maintenance work may be performed through the access hatches, otherwise the cryomodule can be removed, warming up the linac (requirement of minimum two valves separating cold cryogenic lines from the environment). |

- The benchmarking confirms that cryogenic design is one of the dominant drivers of overall sustainability.

### 4.3. Cryomodule Mechanical Design and Integration

While cavity technology is broadly similar (elliptical SRF cavities), integration philosophies differ:

- Variations in cavity supporting schemes, vacuum forces compensation, longitudinal references and alignment strategies have been identified.

| | | |
|---|---|---|
| | | Cavities are supported by the Helium Gas Return Pipe (GRP), which acts as the main structural backbone and an Ivar rod that provides longitudinal anchoring. |
| | | Cavities are supported and aligned with spaceframe at ambient temperature connected to the vacuum vessel. For the SNS cryomodule, a Nitronic rod is added for longitudinal anchoring. |
| | Cavities are supported by a frame anchored to the vacuum vessel through six tie rods. | |
| | Cavities are supported by two support posts (one fixed, one sliding) mounted on a strongback support at ambient temperature. | |

- Helium tank design influences mechanical stiffness and microphonics sensitivity.

- Variations in Magnetic shielding implementation (cold vs warm, single vs multi-layer) with impact on achievable $Q_0$ have also been identified

| | | | | | |
|---|---|---|---|---|---|
| | | | | | The introduction of a cold magnetic shield outside the helium tank is a common practice for all the facilities using Niobium cavities. LCLS-II cavities have two layers of cold magnetic shielding. |
| | | SNS and PIP-II cryomodules have also a warm global magnetic shield. | | | |
| | | LCLS-II and PIP-II cavities rely as well on fast cooldown across the Nb transition temperature to sweep out the trapped magnetic flux and reach the required $Q_0$. | | | |
| | LHC cavities in copper with Nb coating are not sensitive to magnetic field. | | | | |

- Some facilities reported design compromises made to facilitate assembly, reduce cost, or drivers related to schedule, with related impact on operational performance. These trade-offs will continue to be analysed in WP5.

- The integration of a second thermal shield, when in presence of the 8-10K heat intercept, was evaluated but not always considered convenient in comparison to its required capital cost. However, including operational costs and sustainability considerations hints at reconsidering the advantage of a second shield. This justifies further studies.

| | | |
|---|---|---|
| | | For LCLS-II and PIP-II, given the availability of a low temperature heat intercept, it was evaluated the possibility of having a second thermal shield. Although, the effective reduction of the heat load to the 2K was not considered sufficient to justify the cost, complexity and space occupation of a second shield. |
| | In the XFEL cryomodule instead, there are two thermal shields, resulting in very minimal heat loads to the cavity bath. | |

- The evolution toward maintainability is evidenced by the inclusion of access hatches, specifically to allow in-tunnel maintenance of tuners, sensors and connectors. It appears now a standard to include access hatches in the cryomodule design, however, among the facilities considered, the XFEL design does not include access hatches, meaning that any component failure would require removal of the entire module, although this has not been necessary to date.

### 4.4. Commissioning Experience

Important to note is that the facilities used for benchmarking find themselves at very different phases. Some have not fully entered the commissioning phase; others have decades of experience.
An important theme across facilities is the aim to minimise discrepancy between "as-designed" and "as-operated" performance. Factors contributing to performance deviation include:

- Different experiences have been described regarding performance limitations
- Cryoplant limitations during commissioning have been mentioned
- Significant differences in heat loads from design to operation, become complex to manage when the cryoplant design is fixed (e.g. legacy from existing machines, or due to early definition of cryoplant cryogenic capacities).
- Integration aspects and mismatching interfaces, with emphasis in continuous quality controls, have been identified as important lessons to reduce re-work.
- Commissioning strategies, including various approaches for RF conditioning and automation, in conjunction with experts' oversight, significantly reduce machine downtime.

From the installation phase, it became clear that accurate planning and strong in-house expertise are key to ensuring a smooth sequence of operations and sufficient flexibility to accommodate adjustments in the installation sequence. Maximizing preparation work on the cryomodules prior to installation is highly beneficial, as it minimizes the actions in the tunnel, where space is limited. A carefully designed layout—including cable and services routing and access paths for cryomodule maintainable components—is essential. An integrated design approach, in which the cryomodule and its associated services are developed together, is strongly recommended, such coordination ensures compatibility between systems and avoids late-stage conflicts. In contrast, treating the cryomodule and its services as separate design streams often leads to the accumulation of additional layers of complexity. This can result in suboptimal routing, difficult access conditions, and reduced overall functionality, ultimately impacting both operational reliability and long-term maintainability.

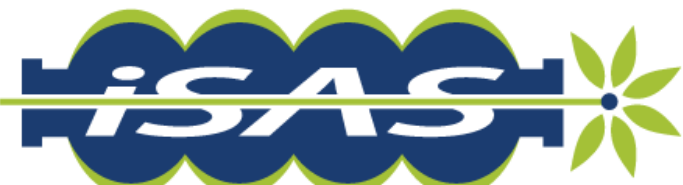

### 4.5. Operational Reliability and Availability

Long-term availability emerged as a critical sustainability metric. Facilities emphasized:

- Redundancy of internal equipment, critical for operation
- Different experiences have been described regarding cavity performance degradation and strategies for compensation with the remaining cavities.
- Modular cryogenic systems and replacement strategies
- Importance of accessibility was emphasized, more specifically for couplers and tuners.
- Failure modes reported include coupler degradation, tuner actuator failure, and vacuum-related incidents.
- The ability or not to isolate a single cryomodule for maintenance without affecting adjacent units strongly impacts overall machine uptime.
- The analysis identified that cryoplant trips remain a major bottleneck for availability; for example, at ESS, a single trip can lead to a 5-7 hour recovery process to re-establish stable 2K conditions.
- To maintain high availability, benchmarking shows that facilities now design for extra RF capacity (e.g. SNS, LCLS-II) to compensate for individual cavity failures without beam degradation.
- The analysis also identified that the cryoplant design includes safety margins to compensate for uncertainty in the estimation of cryomodule static and dynamic heat loads, as a result operating machines are running with cryoplants with considerable operating margins.

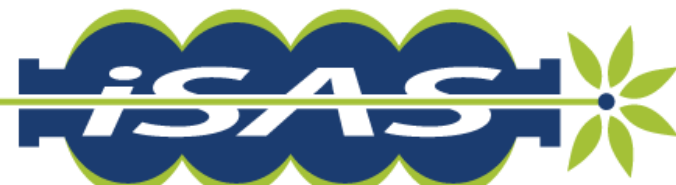

# 5. CONCLUSION

This benchmarking exercise has provided a dual perspective by bridging practical European implementation experience with international best practices. The findings reinforce that beam and RF parameters are the fundamental design drivers, while cryogenic efficiency remains the dominant determinant of overall sustainability. Common trends, such as the adoption of multi-level temperature architectures and the importance of modular cryogenic segmentation for improved maintenance and failure recovery, have emerged as industry standards.

The study also underscores that high availability and long-term reliability are primary metrics for sustainability, often requiring deliberate design-for-maintainability and the inclusion of redundancy for critical components. While unavoidable trade-offs exist between simplicity and optimization, the data gathered through this "fruitful" and transparent collaboration with global experts now forms the strategic and technical backbone for the WP5 roadmap. Ultimately, these insights establish a consolidated knowledge base that will guide the development of a new generation of high-efficiency, resilient, and future-proof cryomodules.

# 6. ACKNOWLEDGMENTS

The authors wish to extend their sincere gratitude to everyone who participated in the benchmarking process, specifically the experts from world-leading facilities across the globe who contributed to this endeavour. It is noteworthy that while most of these institutions are not formal members of the iSAS project, their involvement was secured through the dedicated effort and collaborative spirit of the iSAS WP5 team. The amount of information shared by the participating facilities have exceeded our expectations, triggering many discussions of great importance to be also included in scope of the study.

CERN (LHC, Geneva, Switzerland)

- Vittorio Parma
- Karin Canderan
- Katarzyna Turaj
- Krzysztof Brodzinski
- David Smekens

ESS (European Spallation Source, Lund, Sweden)

- Paolo Pierini
- Nuno Elias
- Emma Simpson
- Henry Przibliski
- Marek Skiba

SNS (Spallation Neutron Source, Oak Ridge, USA)

- John Mammouser
- Paolo Pizzol

XFEL (European X-ray Free Electron Laser, Hamburg, Germany)

- Serena Barbanotti

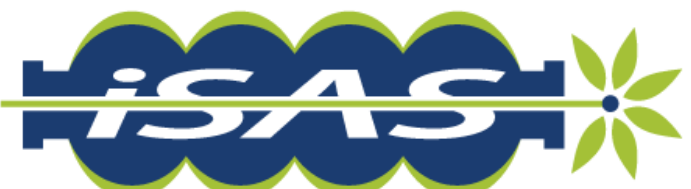

- Kay Jensch
- Julien Branlard
- Joshua Einstein-Curtis

LCLS-II (Linac Coherent Light Source II, SLAC, California, USA)

- Thomas Petersen
- Sebastian Aderhold

PIP-II (Fermi National Laboratory, Batavia, Illinois, USA)

- Donato Passarelli
- Vincent Roger

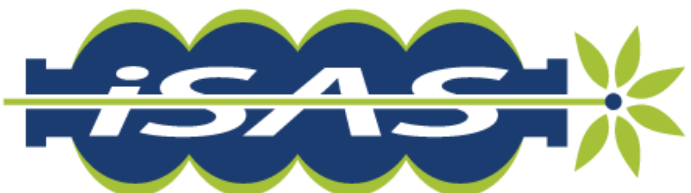